
\font\elevenrm=cmr10 scaled \magstephalf
\font\eleveni=cmmi10 scaled \magstephalf
\font\elevensy=cmsy10 scaled \magstephalf
\font\elevenbf=cmbx10 scaled \magstephalf
\font\elevensl=cmsl10 scaled \magstephalf
\font\eleventt=cmtt10 scaled \magstephalf
\font\elevenit=cmti10 scaled \magstephalf
\skewchar\eleveni='177
\skewchar\elevensy='60

\def\elevenpoint{%
   \def\rm{\fam0\elevenrm}%
   \textfont0=\elevenrm \scriptfont0=\sevenrm \scriptscriptfont0=\fiverm
   \textfont1=\eleveni  \scriptfont1=\seveni  \scriptscriptfont1=\fivei
   \textfont2=\elevensy \scriptfont2=\sevensy \scriptscriptfont2=\fivesy
   \textfont3=\tenex    \scriptfont3=\tenex   \scriptscriptfont3=\tenex
   \textfont\itfam=\elevenit  \def\it{\fam\itfam\elevenit}%
   \textfont\slfam=\elevensl  \def\sl{\fam\slfam\elevensl}%
   \textfont\ttfam=\eleventt  \def\tt{\fam\ttfam\eleventt}%
   \textfont\bffam=\elevenbf \scriptfont\bffam=\sevenbf
      \scriptscriptfont\bffam=\fivebf   \def\bf{\fam\bffam\elevenbf}%
   \smallskipamount=3pt plus1pt minus1pt
   \medskipamount=6pt plus2pt minus2pt
   \bigskipamount=13.4pt plus4pt minus4pt
   \normalbaselineskip=13.4pt
   \setbox\strutbox=\hbox{\vrule height9.4pt depth4pt width0pt}%
   \normalbaselines  \rm}


\font\twelverm=cmr12
\font\eightrm=cmr8
\font\sixrm=cmr6
\font\twelvei=cmmi12
\font\eighti=cmmi8
\font\sixi=cmmi6
\font\twelvesy=cmsy10 scaled \magstep1    
\font\eightsy=cmsy8
\font\sixsy=cmsy6
\font\twelvebf=cmbx12
\font\eightbf=cmbx8
\font\sixbf=cmbx6
\font\twelvesl=cmsl12
\font\twelvett=cmtt12
\font\twelveit=cmti12
\skewchar\twelvei='177 \skewchar\eighti='177 \skewchar\sixi='177
\skewchar\twelvesy='60 \skewchar\eightsy='60 \skewchar\sixsy='60

\def\twelvepoint{%
   \def\rm{\fam0\twelverm}%
   \textfont0=\twelverm \scriptfont0=\eightrm \scriptscriptfont0=\sixrm
   \textfont1=\twelvei  \scriptfont1=\eighti  \scriptscriptfont1=\sixi
   \textfont2=\twelvesy \scriptfont2=\eightsy \scriptscriptfont2=\sixsy
   \textfont3=\tenex    \scriptfont3=\tenex   \scriptscriptfont3=\tenex
   \textfont\itfam=\twelveit  \def\it{\fam\itfam\twelveit}%
   \textfont\slfam=\twelvesl  \def\sl{\fam\slfam\twelvesl}%
   \textfont\ttfam=\twelvett  \def\tt{\fam\ttfam\twelvett}%
   \textfont\bffam=\twelvebf \scriptfont\bffam=\eightbf
      \scriptscriptfont\bffam=\sixbf   \def\bf{\fam\bffam\twelvebf}%
   \smallskipamount=4pt plus1pt minus1pt
   \medskipamount=8pt plus3pt minus3pt
   \bigskipamount=15pt plus5pt minus5pt
   \normalbaselineskip=15pt
   \setbox\strutbox=\hbox{\vrule height10.3pt depth4.7pt width0pt}%
   \normalbaselines  \rm}

\def\tenpoint{%
   \def\rm{\fam0\tenrm}%
   \textfont0=\tenrm \scriptfont0=\sevenrm \scriptscriptfont0=\fiverm
   \textfont1=\teni  \scriptfont1=\seveni  \scriptscriptfont1=\fivei
   \textfont2=\tensy \scriptfont2=\sevensy \scriptscriptfont2=\fivesy
   \textfont3=\tenex    \scriptfont3=\tenex   \scriptscriptfont3=\tenex
   \textfont\itfam=\tenit  \def\it{\fam\itfam\tenit}%
   \textfont\slfam=\tensl  \def\sl{\fam\slfam\tensl}%
   \textfont\ttfam=\tentt  \def\tt{\fam\ttfam\tentt}%
   \textfont\bffam=\tenbf \scriptfont\bffam=\sevenbf
      \scriptscriptfont\bffam=\fivebf   \def\bf{\fam\bffam\tenbf}%
   \smallskipamount=3pt plus1pt minus1pt
   \medskipamount=6pt plus2pt minus2pt
   \bigskipamount=12pt plus4pt minus4pt
   \normalbaselineskip=12pt
   \setbox\strutbox=\hbox{\vrule height8.5pt depth3.5pt width0pt}%
   \normalbaselines  \rm}


\font\ninerm=cmr9
\font\sixrm=cmr6
\font\ninei=cmmi9
\font\sixi=cmmi6
\font\ninesy=cmsy9
\font\sixsy=cmsy6
\font\ninebf=cmbx9
\font\sixbf=cmbx6
\font\ninesl=cmsl9
\font\ninett=cmtt9
\font\nineit=cmti9
\skewchar\ninei='177 \skewchar\sixi='177
\skewchar\ninesy='60 \skewchar\sixsy='60

\def\ninepoint{%
   \def\rm{\fam0\ninerm}%
   \textfont0=\ninerm \scriptfont0=\sixrm \scriptscriptfont0=\fiverm
   \textfont1=\ninei  \scriptfont1=\sixi  \scriptscriptfont1=\fivei
   \textfont2=\ninesy \scriptfont2=\sixsy \scriptscriptfont2=\fivesy
   \textfont3=\tenex    \scriptfont3=\tenex   \scriptscriptfont3=\tenex
   \textfont\itfam=\nineit  \def\it{\fam\itfam\nineit}%
   \textfont\slfam=\ninesl  \def\sl{\fam\slfam\ninesl}%
   \textfont\ttfam=\ninett  \def\tt{\fam\ttfam\ninett}%
   \textfont\bffam=\ninebf \scriptfont\bffam=\sixbf
      \scriptscriptfont\bffam=\fivebf   \def\bf{\fam\bffam\ninebf}%
   \smallskipamount=3pt plus1pt minus1pt
   \medskipamount=6pt plus2pt minus2pt
   \bigskipamount=11pt plus4pt minus4pt
   \normalbaselineskip=11pt
   \setbox\strutbox=\hbox{\vrule height7.8pt depth3.2pt width0pt}%
   \normalbaselines  \rm}

%
%
%
%
%
\catcode`\@=11\relax
\newwrite\@unused
\def\typeout#1{{\let\protect\string\immediate\write\@unused{#1}}}
\typeout{psfig: version 1.1}

%
%
\def\@nnil{\@nil}
\def\@empty{}
\def\@psdonoop#1\@@#2#3{}
\def\@psdo#1:=#2\do#3{\edef\@psdotmp{#2}\ifx\@psdotmp\@empty \else
    \expandafter\@psdoloop#2,\@nil,\@nil\@@#1{#3}\fi}
\def\@psdoloop#1,#2,#3\@@#4#5{\def#4{#1}\ifx #4\@nnil \else
       #5\def#4{#2}\ifx #4\@nnil \else#5\@ipsdoloop #3\@@#4{#5}\fi\fi}
\def\@ipsdoloop#1,#2\@@#3#4{\def#3{#1}\ifx #3\@nnil 
       \let\@nextwhile=\@psdonoop \else
      #4\relax\let\@nextwhile=\@ipsdoloop\fi\@nextwhile#2\@@#3{#4}}
\def\@tpsdo#1:=#2\do#3{\xdef\@psdotmp{#2}\ifx\@psdotmp\@empty \else
    \@tpsdoloop#2\@nil\@nil\@@#1{#3}\fi}
\def\@tpsdoloop#1#2\@@#3#4{\def#3{#1}\ifx #3\@nnil 
       \let\@nextwhile=\@psdonoop \else
      #4\relax\let\@nextwhile=\@tpsdoloop\fi\@nextwhile#2\@@#3{#4}}
\def\psdraft{
	\def\@psdraft{0}
}
\def\psfull{
	\def\@psdraft{100}
}
\psfull
\newif\if@prologfile
\newif\if@postlogfile
\newif\if@bbllx
\newif\if@bblly
\newif\if@bburx
\newif\if@bbury
\newif\if@height
\newif\if@width
\newif\if@rheight
\newif\if@rwidth
\newif\if@clip
\def\@p@@sclip#1{\@cliptrue}
\def\@p@@sfile#1{
		   \def\@p@sfile{#1}
}
\def\@p@@sfigure#1{\def\@p@sfile{#1}}
\def\@p@@sbbllx#1{
		\@bbllxtrue
		\dimen100=#1
		\edef\@p@sbbllx{\number\dimen100}
}
\def\@p@@sbblly#1{
		\@bbllytrue
		\dimen100=#1
		\edef\@p@sbblly{\number\dimen100}
}
\def\@p@@sbburx#1{
		\@bburxtrue
		\dimen100=#1
		\edef\@p@sbburx{\number\dimen100}
}
\def\@p@@sbbury#1{
		\@bburytrue
		\dimen100=#1
		\edef\@p@sbbury{\number\dimen100}
}
\def\@p@@sheight#1{
		\@heighttrue
		\dimen100=#1
   		\edef\@p@sheight{\number\dimen100}
}
\def\@p@@swidth#1{
		\@widthtrue
		\dimen100=#1
		\edef\@p@swidth{\number\dimen100}
}
\def\@p@@srheight#1{
		\@rheighttrue
		\dimen100=#1
		\edef\@p@srheight{\number\dimen100}
}
\def\@p@@srwidth#1{
		\@rwidthtrue
		\dimen100=#1
		\edef\@p@srwidth{\number\dimen100}
}
\def\@p@@sprolog#1{\@prologfiletrue\def\@prologfileval{#1}}
\def\@p@@spostlog#1{\@postlogfiletrue\def\@postlogfileval{#1}}
\def\@cs@name#1{\csname #1\endcsname}
\def\@setparms#1=#2,{\@cs@name{@p@@s#1}{#2}}
%
%
\def\ps@init@parms{
		\@bbllxfalse \@bbllyfalse
		\@bburxfalse \@bburyfalse
		\@heightfalse \@widthfalse
		\@rheightfalse \@rwidthfalse
		\def\@p@sbbllx{}\def\@p@sbblly{}
		\def\@p@sbburx{}\def\@p@sbbury{}
		\def\@p@sheight{}\def\@p@swidth{}
		\def\@p@srheight{}\def\@p@srwidth{}
		\def\@p@sfile{}
		\def\@p@scost{10}
		\def\@sc{}
		\@prologfilefalse
		\@postlogfilefalse
		\@clipfalse
}
%
%
\def\parse@ps@parms#1{
	 	\@psdo\@psfiga:=#1\do
		   {\expandafter\@setparms\@psfiga,}}
%
%
\newif\ifno@bb
\newif\ifnot@eof
\newread\ps@stream
\def\bb@missing{
	\typeout{psfig: searching \@p@sfile \space  for bounding box}
	\openin\ps@stream=\@p@sfile
	\no@bbtrue
	\not@eoftrue
	\catcode`\%=12
	\loop
		\read\ps@stream to \line@in
		\global\toks200=\expandafter{\line@in}
		\ifeof\ps@stream \not@eoffalse \fi
		\@bbtest{\toks200}
		\if@bbmatch\not@eoffalse\expandafter\bb@cull\the\toks200\fi
	\ifnot@eof \repeat
	\catcode`\%=14
}	
\catcode`\%=12
\newif\if@bbmatch
\def\@bbtest#1{\expandafter\@a@\the#1
\long\def\@a@#1
\long\def\bb@cull#1 #2 #3 #4 #5 {
	\dimen100=#2 bp\edef\@p@sbbllx{\number\dimen100}
	\dimen100=#3 bp\edef\@p@sbblly{\number\dimen100}
	\dimen100=#4 bp\edef\@p@sbburx{\number\dimen100}
	\dimen100=#5 bp\edef\@p@sbbury{\number\dimen100}
	\no@bbfalse
}
\catcode`\%=14
\def\compute@bb{
		\no@bbfalse
		\if@bbllx \else \no@bbtrue \fi
		\if@bblly \else \no@bbtrue \fi
		\if@bburx \else \no@bbtrue \fi
		\if@bbury \else \no@bbtrue \fi
		\ifno@bb \bb@missing \fi
		\ifno@bb \typeout{FATAL ERROR: no bb supplied or found}
			\no-bb-error
		\fi
		\count203=\@p@sbburx
		\count204=\@p@sbbury
		\advance\count203 by -\@p@sbbllx
		\advance\count204 by -\@p@sbblly
		\edef\@bbw{\number\count203}
		\edef\@bbh{\number\count204}
}
%
%
\def\in@hundreds#1#2#3{\count240=#2 \count241=#3
		     \count100=\count240	
		     \divide\count100 by \count241
		     \count101=\count100
		     \multiply\count101 by \count241
		     \advance\count240 by -\count101
		     \multiply\count240 by 10
		     \count101=\count240	
		     \divide\count101 by \count241
		     \count102=\count101
		     \multiply\count102 by \count241
		     \advance\count240 by -\count102
		     \multiply\count240 by 10
		     \count102=\count240	
		     \divide\count102 by \count241
		     \count200=#1\count205=0
		     \count201=\count200
			\multiply\count201 by \count100
		 	\advance\count205 by \count201
		     \count201=\count200
			\divide\count201 by 10
			\multiply\count201 by \count101
			\advance\count205 by \count201
		     \count201=\count200
			\divide\count201 by 100
			\multiply\count201 by \count102
			\advance\count205 by \count201
		     \edef\@result{\number\count205}
}
\def\compute@wfromh{
		\in@hundreds{\@p@sheight}{\@bbw}{\@bbh}
		\edef\@p@swidth{\@result}
}
\def\compute@hfromw{
		\in@hundreds{\@p@swidth}{\@bbh}{\@bbw}
		\edef\@p@sheight{\@result}
}
\def\compute@handw{
		\if@height 
			\if@width
			\else
				\compute@wfromh
			\fi
		\else 
			\if@width
				\compute@hfromw
			\else
				\edef\@p@sheight{\@bbh}
				\edef\@p@swidth{\@bbw}
			\fi
		\fi
}
\def\compute@resv{
		\if@rheight \else \edef\@p@srheight{\@p@sheight} \fi
		\if@rwidth \else \edef\@p@srwidth{\@p@swidth} \fi
}
%
\def\compute@sizes{
	\compute@bb
	\compute@handw
	\compute@resv
}
%
%
\def\psfig#1{\vbox {
	%
	\ps@init@parms
	\parse@ps@parms{#1}
	\compute@sizes
	\ifnum\@p@scost<\@psdraft{
		\typeout{psfig: including \@p@sfile \space }
		\special{ps::[begin] 	\@p@swidth \space \@p@sheight \space
				\@p@sbbllx \space \@p@sbblly \space
				\@p@sbburx \space \@p@sbbury \space
				startTexFig \space }
		\if@clip{
			\typeout{(clip)}
			\special{ps:: \@p@sbbllx \space \@p@sbblly \space
				\@p@sbburx \space \@p@sbbury \space
				doclip \space }
		}\fi
		\if@prologfile
		    \special{ps: plotfile \@prologfileval \space } \fi
		\special{ps: plotfile \@p@sfile \space }
		\if@postlogfile
		    \special{ps: plotfile \@postlogfileval \space } \fi
		\special{ps::[end] endTexFig \space }
		\vbox to \@p@srheight true sp{
			\hbox to \@p@srwidth true sp{
				\hfil
			}
		\vfil
		}
	}\else{
		\vbox to \@p@srheight true sp{
		\vss
			\hbox to \@p@srwidth true sp{
				\hss
				\@p@sfile
				\hss
			}
		\vss
		}
	}\fi
}}
\catcode`\@=12\relax

\elevenpoint
\font\ninerm=cmr9
\font\tenrm=cmr10
\font\scaps=cmcsc10
\font\bbbf=cmbx10 scaled\magstep2
\font\bbrm=cmr10 scaled\magstep2
\font\bbf=cmbx12
\vsize = 23.2 cm
\voffset 1.2cm
\hfuzz=1pt
\vfuzz=1pt

\def\ts{$\,$}
\def\n{\noindent}
\def\s{\smallskip}
\def\m{\medskip}
\def\b{\bigskip}
\long\def\fussn#1#2{$\!${\baselineskip=11pt\parindent=5pt
      \setbox\strutbox=\hbox{\vrule height 7pt depth 2pt width 0pt}%
      \ninepoint
      \everypar{\hangindent=\parindent}
      \footnote{#1}{#2}\everypar{}\elevenpoint}}
\def\eps{\varepsilon}
\def\ms{m$\cdot$s$^{-1}$}
\def\e#1{Eq.\ts(#1)}
\def\es#1{Eqs.\ts(#1)}
\def\dex#1#2{$#1\cdot10^{#2}$}
\def\cz{convection zone }
\def\lu{\ifmmode{\hat {\bf l}}\else ${\hat {\bf l}}$\fi}
\def\nnu{\ifmmode{\hat {\bf n}}\else ${\hat {\bf n}}$\fi}
\def\nnun{\ifmmode{{\hat {\bf n}}_0}\else ${\hat {\bf n}}_0$\fi}
\def\lun{\ifmmode{{\hat {\bf l}}_0}\else ${\hat {\bf l}}_0$\fi}
\def\bu{\ifmmode{\hat {\bf b}}\else ${\hat {\bf b}}$\fi}
\def\hu{\ifmmode{\hat {\bf h}}\else ${\hat {\bf h}}$\fi}
\def\zu{\ifmmode{\hat {\bf z}}\else ${\hat {\bf z}}$\fi}
\def\ve{\ifmmode{{{\bf v}}_e}\else ${{\bf v}}_e$\fi}
\def\gu{\ifmmode{{\hat {\bf g}}_0}\else ${\hat {\bf g}}_0$\fi}
\def\gul{(\gu\cdot\lun)}
\def\gun{(\gu\cdot\nnun)}
\def\rb{{\bf r}}
\def\Rn{\ifmmode{R_0}\else $R_0$\fi}
\def\ub{{\bf u}}
\def\cb{{\bf c}}
\def\vb{{\bf v}}
\def\gb{{\bf g}}
\def\dt#1{{d #1 \over d t}}
\def\dz#1{{d #1 \over d z}}
\def\Dl#1{{d #1 \over d l}}
\def\dl#1{{{\partial #1}\over{\partial l}}}
\def\dln#1{{{\partial #1}\over{\partial l_0}}}
\def\deps{\dln{\eps}}
\def\deta{\dln{\eta}}
\def\qsq{\quad = \quad}
\def\sgs{\; = \;}
\def\sms{\; - \;}
\def\sbs{\, + \,}
\def\sps{\; + \;}
\def\qaq{\quad \approx \quad}
\def\Bb{{\bf B}}
\def\ivp{{1 \over 4\pi}}
\def\sgn#1{\hbox{sgn}(#1)}
\def\tz{\tilde{z}}
\def\al{{\hat \alpha}}
\def\tal{{\tilde \alpha}}
\def\ro{\rho_0}
\def\roe{\rho_{e0}}
\def\EREL{\left( {\eta\over R_0} + \deps \right)}
\def\hpe{H_{pe}}
\def\hv{H_{v0}}
\def\hp{H_{p0}}
\def\sd{s\Delta}
\def\delzh{{z_1\over\hpe}}
\def\ven{(\ve_0\cdot\nnun)}
\def\ai{\alpha_1}
\def\aii{\alpha_2}
\def\bq{B_0^{\,2}}
\def\rq{R_0^{\,2}}
\def\betro{\beta\left(1-{\roe\over\ro}\right)}
\def\gami{{1\over\gamma}}
\def\gamiz{{2\over\gamma}}
\def\homeg{{\hat\omega}}
\def\vt{\vartheta}
\def\brj{\beta r j}
\def\brm{\beta r m}
\def\eg{e.g.\ts\ts}
\def\ie{i.e.\ts\ts}
\def\cf{cf.\ts}
\def\lhs{l.h.s.\ts\ts}
\def\rhs{r.h.s.\ts\ts}
\def\sec#1{Sec.\ts\ts{#1}}
%
 
\def\ref{\par \noindent \hangindent 20pt\ignorespaces} 
\def\aap #1,{, {\it Astron.\ Astrophys.} {\bf #1},}  
\def\apj #1,{, {\it Astrophys.\ J.} {\bf #1},}  
\def\apjs #1,{, {\it Astrophys.\ J. Suppl.} {\bf #1},}  
\def\mnras #1,{, {\it Mon.\ Not.\ Roy.\ Astron.\ Soc.} {\bf #1},}  
\def\sp #1,{, {\it Solar Phys.} {\bf #1},}
\def\jfm #1,{, {\it J. Fluid Mech.} {\bf #1},}
\def\nat #1,{, {\it Nature} {\bf #1},}

{\nopagenumbers

\def\c{\centerline}

\vglue 2.5cm

\c {\bbbf Comments on the structure and dynamics}

\b

\c {\bbbf of magnetic fields in stellar convection zones}

\vglue 3.5cm

\c {Habilitationsschrift}
\m
\c {zur Erlangung der Lehrbefugnis im Fachgebiet}
\m
\c {Astronomie und Astrophysik}

\b\b\b\b

\c {vorgelegt dem Fachbereich Physik in der}
\m
\c {Mathematisch-Naturwissenschaftlichen Fakult\"at}
\m
\c {der Georg-August-Universit\"at zu G\"ottingen}

\b\b\b\b

\c {von}
\m
\c {Manfred Sch\"u\ss ler}
\m
\c {aus Weinheim}

\vfill

\c {G\"ottingen 1990}

\vfill\eject

\vglue 2cm

\n {\bbbf Preliminary remark }

\b\m

\n 
This paper is a ``Habilitationsschrift'', a second thesis required until
recently by universities in Germany and in a few other countries to
obtain the right to lecture. It was accepted by the University of
G{\"o}ttingen in 1990 after review by a number of german and
international experts. Although the introduction and the references
represent the state of research in 1990, most of the remaining content
is still relevant and has never been published elsewhere. The most
important part is the derivation of a linear stability formalism for
thin magnetic flux tubes following an arbitrary path in a
gravitationally stratified medium with a stationary velocity. It was
found later (Ferriz-Mas \& Sch{\"u}ssler, Geophys. Astrophys. Fluid
Dyn. vol. 72, 209; 1995) that, for consistency, the inertial term in the
equation of motion for the external medium should be included in
Eq. (3.24), which leads to an additional term in the stability
equations in the case of a spatially varying external velocity. This
term is missing in the present text, but can be easily introduced into
the formalism.

\vfill\eject

\vglue 2cm

\n {\bbbf Summary}

\b\m

\n Some aspects of magnetic fields in stellar convection zones are investigated
in this contribution. Observational and theoretical results are discussed
which support the conjecture that the magnetic field structure in a convection
zone is intermittent with most of the magnetic flux being concentrated in small
filaments or `flux tubes' surrounded by field-free plasma. These kind of
structures can be mathematically described with aid of the `approximation
of slender flux tubes' whose general form for flux tubes which follow an
arbitrary path in space is rederived and discussed.

The approximation is applied to equilibrium structures of
flux tubes determined by hydrostatic equilibrium along the magnetic field
lines and by a balance of buoyancy, curvature and drag forces (exerted by
external velocity fields like convection, rotation and large-scale flows) 
perpendicular to the field. 
Some general properties of static equilibria (without drag forces) are 
derived and it is shown that such structures are incompatible with the observed 
properties of solar magnetic fields. We discuss methods to determine equilibrium
flux tubes in practice and give analytical examples of stationary tubes
in a horizontal velocity field.

In the central part of the contribution we present a linear stability 
analysis of general flux tube equilibria including arbitrary external velocity 
fields. The tube may follow a curved path in space, gradients of external 
velocity and gravitational acceleration are included. The general equations
for Lagrangian displacements are derived and for the application to stellar
convection zones we give a suitably non-dimensionalized, approximate
form for large values of the plasma parameter $\beta$ which
represents the ratio of gas pressure to magnetic pressure. For static equilibria
a symmetric form of the equations is obtained which allows the application of
variational methods and a simplified numerical treatment. It also serves as 
a consistency check for the (somewhat lengthy) algebra. The formalism is applied
to analytically tractable cases, namely horizontal flux tubes and 
symmetric loops with horizontal tangent at the point of extremum 
(maximum or minimum). Numerical examples on the basis of the properties
of the solar convection zone indicate that in a superadiabatically 
stratified environment and in the 
absence of an external flow all these structures are monotonically 
unstable. Stabilizing external velocity fields can be constructed, but they
do not seem to be of much practical importance for the case of a stellar
convection zone. Additionally, we have found that overstable modes can be 
excited by an external velocity field in which case the
drag force conspires with the curvature force with the result of oscillations
with growing amplitude. Overstable modes cannot be stabilized by 
stratification; they appear also in subadiabatic regions like a layer of
overshooting convection.

Finally, we attempt to summarize our present state of understanding of 
magnetic fields in stellar convection zones and, in particular, the convection
zone of the Sun. We favor the picture of a
strongly fragmented, intermittent field structure. While in the deep parts
the individual magnetic filaments are passive with respect to large-scale 
velocity fields, surface fields exhibit a peculiar nature owing to thermal
effects and the dominance of buoyancy. As to the dynamo problem,  
we find that observational and theoretical evidence
is in favor of a `boundary layer dynamo' operating in an overshoot region
below the superadiabatic parts of a convection zone.

\vfill\eject

\vglue 3cm

\parindent=0pt
\n {\bbbf Acknowledgements}

\b\m

\n Many people have contributed to this work, be it by discussion and
criticism, by moral and technical support, or by their own work which laid the
foundations on which I could build by adding a small brick to the walls of
an unfinished building. Let me thank in particular 
\m

V. Anton for the good time we spent calculating flux tube models, 

W. Deinzer for his encouragement and support, 

H. D\"uker for help with technical problems,

A. Ferriz-Mas for his scrutinity with mathematical details and comments on Ch.\ts{3},

M. Kn\"olker for sharing his knowledge on eigenvalue problems, 

D. Schmitt for advice on stability problems and the energy principle, 

M. Stix for discussions about ways to determine the stability of loops, and

H. Spruit for clarifying answers concerning the approximation of slender 
flux tubes.
\m

This work has been carried out and written down partly at the 
Kiepenheuer-Institut f\"ur Sonnenphysik, Freiburg, and partly at the
Universit\"ats-Sternwarte, G\"ottingen. I like to thank my colleagues at
both institutions for their support and for a climate of cooperation and
kindness. 
Moreover, I am grateful to U. Grossmann-Doerth, R. Hammer and H. Schleicher 
for having taken the burden of additional work in a critical phase of
 installation
and management of the computer network at the Kiepenheuer-Institut while I was
on leave in G\"ottingen or absorbed in writing.

\m
Finally, I want to thank E.N. Parker for his life-long work on solar and stellar
magnetic fields which revealed so many basic mechanisms, opened so many doors
and possibilities for investigation and understanding -- and for continuously 
stirring up inveterate ways of thinking by pointing out inconsistencies and
throwing new ideas into the arena.

\vfill\eject

\def\leaderfill{\leaders\hbox to 1em{\hss.\hss}\hfill}
\def\itemitemitem{\par\indent\indent\indent
                  \hangindent4\parindent
                  \textindent}
\parindent=22pt

\n {\bbbf Contents}
\b\b
\n {\bf 1.\quad Introduction}               \quad \leaderfill  {\bf 1} \quad
\b
\n {\bf 2.\quad Formation of structures}     \quad \leaderfill {\bf 5} \quad
\m
\itemitem{2.1} Rayleigh-Taylor instability of a magnetic layer
               \quad \leaderfill \quad 5 \quad
\m
\itemitem{2.2} Flux expulsion              \quad \leaderfill \quad 8 \quad
\m
\itemitem{2.3} Instabilities and fragmentation of single flux tubes
               \quad \leaderfill 10 \quad
\b
\n {\bf 3.\quad The approximation of slender flux tubes}
                \quad \leaderfill {\bf 14} \quad
\b
\n {\bf 4.\quad Flux tubes in equilibrium}       \quad \leaderfill {\bf 21} \quad
\m
\itemitem{4.1} Static equilibrium
               \quad \leaderfill 21 \quad
\m
\itemitem{4.2} Stationary equilibrium
               \quad \leaderfill 26 \quad
\b
\n {\bf 5.\quad Stability of flux tubes}          \quad \leaderfill {\bf 31} \quad
\m
\itemitem{5.1} Previous work
               \quad \leaderfill 31 \quad
\m
\itemitem{5.2} Equilibrium
               \quad \leaderfill 33 \quad
\m
\itemitem{5.3} Perturbation equations
               \quad \leaderfill 34 \quad
\m
\itemitem{5.4} Non-dimensionalization and the case $\beta\gg 1$
               \quad \leaderfill 40 \quad
\m
\itemitem{5.5} Symmetric form for static equilibrium
               \quad \leaderfill 44 \quad
\m
\itemitem{5.6} Horizontal tubes with vertical external flow
               \quad \leaderfill 46 \quad
\m
\itemitem{5.7} Symmetric loops with vertical external flow
               \quad \leaderfill 54 \quad
\m
\parindent=19pt
\itemitemitem{5.7.1} Local analysis
               \quad \leaderfill 54 \quad
\m
\itemitemitem{5.7.2} Constant vertical displacement
               \quad \leaderfill 58 \quad
\m
\itemitemitem{5.7.3} Heuristic approach for perturbations with large wavelength
               \quad \leaderfill 61 \quad
\m
\parindent=22pt
\itemitem{5.8} Summary of the stability properties
               \quad \leaderfill 67 \quad
\b
\n {\bf 6.\quad Dynamics of flux tubes in a convection zone}
                \quad \leaderfill {\bf 69} \quad
\m
\itemitem{6.1} Size distribution
               \quad \leaderfill 69 \quad
\m
\itemitem{6.2} The relation of the basic forces
               \quad \leaderfill 72 \quad
\m
\itemitem{6.3} The peculiar state of the surface fields
               \quad \leaderfill 74 \quad
\m
\itemitem{6.4} Consequences for the dynamo problem
               \quad \leaderfill 76 \quad
\b
\n {\bf 7.\quad Outlook}                         \quad \leaderfill {\bf 79} \quad
\b
\n {\bf \hskip 23pt References}                  \quad \leaderfill {\bf 81} \quad

\parindent=20pt
\vfill\eject

\vfill\eject}

\topskip = 24pt
\headline={\vbox{\line{\tenit\ifodd\pageno\global\hoffset=0mm\hss 
      \otopline 
      \hss\folio\else\global\hoffset=-5mm\folio\hss
      \etopline
      \hss\fi}\smallskip\hrule}\hss}
\footline = {\hss}

\pageno=1

\vglue 2cm

\def\otopline{ 1. Introduction }

\def\etopline{ 1. Introduction }

\n {\bbbf 1. Introduction}

\b\m

\n The activity of the Sun and other stars with outer convection zones and the 
origin of their hot chromospheres, coronae, and winds is intimately related to 
the existence of magnetic fields in their atmospheres. Most probably, the source 
of the magnetic flux which is observed to pervade photospheres of late-type stars 
is the underlying  convection zone. In the case of the Sun, magnetic flux
is directly observed to emerge from the convection zone. On the other hand,
the 22-year period of the solar magnetic cycle leads to a very small (skin) 
depth to which the alternating magnetic fields can penetrate within the 
electrically well-conducting, quiescent radiative region below the solar 
convection zone. 

A complete theory of the structure, dynamics and evolution of magnetic
fields in stellar convection zones does not exist -- and this work does
not attempt to provide one. This lack of a consistent description
of a basic astrophysical situation is due to a) the impossibility of direct
measurements and b) our unsufficient comprehension of turbulent flows.

In the case of the Sun, although the physics of the photosphere is much 
more complicated, thanks to a wealth of observational data
our understanding of photospheric magnetic fields is much more advanced than the
state of theory for the fields within the convection zone. Only indirect
observational evidence about their state is available through photospheric
observations -- and due to the peculiar nature of this layer the results
are not necessarily representative for the deeper parts (\cf Sec.\ts6.3).
The situation can be illustrated by the following simile: Imagine a person
who is unfamiliar with the purpose and operation of clocks. Imagine further
that this person is confronted with a mechanical clock and the task to
analyze the internal mechanism without opening it, just from the visual
appearance of the dial, the motion of the hands and a spectral analysis of
the ticking. Good luck\ts! A person who tries to understand the structure and 
dynamics of the convection zone and its magnetic fields is in a similar
situation: Only a shallow surface layer can be observed directly (dial and
hands) while global oscillations (ticking) supply some indirect information
from the deeper layers.

The large spatial extension of a stellar convection zone and the 
small viscosity of a stellar plasma lead to enormous values of the Reynolds
number $Re = U L/\nu$ ($U$, $L$: velocity and spatial scale of the dominant 
convective flow, $\nu$: kinematic viscosity) which describes the ratio of
the magnitudes of inertial force and viscous force. For the granular velocity 
field observed in the solar photosphere ($U \approx 1$~k\ms, $L \approx 10^3$~km,
$\nu \approx 10^{-3}$~m$^2\cdot$s$^{-1}$) we find $Re \approx 10^{12}$.
Consequently, the nonlinear inertial forces dominate and a turbulent cascade of 
kinetic energy to larger spatial wavenumbers ensues until scales of less than
1 cm are reached at which disspation by molecular viscosity effectively takes 
place. Stellar convection zones thus span a huge range of scales which reach
from their global dimensions and time scales (of the order of $10^5$~km and
$10^6$ s, respectively) to the disspation range of about 1 cm with a related
dynamical time scale of less than 1 s. The problem of describing turbulent
convection in a stellar convection zone is rendered even more difficult by
the influences of rotation and of stratification due to gravity.

The {\it magnetic\/} Reynolds number $Re = U L/\eta$ where $\eta$ denotes
the magnetic diffusivity is also very large. For the solar convection zone it
increases from about $10^3$ in the upper layers to about $10^9$ near the bottom
(\cf Stix, 1976). Consequently, field line advection and stretching is much more
important than Ohmic dissipation on the dominant scale of convective flows. 
Similar to the fluid motions, the magnetic field spans large ranges of spatial
and temporal scales which extend from the global convective to the dissipative
scales. 

Given our insufficient understanding of turbulent convection it may seem futile
to complicate things even more by introducing magnetic fields or, more precisely,
by taking account of the large electrical conductivity of the stellar plasma.
Besides the fact that the very presence of magnetic fields in the solar and 
stellar
atmospheres forces us to do so, the additional complication brought about by 
including magnetic fields must not necessarily be prohibitive since there is
a number of indications that magnetic fields do not significantly influence
the global structure and dynamics of stellar convection zones. 
One piece of evidence is
the success of the theory of stellar structure and evolution which has been
developed without taking account of magnetic fields while the other line of 
arguments is provided by the comparatively small magnitude of observed variations 
of the Sun during the activity cycle. 

The reversals of the polar fields, the polarity rules for active regions, 
and the strong variation of the frequency of occurrence of large active regions 
indicate a major change of the magnetic structure in the convection zone during 
the activity cycle. On the other hand, the change of the solar convection zone 
is much less significant: Variations of the surface rotation rate of both plasma 
and magnetic structures (at a given heliographic latitude) are smaller
than a few percent (Howard, 1984; Schr\"oter, 1985) while a velocity structure
associated with the activity belts (misnamed `torsional oscillation' by
its discoverers, \cf LaBonte and Howard, 1982) has an amplitude of less than
1\% of the differential rotation in latitude. Convective flow patterns do not
show a significant change during the cycle either, apart from slight variations
of the size distribution of granules (M\"uller and Roudier, 1984) and, possibly,
of their temperature structure (Livingston and Holweger, 1982). It is
improbable that much larger variations in the deep layers of the convection
are effectively `screened' by the surface layers since perturbations of
temperature and velocity are transmitted from the bottom of the convection
zone to the surface without being significantly attenuated (Stix, 1981b).
Solar cycle variations of the solar radius are smaller than 
$200$ km (Wittmann et al., 1981; Parkinson, 1983). The short-term 
luminosity variation is of the order of $10^{-3}$ and corresponds 
directly to the fraction of the surface covered by sunspots while a variation
of about the same magnitude on the time scale of the cycle is indicated 
(Willson, 1984). However, larger heat flux variations on the time scale of the
solar cycle in the deep layers of a convection zone may be efficiently
screened due to its large thermal conductivity and heat capacity (Spruit, 1977a,
1982; Stix, 1981b). 

All these results support the thesis that magnetic fields do not significantly
modify the convection zone on a global scale. However,
one must not conclude that magnetic forces can as well be neglected locally,
\ie at any given location. The contrary seems to be the case: A global
equipartition of magnetic and kinetic energy which would lead to major changes
of the convection zone during the magnetic cycle is avoided by concentration
of the magnetic flux into filaments of strong field with large regions of 
non-magnetic, undisturbed convection between (Parker, 1984a). Such a `phase
separation' in a convecting and a magnetic phase is observed for the case
of the solar photosphere: Most of the observable magnetic flux (outside sunspots)
is in the form of concentrated structures of high flux density which are
arranged in a network defined by the downflow regions of granulation and 
supergranulation, the dominant convective structures (Stenflo, 1989; 
Solanki, 1990). 

Under the conditions prevailing in a convection zone the flux expulsion process 
(for a discussion and further references see Sec.\ts2.2) which is responsible
for this separation in magnetic and convecting regions leads to {\it local\/}
equipartition of the magnetic and kinetic energy {\it densities\/} which gives
a flux density of about $10^4$ Gauss in the deep parts of the solar convection
zone. Since the amount of magnetic flux which emerges during one half cycle
(11 years) is about $10^{24}$ mx (Howard, 1974) it fills only about 1\% of 
convection zone volume given such a flux density. Consequently, the total
magnetic energy of $E_{mag}\approx$ \dex{3}{35} erg amounts to about 1\% of the 
total kinetic energy ($E_{con}$) of the convective flows. The total energy of 
differential rotation  ($E_{dr}$) is of the same order of magnitude as $E_{con}$ 
since the velocity differences due to differential rotation
in depth and in latitude (Duvall et al., 1984, 1987) are of the same order of
magnitude as the convective velocities in the deeper parts of the convection
zone. Consequently, we find the following scaling of total energies within
the convection zone:

$$ E_{con} \; \approx \; E_{dr} \; \approx \; 10^2\, E_{mag}. $$

\n These relations are consistent with the observed percent-level variations of 
the properties of the solar convection zone during the activity cycle. In the 
light of these considerations it seems adequate to take the convective flows as 
given and undisturbed by the magnetic fields for all scales
which are large compared to the typical size of the magnetic flux concentrations.
Small-scale flows locally are strongly affected by the presence of the
field and probably convective heat exchange between a flux concentration and
its environment is largely suppressed.

While a complete theory is lacking, a variety of more or less satisfactory 
physical descriptions and models of certain aspects of the complicated 
thermodynamical and (magneto)hydrodynamical system represented by a stellar
convection zone can be found in the literature. We do not attempt to give a
complete overview but we may roughly classify
them in two complimentary groups, namely the {\it mean field approach\/}
and the {\it model problem approach.}  The contributions belonging to the
first group attempt to describe the large-scale structure and dynamics 
of a stellar convection zone with the aid of parametrized model equations 
for appropriately averaged quantities which vary on large scales. 
Such equations typically contain parameters or functions which
represent conjectures about small-scale processes. Prominent examples are 
Prandtl's mixing length formalism which has been used successfully in
the theory of stellar structure and evolution and the mean-field 
approach for magnetic fields (\eg Parker, 1979a; Krause and R\"adler, 1980) 
which led to the presently most developed theory of the solar cycle, the theory 
of turbulent dynamo action. 

Attempts towards a numerical simulation of the 
hydrodynamical and magnetic structure of the convection zone 
(Gilman and Miller, 1981; Gilman, 1983; Glatzmaier 1984, 1985a,b;
Brandenburg et al., 1990)
also fall into this category: Due to limitations of memory capacity and
processor speed of presently available computers only a small part of the range 
of spatial and temporal scales
can be covered by the simulation and the influence of the small scales 
is parametrized by introducing `turbulent' values for viscosity as well as
thermal and magnetic diffusivity. The effective (hydrodynamical and magnetic)
Reynolds numbers for such simulations are therefore many orders of magnitude
smaller than those of the real system.
The results 
are partly in contradiction to observational data: Neither the predicted
uniformity of angular velocity on cylindrical surfaces nor the large amplitude
of large-scale convective flows has been borne out by measurements
(Duvall et al., 1987; Brown et al. 1989; Dziembowski et al., 1989;
LaBonte et al., 1981).
Furthermore, the characteristic features of the solar cycle could not be
reproduced by the simulations. Apparently processes operating on small scales
which have not been resolved play an important r\^ole for the hydrodynamic
and the magnetic structure of the convection zone. 

In the contributions belonging to the second group, {\it model problem 
approaches\/}, 
individual processes are studied in (artificial) isolation and their relevance
for the global behavior of the system is evaluated. This may lead to the
introduction of new terms in model equations and to a more  sensible 
parametrization in numerical simulations. Furthermore, one might attempt to
combine a sample of reasonable well understood processes like
a jigsaw puzzle in order to obtain a description of the whole system. The
work of E.N. Parker (\cf Parker, 1979a) is a prominent example for the
model problem approach. The theory of magnetic flux tubes (\eg Spruit and 
Roberts, 1983), the work on magnetoconvection carried out by N.O. Weiss 
and his colleagues in Cambridge (\cf Proctor and Weiss, 1982), and the simulations
of turbulence with magnetic fields performed by the Nice group around U. Frisch
and A. Pouquet (\eg Meneguzzi et al., 1981; Grappin et al., 1982; Pouquet, 1985;
Meneguzzi and Pouquet, 1989) also fall in the group of model problems.

The work presented here belongs to the same class of contributions. 
Its motivation results from the debate on magnetic flux storage in a
convection zone and the location of the dynamo process which is responsible 
for the solar activity cycle. Parker (1975a) argued that magnetic buoyancy 
leads to rapid flux loss and thus prohibits the storage of magnetic flux within 
the convection zone for time intervals comparable to the cycle period.
This argument was strengthened by Spruit and van Ballegooijen (1982) who showed
that toroidal flux tubes are unstable in a superadiabatically stratified region.
Following earlier proposals (\eg Spiegel and Weiss, 1980; 
Galloway and Weiss, 1981) they suggested that
a slightly subadiabatic region of overshooting convection near the bottom of
the solar convection zone represents a favorable place for the storage of 
magnetic flux and the operation of a dynamo mechanism 
(see also van Ballegooijen 1982a,b; Sch\"ussler, 1983, 1984a; Durney, 1989).

However, are the arguments given so far sufficient to definitely exclude that
the major part of the magnetic flux emerging in the solar activity cycle is stored
within the convection zone proper\ts? For instance, Parker (1987a-d; 1988a-c) has 
argued that `thermal shadows' due to local suppression of convection can keep
large portions of magnetic flux in the deep parts of the convection zone.
Here we consider another possibility and investigate whether flux tubes could 
possibly find a stable equilibrium in the convection zone if they follow a curved 
path in space and/or if the influence of external flows is taken into account.
Of particular interest in this respect are loop structures and sequences of loops 
(`sea serpents'). To this end we reconsider the general form of the approximation
of slender flux tubes (Spruit, 1981a,b), determine general properties of static
and stationary flux tube equilibria, and generalize the approach of Spruit 
and van Ballegooijen (1982, see also van Ballegooijen, 1983; van Ballegooijen
and Choudhuri, 1988) to derive a stability analysis formalism for a flux
tube which follows an arbitrary path in space. This analysis is embedded in a 
general discussion of the structure and dynamics of magnetic fields in a stellar 
convection zone which is necessarily tentative and far from rigorous.

We start by presenting some arguments in 
favor of the conjecture that magnetic fields in stellar convection zones
consist of small, concentrated structures separated by field-free plasma. 
As a basis for this conjecture, a variety of physical processes which form and 
maintain such structures is discussed in Ch.\ts2. The small size of the magnetic
flux concentrations resulting from fragmentation and expulsion processes
permits their description using the approximation of slender flux tubes. 
In Ch.\ts3 we rederive the general form of this approximation for a flux tube
which follows an arbitrary path in space.
We use a somewhat different approach than that taken
by Spruit (1981a,b) in order to elucidate some aspects of the approximation.
The formalism is applied in Ch.\ts4 to obtain some general properties of flux
tubes in static equilibrium given by the balance of buoyancy and curvature force
and of flux tubes in stationary equilibrium for which the drag force exerted
by an external flow field is additionally taken into account. 
In Ch.\ts5, the central part of this 
contribution, a formalism is derived which allows the stability analysis of 
a general static or stationary flux tube equilibrium with arbitrary path in space.
Some basic stability properties are derived by application of this formalism to a 
number of special cases which can be treated analytically, in particular
horizontal tubes and symmetric loops. In Ch.\ts6 the outcome of these 
calculations, the results of other investigations and further considerations are 
tentatively combined in a (hopefully) coherent picture which summarizes the
author's present view of magnetic fields in stellar convection zones. This 
includes also a discussion of
the consequences for the dynamo problem. Finally, an outlook on possibilities for
extension and continuation of this work is given in Ch.\ts7.

\vfill\eject

\def\otopline{ 2. Formation of structures}

\vglue 2cm

\def\etopline{ 2.1 Rayleigh-Taylor instability of a magnetic layer}

\n {\bbbf 2. Formation of structures}

\b\m

\n The work presented here is based on the hypothesis that, similar to the 
observed magnetic fields in the photosphere of the Sun, most of the magnetic 
flux within a stellar 
convection zone is concentrated into filaments or {\it flux tubes}\fussn{*}
{Strictly spoken, the term `flux tube' refers to a cylindrically shaped bundle 
of magnetic field lines. However, in what follows we shall often use this term
more loosely to generally denote a magnetic filament or flux concentration 
of arbitrary shape.}
embedded in nearly field-free 
plasma. Although at present this conjecture can neither be proven theoretically 
in a rigorous way nor undubitably demonstrated by observations
it is supported by a number of observational indications and theoretical results.

Observationally, Hale's polarity rules for solar active regions are known to apply
with very few exceptions (\eg Howard, 1989). This can only be so 
if the magnetic fields in the convection zone are strong
enough to avoid a significant deformation by convective velocity fields, \ie if 
the magnetic energy density is equal to or larger than the kinetic energy density
of convection. Otherwise the magnetic fields would be passively carried around
at random and the erupting active region would not show a preferred orientation.
Consequently, the magnetic field strength at least must be of the
order of the {\it equipartition field strength}

$$ B_e = v_c (4\pi\rho)^{1/2} \eqno (2.1) $$

\n ($v_c$: convective velocity, $\rho$: density).
Throughout the whole convection zone $B_e$ is much larger than the 
average field strength of about 100 Gauss which can be estimated from the total
magnetic flux emerging during one activity cycle. Consequently, if the magnetic
flux in the convection zone has at least equipartition field strength
it is strongly intermittent and fills only a small fraction of the volume.
The observed properties of emerging active regions and the formation of sunspots 
by accumulation of fragments also
indicate that the magnetic flux already is in a concentrated form before it 
appears at the solar surface (Zwaan, 1978; McIntosh, 1981; Garcia de la Rosa, 
1987).

Theoretical arguments for a filamented nature of the magnetic fields
in a stellar convection zone are given in the subsequent sections.

\b\b

\n {\bbf 2.1 Rayleigh-Taylor instability of a magnetic layer}

\b

\n Beginning with Parker (1975a) a number of arguments has been put forward
which support the assertion that the major part of the magnetic flux which is
responsible for the solar activity cycle cannot be kept in the superadiabatic
parts of the convection zone for times comparable to the cycle period. The flux
rather has to be stored below in a region
of overshooting convection where a magnetic layer forms which occasionally 
ejects magnetic flux into the 
convection zone proper (Spiegel and Weiss, 1980). Since the thickness of
such an overshoot layer probably is less than the local pressure scale height 
of about \dex{5}{4} km (Shaviv and Salpeter, 1973; Schmitt et al., 1984; van
Ballegooijen, 1982b; Pidatella and Stix, 1986) 
and thus a lot of toroidal magnetic flux (about $10^{24}$ mx) 
has to be accommodated in a rather small volume the
field there is thought to be densely packed (non-filamented) and,
because the flow velocities are small compared to the sound speed, in
magnetostatic equilibrium.

If the (horizontal or toroidal) magnetic field decreases rapidly enough with
height or even drops to zero discontinuously at some level, this equilibrium
which is basically a balance between gravity and the gradient
of the total (magnetic + gas) pressure becomes unstable with respect to an
interchange of more and less magnetized fluid which lowers the potential
energy of the system. This is the magnetic Rayleigh-Taylor instability\fussn{*} 
{In the case of a discontinuous transition
between a non-magnetic plasma and a vacuum magnetic field the instability is
known as Kruskal-Schwarzschild instability (\cf Cap, 1976, Ch.\ts11).} 
of a layer of magnetic field directed perpendicular to the local direction of 
gravity (Gilman, 1970; Cadez, 1974; Acheson and Gibbons, 1978; Acheson, 1979; 
Parker, 1979b; Schmitt and Rosner, 1983; Hughes, 1985a,b, 1987; 
Schmitt, 1985; Hughes and Cattaneo, 1987).
For this kind of instability the stratification need not necessarily show a
density inversion as in the case of the hydrodynamical Rayleigh-Taylor
instability. The nonlinear evolution of the instability in the case of
a discontinuity of the magnetic field is nicely illustrated by the numerical 
simulation of Cattaneo and Hughes (1988) who showed that secondary 
Kelvin-Helmholtz instabilities lead to the formation of intense vortices
whose dynamical interaction dominates the dynamics after the first phases of
the instability. 

Because of the stabilizing effect of magnetic curvature forces linear stability
analysis shows that the fastest growing perturbations are those with large
wavelength along  and small wavelength perpendicular to the equilibrium magnetic
field.  Consequently, the formation of {\it thin\/} structures at the upper
boundary of a sheet of horizontal magnetic field (where the field strength may
rapidly decrease with height) is favored.
Quantitative results for the case of a stellar convection zone are 
difficult to obtain since most of
the work done so far is restricted to linear analysis under idealized assumptions.
The typical fragment sizes of a destabilized flux sheet depend on diffusive 
effects (thermal, viscous and magnetic) and on the detailed height
dependence of field strength and entropy.  Another important effect to consider
is {\it rotation\/} which may drastically reduce the growth rates or even
entirely suppress the instability (\cf Acheson, 1978; 1979; Roberts and
Stewartson, 1977).

We shall not discuss here the complex variety of instability mechanisms which
grows out of the interaction of differential rotation, magnetic field,
stratification and diffusive effects (see \eg Schmitt and Rosner, 1983). In
order to give a crude estimate of typical temporal and spatial scales we
consider a case which is thought to represent the typical properties of a
magnetic layer in the overshoot region below the solar convection zone:
Adiabatic or weakly subadiabatic temperature stratification, sound speed ($\approx
500$ k\ms) large compared to rotational velocity (1.4 k\ms) which, in turn, is
large compared to the Alfv\'en velocity ($\approx 60$ \ms\ for an equipartition
field of about $10^4$ Gauss). Under these circumstances the magnetic
Rayleigh-Taylor instability evolves near the top of the sheet, \ie in a region
where the field strength, $B$, decreases with height faster than density,
$\rho$, in the form of growing magnetostrophic waves which propagate along the
direction of the equilibrium magnetic field (Acheson, 1979; Schmitt, 1985).
For a magnetic layer of thickness $D$ the fastest growing wave mode is
characterized by the wavenumbers $n$ (perpendicular to both gravity 
and magnetic field) and $k$ (parallel to the field) which are given
by (Acheson, 1979, Ch.\ts3)

$$      n^2D^2 = {\pi \over 2}\; C^{1/2} \left[ -\; {H_p\over\gamma}\;
            {d \over dz}\ln\left( {B\over\rho} \right) \right]
            \eqno (2.2) $$
          
$$     k^2 H_p^2 = {1 \over 2}  \left[ -\; {H_p\over\gamma}\;
            {d \over dz}\ln\left( {B\over\rho} \right) \right]
            \eqno (2.3) $$        

\n ($H_p$: pressure scale height, $\gamma$: ratio of specific heats,
$z$: height coordinate, antiparallel to the direction of gravity).
The quantity $C$ is given by 

$$ C = {v_A^2 \over 2\Omega\eta}\; {D^2 \over H_p^2} \eqno (2.4) $$

\n ($v_A$: Alfv\'en speed, $\Omega$: angular velocity, $\eta$: magnetic
diffusivity). Using values of $B = 10^4$ Gauss, $\Omega= $\dex{2.7}{-6}
s$^{-1}$, the surface equatorial rotation rate, $\eta = 10^4$ cm$^2$s$^{-1}$,
$H_p = 6\cdot10^4$ km and $D = 10^4$ km (\cf Schmitt et al., 1984) we find 
$C \approx 2\cdot10^7$.  Taking the thickness of the layer, $D$, as typical
length scale for the decrease of $B/\rho$ with height, we find

$$   -\;  {d \over dz}\ln\left( {B\over\rho} \right) \approx D^{-1}
          \eqno (2.5) $$
        
\n and from \es{2.2} and (2.3)

$$ \eqalign  { n^{-1} &\approx 6\cdot10^{-3} D \approx 60\; \hbox{km} \cr
   k^{-1} &\approx .75 \; H_p \approx 4.5\cdot10^4\; \hbox{km}\; .\cr} $$

\n Thus the fastest growing wave has a longitudinal wavelength of about a
scale height and a much smaller transversal scale. With increasing amplitude
the wave penetrates into the convection zone proper and its filaments
become subject to the convective flows. Furthermore, they are liable to other 
instabilities and fragmentation processes (see Sec.\ts2.3). We conclude that
the magnetic Rayleigh-Taylor instability of a magnetic layer in an overshoot
region leads to magnetic fragments in the convection zone which have a 
transversal scale of the order of $100$ km or less. 

The growth time, $\tau$, of the instability is given by

$$ \eqalignno{
             \tau \quad &\approx \quad {4\Omega H_p^2\over v_A^2}\;
            \left[ -\; {H_p\over\gamma}\;
            {d \over dz}\ln\left( {B\over\rho} \right) \right]^{-1}&\cr&&\cr
            &\approx \quad {4\gamma\Omega H_p D\over v_A^2}
            \quad \approx \quad 35\; \hbox{days} \,. &(2.6) } $$

\n This rather large time scale reflects the stabilizing influence of rotation.
The instability cannot be suppressed by the stable, subadiabatic stratification
of the overshoot layer since the radiative thermal diffusivity $\kappa \approx$
\dex{2}{7} cm$^2$s$^{-1}$ equalizes temperature differences
between a structure with $l = 60$ km diameter and its environment in a
timescale $l^2/\kappa \approx 41$ days which is comparable to the growth time
of the instability given by \e{2.6}. The flux loss of a magnetic layer in
the overshoot region due to the magnetic Rayleigh-Taylor instability is
possibly limited by the `turbulent diamagnetism' of the convection zone
which is briefly discussed in Sec.\ts2.2.

The conclusions drawn above are not significantly changed if an appropriate
``turbulent'' value for the magnetic diffusivity is taken instead of the
molecular value $\eta = 10^4$ cm$^2$s$^{-1}$. This is true even if the
suppression of motions by the strong magnetic field in the layer is ignored. We
take an appropriate ``microscale'' $\delta = 100$ km for the motions to be
described by the turbulent diffusivity $\eta_t$ which is given by

$$ \eta_t \; \approx \; 0.1 \, u(\delta) \, \delta  \eqno (2.7) $$
   
\n where $u(\delta)$ is the turbulent velocity on the spatial scale $\delta$.
Assuming a Kolmogorov spectrum with an external scale $L = 10^{10}$ cm
and $u(L) = 10^4$ c\ms\ as typical for the convective flows in the deep
convection zone we find

$$ u(\delta) \; = \; u(L) 
   \left( {\delta\over L} \right)^{1/3} \approx \; 10^3 \;
   \hbox{cm}\cdot\hbox{s}^{-1} \eqno (2.8) $$

\n and using \e{2.7} we have $\eta_t \approx 10^9$ cm$^2$s$^{-1}$. Hence, the
quantity $C$ is decreased by a factor $10^5$ (\cf Eq.\ts2.4) and we see from
\e{2.2} that the transversal length scale $n^{-1}$ is increased by a factor
$10^{5/4}$ to about 1000 km which confirms a posteriori our choice of the
microscale in \e{2.7}.  The fragment size is still much smaller than the local
scale height and the typical length scales of the convective motions. While the
growth time of the instability given by \e{2.6} is not changed, the thermal
diffusion timescale is now about 30 years and radiative heating cannot
remove the stabilizing effect of a subadiabatic temperature stratification.

In a much more detailed study Schmitt and Rosner (1983) come to essentially
the same conclusions. The magnetic Rayleigh-Taylor instability is 
of ``double-diffusive'' nature:
For molecular magnetic diffusivity we have $\eta/\kappa \ll 1$, \ie the
stabilizing effect of the subadiabatic stratification is connected with
a larger diffusivity ($\kappa$) than the destabilizing magnetic field
gradient ($\eta$) and instability ensues. On the other hand, if $\eta = \eta_t$
we have $\eta > \kappa$ and a sufficiently subadiabatic stratification like
the one used by Schmitt and Rosner (1983) removes the magnetic instability
while an adiabatic stratification as assumed by Acheson (1979) still leads
to instability. Consequently, the stability properties of a magnetic layer
depend sensitively on the entropy gradient within the overshoot region.
Model results for this quantity have been presented, among others, by
Shaviv and Salpeter (1973), Schmitt et al. (1984) and Pidatella and Stix
(1986).  All these authors find that the subadiabaticity is rather small
($\nabla - \nabla_{ad} \approx -10^{-6} ... -10^{-7}$) in the overshoot layer.

Schmitt and Rosner (1983) propose the following scenario: During a first phase
magnetic flux is accumulated and amplified by dynamo processes but the field
strength stays below the equipartition value and the turbulent motions are not
strongly affected by the magnetic field. Consequently, turbulent diffusivities
are appropriate and, given a sufficiently large subadiabaticity, the
configuration is stable. As the field strength increases (\eg by differential
rotation) the turbulent motions are more and more affected by the field and the
magnetic diffusivity approaches its molecular value. As we have seen above,
this quenches the stabilizing effect of the stratification, the magnetic layer 
sheet becomes unstable and is ejected into the convection zone in the form of 
small filaments. 

The important point for this work which mainly deals with the
magnetic structure {\it within\/} the convection zone proper is that in any
case very small structures with diameters between 100 and 1000 km are formed
when the instability sets in.

\b\b
\def\etopline{ 2.2 Flux expulsion }

\n {\bbf 2.2 Flux expulsion}

\b

\n Independent of having entered from below or being generated in place magnetic
flux within a convection zone interacts with the convective flows, a situation
which the theory of magnetoconvection attempts to describe (see Proctor 
and Weiss, 1982, for a review). An important result of this interaction is
{\it flux expulsion\/}, a process which has first 
been demonstrated in the kinematical case (passive magnetic field)
by Parker (1963). He showed that in an electrically well-conducting plasma 
with a stationary velocity field a magnetic field is excluded from the regions 
of closed streamlines. Starting with the work of Weiss (1966) a number of 
numerical studies have been performed (\eg Galloway et al., 1978; Weiss, 1981a,b).
They showed that 
in a convecting medium at high magnetic Reynolds number $R_m=U L/\eta$
($U$, $L$: Velocity and size of the dominating convective cell, $\eta$: 
magnetic diffusivity) permeated by a magnetic field, the magnetic flux
is concentrated into filaments between the convection cells. This effect
is related to the phenomenon of `intermittency' for magnetic fields in
turbulent flow (\eg Kraichnan, 1976; Orszag and Tang, 1979; Meneguzzi et al.,
1981). 

In a compressible,
stratified fluid the magnetic flux concentrations formed by flux expulsion
are found in the convective downflow regions (Nordlund, 1983; 1986; Hurlburt
et al., 1984; Hurlburt and Toomre, 1988). Observations demonstrate that the
flux expulsion process operates in the solar (sub-)photosphere: On both the 
granular (Title et al., 1987) and the supergranular scale (the well-known
network fields) magnetic flux is predominantly located and concentrated in the 
downflow regions.

The nonlinear back-reaction of the magnetic field on the convective flows via the 
Lorentz force 
limits the flux density which can be achieved by flux expulsion to a value
which is roughly given by the equipartition of magnetic and kinetic energy
density. This limit may be modified by compressibility, diffusive and thermal 
effects (\eg Galloway et al., 1978; Sch\"ussler, 1990). 
Furthermore, motion is excluded from strong flux concentrations
and a kind of ``phase separation'' between field-free, convecting fluid and
magnetic, almost stagnant regions evolves. Such a situation 
seems to be favored energetically (Parker,
1984): The interference of the magnetic field with the convective energy
transport is minimized and the total energy is smaller
than for a state with a diffuse field and the same convective energy flux.
The exclusion of motion for fields stronger than equipartition suppresses
the convective heat exchange between the magnetic structure and its surroundings.
Thermal interaction with the environment is reduced to radiative energy transport.

The properties of nonlinear magnetoconvection are much more complicated
than can be discussed here (see Proctor and Weiss, 1982). For our purposes
it suffices to state that a magnetic field permeating a convecting
fluid at high magnetic Reynolds number inevitably is concentrated into structures 
of about equipartition field strength. In the convection zone the magnetic
Reynolds number for the dominating convective flows is everywhere large: It
increases from about $10^3$ in the upper layers to about $10^9$ near the bottom
(\cf Stix, 1976). Consequently, flux expulsion is relevant throughout the whole
convection zone and we have to expect a concentrated, filamented magnetic
field structure.

The well-known formal analogy between the MHD induction equation and the 
equation which determines the time evolution of {\it vorticity\/} supports the
conjecture that the expulsion effect operates also for vorticity and leads
to the formation of intense whirls or vortices. An example in cylindrical
geometry has been given by Galloway (1978) while Sch\"ussler (1984a) showed
that a full analogy between {\it kinematic} expulsion of magnetic field and 
vorticity holds for two-dimensional flow in cartesian geometry. Numerical 
simulations of turbulence (\eg McWilliams, 1984) and  laboratory experiments
in rotating, turbulent fluids (McEwan, 1973; 1976; Hopfinger et al., 1982) have
clearly demonstrated vorticity expulsion for rotationally dominated flows
(\ie flows with small Rossby number $U/(2\Omega L)$ with $U$, $L$: velocity 
and size of the dominating eddy, $\Omega$: angular velocity).
A similar effect occurs in
the simulations of solar granulation carried out by Nordlund (1984b, 1985a)
who found that narrow granular `fingers' come into rapid rotation.
At least in the surface layers of the Sun magnetic field and vorticity are
concentrated in the same locations such that magnetic flux concentrations
become surrounded by rapidly rotating, descending whirl flows.

In a general context flux expulsion is related to the idea of {\it turbulent
diamagnetism\/} which goes back to Zel'dovich (1957) and Spitzer (1957).
This means a transport of magnetic field antiparallel to the gradient of
turbulent intensity in inhomogeneous turbulence. The effect tends to expel
a magnetic field into the boundary parts of a confined turbulent
region (like a convection zone). Adopting a two-scale approach turbulent 
diamagnetism has been studied by  R\"adler (1968), Moffatt (1983) and
Cattaneo et al. (1988). Again, we may conjecture that the effect
operates in a similar way for vorticity. For the illustrative case
of spatially periodic velocity field in two dimensions with a perpendicular
shear flow Sch\"ussler (1984a) applied a mean-field treatment for the kinematic
case and showed that vorticity is transported and expelled in the same way as a 
magnetic field. He also gave an estimate for the effect of this mechanism on the
depth dependence of rotation in the lower half of the solar convection
zone where rotation is dominant (Rossby number $\approx 0.3$). A stationary
profile of the angular velocity $\Omega$ could be determined by assuming a
balance between the `negative viscosity' effect of 
vorticity expulsion and normal (turbulent) viscosity. 

For the solar and stellar convection zones we may conjecture that flux expulsion
has two major effects in a stellar convection zone: It leads to a filamentary 
state of magnetic fields by generating local concentrations of magnetic flux 
and vorticity {\it and\/} it pushes both magnetic field and vorticity (angular 
momentum in a rotating system) to the boundaries. This generates a magnetic shear 
layer at the bottom of the convection zone, a region which is favorable for the 
operation of a dynamo mechanism (see Sec.\ts{6.4}). 

\b\b
\def\etopline{ 2.3 Instabilities and fragmentation of single flux tubes}

\n {\bbf 2.3 Instabilities and fragmentation of single flux tubes}

\b

\n Having shown some evidence in favor of a concentrated and filamented state of 
magnetic fields in a stellar convection zone we may ask for the typical size of
a flux concentration. The size distribution of magnetic structures
depends not only on the initial sizes of the flux tubes injected into the
region but also on fragmentation, accumulation and coagulation processes
operating in the convection zone itself (Bogdan, 1985; Bogdan and Lerche,
1985). Large-scale convective flows tend to accumulate magnetic flux by way of 
the flux expulsion mechanism discussed in the preceding section but a number 
of processes counteracts this tendency to form large structures in the 
form of single, coherent flux tubes.

A first mechanism to mention is, again, the {\it magnetic Rayleigh-Taylor 
instability:}
If a magnetic structure is oriented mainly horizontally, it may be fragmented
by this instability if it is larger than the scale given by \e{2.2}.  However,
the rather large growth time (Eq.\ts2.6) must be compared with the time scales
of other processes in order to evaluate the relevance of this instability. Of
particular importance in this connection are the hydromagnetic interchange
instability and a special form of the Kelvin-Helmholtz instability.

The {\it interchange instability\/} is well known from laboratory plasmas (\eg
Krall and Trivelpiece, 1973, Ch.\ts5; Cap, 1976, Ch.\ts11). As a simple example
consider the interface between a non-magnetic plasma and a vacuum magnetic
field where in equilibrium gas pressure and magnetic pressure are equal. This
equilibrium is unstable if, looking from the plasma side, the boundary is
concave: the potential energy of the system can be diminished by
exchanging magnetic and non-magnetic volume elements because this procedure 
decreases the magnetic
tension. Conversely, if the boundary is convex with respect to the plasma, the
equilibrium is stable. These considerations apply in a similar way to a
non-vacuum magnetic field. For cylindrical or rotationally symmetric
configurations the instability is often referred to as {\it fluting\/} 
or {\it flute instability\/} since the most rapidly growing perturbations are 
reminiscent to the flutes of classical columns.

The interchange instability is important for magnetic fields in the convection
zone in at least three respects. Firstly, it increases the efficiency of
magnetic field line reconnection (Parker, 1979a, Ch.\ts15); secondly, it can
lead to fragmentation of vertical magnetic structures in the uppermost
(subphotospheric) layers: Since the gas pressure decreases rapidly with height,
a vertical flux tube (\eg a sunspot) flares out with height. Consequently, the
interface becomes concave with respect to the non-magnetic plasma and the
configuration is liable to the interchange instability (Parker, 1975b;
Piddington, 1975). Meyer et al. (1977) showed that the stratification of the
fluid (due to gravity) stabilizes large flux tubes (magnetic flux larger than
about \dex{5}{19} mx): Interchanging magnetic and non-magnetic fluid entails
lifting dense material above light material which gives a positive contribution
to the potential energy.  Similarly, small flux tubes (magnetic flux below
about \dex{5}{17} mx) may be stabilized by surrounding whirl flows
(Sch\"ussler, 1984b).  In this case, the dynamical stability of the angular
momentum distribution at the boundary compensates the destabilizing effect of
field line curvature.

The third aspect -- which is most important for the discussion here -- is the
interchange instability of deformed magnetic structures in the deep convection
zone. Because of the large electrical conductivity and due to hydrodynamic
coupling, magnetic structures follow the fluid motions (convection,
differential rotation) until the curvature forces have grown strong enough to
resist further deformation.  However, for the curvature force to come into play 
a flux tube must be significantly deformed (\cf Sec.\ts6.2).
As a simple example, consider an initially vertical flux tube
subject to a horizontal, localized jet-like flow. The tube is deformed 
to the shape sketched in Fig.\ts{1} and reaches an equilibrium which, in the
absence of gravity, is determined by the balance of the hydrodynamic drag force
and the magnetic curvature force (\cf Sec.\ts4.2).

\vbox to 6.5cm{\vss
$$\c{\hfill\psfig{figure=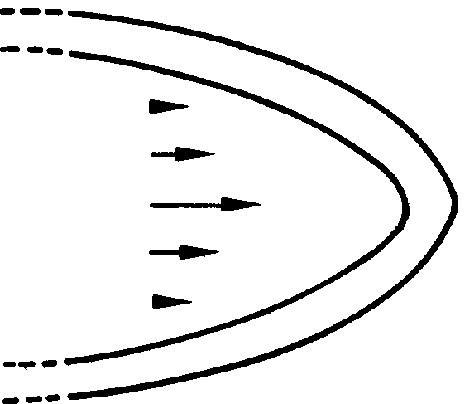,height=5.cm}\hfill}$$
\vss}

\tenpoint
\baselineskip=10pt
\n {\bf Fig.\ts{1}:} Sketch of a flux tube under the influence of a
horizontal, jet-like velocity field. An equilibrium is reached if the
curvature force balances the hydrodynamic drag force.  The interface on the
upstream side is liable to the interchange instability.

\vglue 0.5cm

\elevenpoint

\n The interface between the
flux tube and the surrounding plasma is unstable on the upstream side which
faces the flow and in the absence of stabilizing effects the flux tube splits
up into smaller fragments. Linear stability analysis in the case of vanishing
magnetic diffusivity gives for the growth time, $\tau$, of the instability (\cf
Cap, 1976)

$$ \tau(n) \; = \; \left( 
           {\rho R \over 2\, \Delta p\, n} \right)^{1/2} \eqno (2.9) $$
         
\n ($\Delta p$: gas pressure difference at the interface; $n$: perturbation wave 
number, perpendicular to equilibrium magnetic field; $R$: radius of
curvature of the interface).  Since the gas pressure difference is equal to the
magnetic pressure of the flux tube, \ie $\Delta p = B^2/8\pi$, we find from
\e{2.9}

$$ \tau(n) \; = \; (v_A)^{-1} \left( {R \over n}
           \right)^{1/2} \eqno (2.10) $$

\n where $v_A = B/(4\pi\rho)^{1/2}$ is the Alfv\'en velocity. An upper
limit for the growth time can be obtained by specifying a lower limit for $n$,
\ie $n \ge (2a)^{-1}$ where $a$ is the radius of the flux tube. This leads to

$$ \tau \; \le \; {\left(2 R\, a\right)^{1/2} \over v_A}
   \eqno (2.11) $$
   
\n Consequently, the growth time is smaller than the travel time of an
Alfv\'en wave over a distance given by the geometric mean of flux tube
diameter and its radius of curvature. For a strongly deformed tube with $R
\approx a$ we find $\tau \approx a/v_A$. As an example consider a large tube, $a =
10^4$~km, in the lower convection zone of the Sun with $B = B_e$, 
$v_A = 100$ \ms\ which
is moderately deformed, $R \approx H_p \approx 6\cdot10^4$
km, by convective flows or differential rotation. According to \e{2.11} we find
a growth time of $\tau \approx 3$ days. Hence, even a moderately curved
structure fragments within a few days.

Since the growth time decreases for smaller perturbation wavelength (\cf
Eq.\ts{2.10}), equality with the time scale of magnetic diffusion is reached
only for very small fragment size. For this size, $d = 1/n_0$, inhomogeneities
are as rapidly smoothed out by diffusion as they are formed due to interchange
instability.  $d$ can be estimated by equating $\tau(n_0)$ with the diffusion time
$\tau_d = d^2\eta^ {-1}$ which gives

$$ d \sgs \left( {2 R \eta^2\over v_A^2} \right)^{1/3}. \eqno (2.12) $$
           
\n For the large flux tube discussed above we find $d \approx 40$ km even if we
use the turbulent diffusivity derived after \e{2.8}. The growth time for
such a structure is only a few hours. Consequently, the interchange instability
constitutes a very efficient fragmentation mechanism which leads to splitting of 
even moderately deformed flux tubes in a time scale of hours to days. The resulting
fragment sizes are of the same order of magnitude as those generated by the 
magnetic Rayleigh-Taylor instability at the upside of a horizontal flux tube
but the growth time of the latter is much larger.

Another important fragmentation mechanism is related to the {\it
Kelvin-Helmholtz instability\/} (Tsin\-ga\-nos, 1980). Assume a flux tube
embedded in an external velocity field as sketched in Fig.\ts{2a}.
A small perturbation near the stagnation point leads to the flow geometry
sketched in Fig.\ts{2b}. The centrifugal force due to the curved streamlines
near the stagnation point leads to a pressure gradient which causes a local
pressure maximum at the interface. Very similar to the interchange instability,
this causes growth of the perturbation and fragmentation of the flux tube.
This process is nicely demonstrated by laboratory experiments with rising gas
bubbles in liquids (see the photographs reproduced in Tsinganos, 1980) and is
also visible in a numerical simulation of buoyant, rising flux tubes
(Sch\"ussler, 1979).

\vbox to 5.5cm{\vss
$$\c{\hfill\psfig{figure=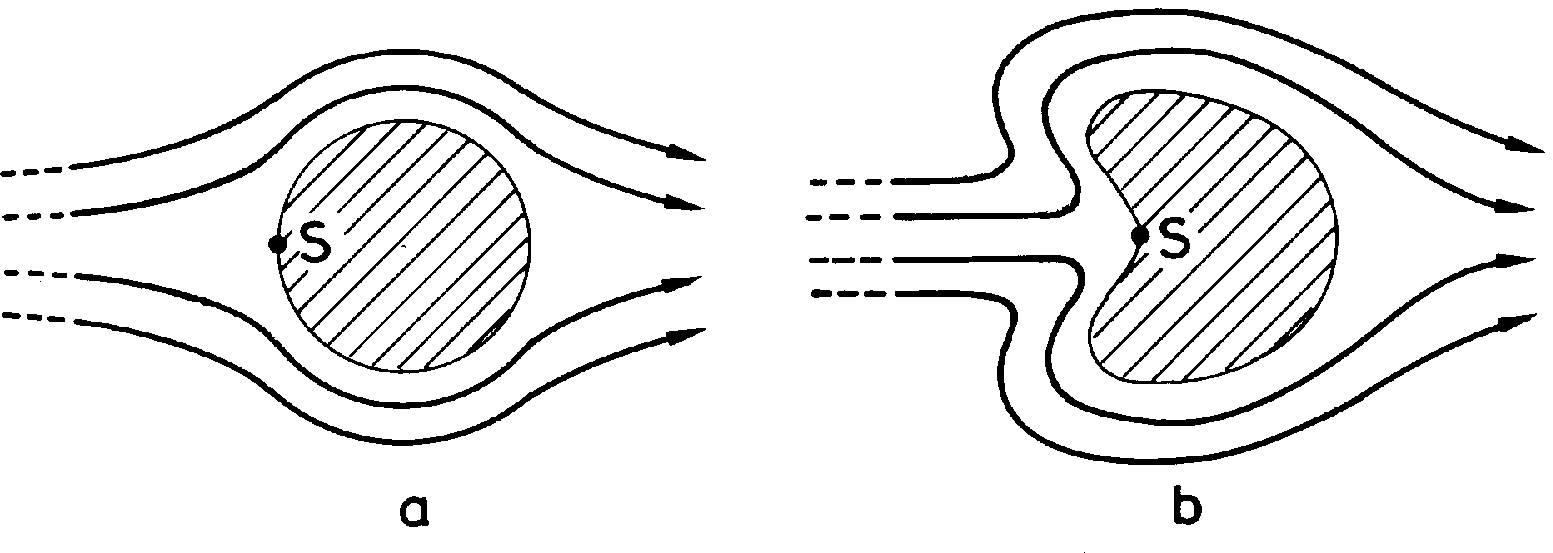,height=5.0cm}\hfill}$$
\vss}


\tenpoint
\baselineskip=10pt
\n {\bf Fig.\ts{2}:} Dynamical fragmentation of a flux tube. {\bf a}: Flux
tube embedded in an external flow field. A small perturbation near the
stagnation point $S$ leads to the situation sketched in {\bf b}: The the
centrifugal force causes a pressure gradient such that a local pressure maximum
evolves at the interface. This leads to growth of the perturbation and
fragmentation of the tube.

\vfill\eject

\elevenpoint

\n All these considerations lead us to expect strongly fragmented magnetic
structures of the size of a few tens of km in the solar convection zone.
Converging flows could possibly accumulate them in loose bundles but they are
unable to produce large, monolithic structures since we have seen that the
flows themselves cause fragmentation (see also Sch\"ussler, 1984b, and the
discussion in Sec.\ts6.1). 

However, could possibly a {\it twisted\/} flux tube escape from being fragmented?
With an azimuthal component, $B_\phi$, of the
magnetic field the internal pressure gradient is changed and, furthermore, 
perturbations which do not bend magnetic field lines are no longer possible. A 
sufficiently large $B_\phi$ may suppress all fragmentation processes discussed so
far. For the interchange instability the necessary magnitude can be estimated
by equating the (stabilizing) tension force due to $B_\phi$ with the
(destabilizing) curvature force exerted by the longitudinal field $B_z$

$$ {B_\phi^2 \over 4\pi a} \quad \approx \quad {B_z^2 \over 4\pi R} $$

\n which gives for the ratio of the field components

$$ {B_\phi \over B_z} \quad \approx \quad 
   \left( {a \over R} \right)^{1/2} . \eqno (2.13) $$

\n For the large flux tube discussed above \e{2.13} gives a $B_\phi/B_z \approx
0.4$, \ie the azimuthal field must be of the order of the longitudinal field
in order to stabilize a moderately deformed flux tube ($R = H_p$)
with respect to the interchange instability. With such large azimuthal fields,
however, kink instabilities become relevant at perturbation wavelengths along
the flux tube which are comparable to the diameter of the structure, \ie the
flux tubes buckles, reconnects and evolves rapidly on a dynamical time scale.
Presumably this instability either leads directly to fragmentation or it
removes most of the twist and leaves the flux tube unprotected against the
other instabilities discussed above.  A twisted flux tube is kink unstable for
longitudinal wavelengths $\lambda$ which satisfy the inequality (Cap, 1976,
Ch.\ts11)

$$ {\lambda \over a} \quad > \quad {B_z \over B_\phi}\,. \eqno(2.14)  $$

\n Using this and \e{2.13} we find that a flux tube which is stabilized
against interchanging by an azimuthal field is kink unstable for

$$ \lambda \quad > \quad \left( a\,R \right) ^{1/2} . \eqno(2.15) $$

\n In our example we have $\lambda > 2.4\cdot10^4$ km. The time scale for the
kink instability is of the same order of magnitude as that of the interchange
instability. On the other hand, slightly deformed {\it small\/} flux tubes can
be stabilized by a much smaller amount of twist. For $a = 100$ km and $R =
H_p = 6\cdot10^4$ km, a ratio $B_\phi/B_z = 0.04$ is sufficient. The helical
kink instability would set in on a scale of $\lambda \approx 2500$ km which is
much larger than the flux tube diameter and could possibly, as conjectured by
Parker (1979a, Ch.\ts9.2), saturate in a stable, cork-screw shaped form of the
flux tube. In the uppermost layers of the convection zone, such a modest amount
of twisting could conceivably be produced by a surrounding whirl flow which
itself exerts a stabilizing influence (Sch\"ussler, 1984b).

\vfill\eject

\def\otopline{ 3. The approximation of slender flux tubes }

\vglue 2cm

\def\etopline{ 3. The approximation of slender flux tubes }

\n {\bbbf 3. The approximation of slender flux tubes}

\b\m

\n The considerations presented in the preceding chapter support the view that
the magnetic flux in a \cz consists of an ensemble of thin (diameter $< 100$ 
km $\ll H_p$), concentrated (at least equipartition field strength) filaments.
Such structures allow a simplification of the MHD equations if all
quantities do not vary significantly within each cross section 
and if the spatial scales of variations {\it along\/} the filament are large
compared to its diameter. In particular, this approximation requires that the
diameter is small compared to pressure scale height, radius of curvature, and
to the longitudinal length scales of all dynamical processes (\eg
longitudinal wavelengths, scale of variation of flows).  Under these
circumstances the global statics and dynamics of a filament (excluding processes 
like body waves which involve a significant structure within a cross-section, 
\cf Ferriz-Mas et al., 1989) can be described using truncated Taylor
expansions of the lateral variation of all quantities. A truncation of the
resulting set of equations at {\it zeroth\/} order leads to the {\it approximation
of slender flux tubes\/} (ASF) which involves only the values of
the quantities along a representative curve (\eg the axis of a flux tube with
circular cross section) or averages over the cross section.

For {\it vertical, axisymmetric\/} flux tubes with straight axis the ASF has
been systematically derived by Roberts and Webb (1978; see also Defouw, 1976).
For this case, the general properties of the expansion procedure are discussed
by Ferriz-Mas and Sch\"ussler (1989). Spruit (1981a,b) gives a general
form of the ASF for flux tubes with curved axis. 
For the application to extragalactic jets, Achterberg (1982, 1988) has derived
a similar approximation which he calls the `firehose limit'.
In what follows we shall rederive the ASF 
in a way somewhat different from Spruit's approach in order to introduce
the formalism and notation to be used in the subsequent chapters 
and hopefully also to elucidate the nature of the approximation.


\vbox to 7.5cm{\vss
\c{\hss\psfig{figure=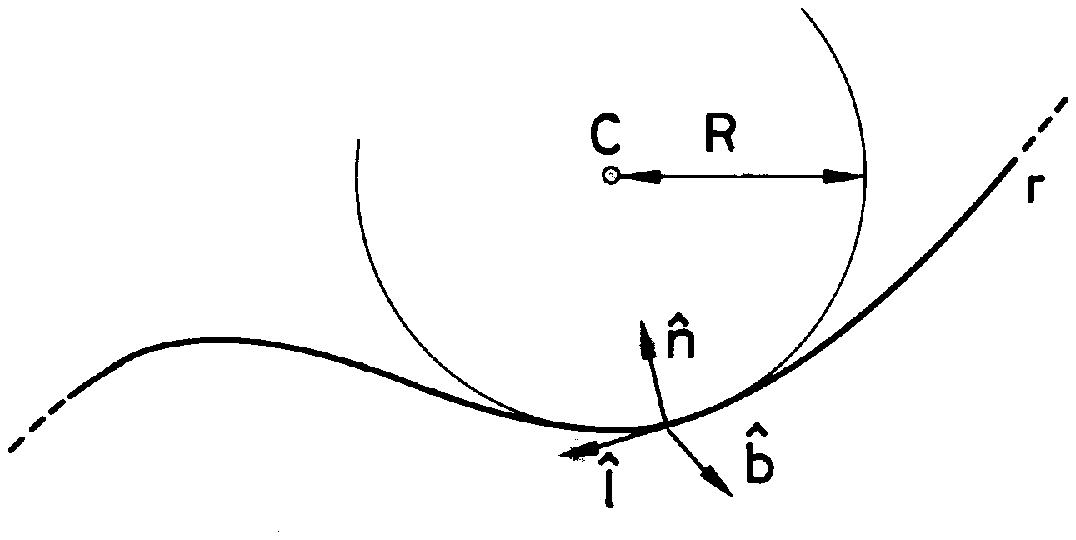,height=6.0cm}\hss}
\vss}

\tenpoint
\baselineskip = 10pt

\n {\bf Fig.\ts{3}:} Sketch of a space curve \rb($l$) described in
each point $P$ by the orthogonal unit
vectors \lu\ (tangent), \nnu\ (normal), \bu\ (binormal, perpendicular to the
plane of the paper), and by the radius of curvature, $R$.

\vfill\eject
\elevenpoint

\n Consider a space curve whose path is described by the variation of the radius 
vector $ \rb = \rb(l)$ with the arc length, $l$, along the curve.\fussn{*}
{In general, the path and all other quantities depend explicitly on time. We
take this into account by writing all spatial derivatives partial but we do not
especially indicate the time dependence unless it actually matters.}
As indicated in Fig.\ts{3} the curve is described at every point, $P$, by the
triad of unit vectors \lu\ ({\it tangent\/}), \nnu\ 
({\it normal\/}), and \bu\ ({\it binormal\/})
which are given by

$$ \eqalignno {
   &{\lu} \; = \; \dl{\,\rb} &(3.1) \cr&&\cr
   &{\nnu} \; = \; R \, \dl{\,\lu} &(3.2) \cr&&\cr 
   &{\bu} \; = \; \lu\; \times\; \nnu &(3.3) \cr } $$

\n The {\it radius of curvature\/} is given by

$$ R \; = \; \left\vert\; \dl{\lu}\; \right\vert^{-1} . \eqno (3.4) $$

\n The vectors \lu\ and \nnu\ define the local plane of the curve (plane of the
paper in Fig.\ts{3}) which contains the local center of curvature, $C$. The
change of the local plane of the curve is described by the derivative of the
binormal unit vector which defines the {\it radius of torsion\/}, $R_t$:

$$ \dl{\bu} \qsq \lu\; \times \; \dl{\nnu} \qsq R_t^{-1} \, \nnu \;.
   \eqno (3.5) $$

\n In the ASF we consider a flux tube as a coherently moving bundle of magnetic
field lines whose paths can be represented by one single space curve ${\bf
r}(l,t)$. For such a description to be valid, the variation of tangent, \lu,
radius of curvature, $R$, and all physical quantities within the cross section
of the tube in the plane perpendicular to \lu\ has to be sufficiently small.
Since it allows a better readable presentation, in what follows we assume a 
circular cross section of the flux tube and take its axis 
as representative space curve and as origin of Taylor expansions. However, the
following considerations are valid for any shape of the cross section and can
easily be generalized as long as the flux tube remains sufficiently thin in
any direction perpendicular to its local tangent. Note in particular that we
do {\it not\/} (and in general are not allowed to) assume axial symmetry.

The natural coordinates within any cross section of the tube are defined
along the normal and binormal directions with the axis as origin. 
Using $\xi$ for the coordinate in the direction of the local normal, \nnu,  we
may write for the differences $\Delta R$ and $\vert\Delta \lu\vert$ between
axis ($\xi = 0$) and boundary ($\xi = a$) of the flux tube in the normal 
direction

$$ \eqalignno {
   {\vert\Delta R\vert \over R} \; &= \; \left\vert {\partial R \over
   \partial \xi} \right\vert_{\xi = 0}  {a\over R} &(3.6) \cr&&\cr
   \vert\Delta \lu\vert  \; &= \; \left\vert {\partial \lu \over
   \partial \xi} \right\vert_{\xi = 0}  a \;. &(3.7) \cr } $$

\n The quantities on the left of \es{3.6} and (3.7), respectively, can be
made arbitrarily small if $a$ is chosen small enough, in particular if the
diameter of the flux tube is everywhere small compared to the local radius of
curvature of the axis.  Similar relations are valid for the
variation in the binormal direction and for the variation of torsion which
are satisfied also by a sufficiently thin flux tube.  

For the (zeroth order) ASF we consider the values of the variables (and
possibly their first spatial derivatives, see below) on the axis of the tube.
The combined equations of induction and continuity (Wal\'en equation for zero
resistivity) and the equation of motion for the fluid {\it within\/} the flux
tube in Lagrangian form read

$$ \eqalignno {
  &{d \over dt} \left( {\Bb \over \rho} \right) \; = \; \left( {\Bb \over \rho}
      \cdot \nabla \right) \ub  &(3.8)
      \cr&&\cr
   &\rho\, \dt{\ub} \; = \; - \nabla p \; + \; \rho\,{\bf g} \; + \;
        \ivp ( \nabla \times \Bb ) \times \Bb  \; + \; {\bf F}_D
   &(3.9) \cr } $$

\n where \ub\ denotes fluid velocity, $\rho$ density, $p$ gas pressure,
{\bf g} gravitational acceleration, and \Bb\ the magnetic flux density vector.
${\bf F}_D$ is the drag force which results from the motion of the flux tube
relative to the surrounding, non-magnetic fluid.  For a sufficiently thin tube
we may write for the flux density on the axis (the representative space curve)

$$ \Bb(l) \; = \; B(l) \, \lu \eqno (3.10) $$

\n and \e{3.8} can be rewritten

$$ {B \over \rho} \dt{\,\lu}\; +\; \lu\, {d \over dt}\left(
   {B \over \rho} \right) \; = \; {B \over \rho} ( \lu \cdot \nabla )
   \ub \; \equiv \; {B \over \rho} \dl{\ub} \eqno (3.11) $$

\n By scalar multiplication with \lu\ and noting that a unit vector is 
perpendicular to its derivative we find from \e{3.11}

$$ {d \over dt}\left( {B \over \rho} \right) \; = \; {B \over \rho}
   \; \lu \cdot \dl{\ub} \; = \; {B \over \rho} \left( \dl{\ub\cdot\lu}
   - \ub \cdot \dl{\,\lu} \right). \eqno (3.12) $$

\n We write $\ub\cdot\lu \equiv u_l, \, \ub\cdot\nnu \equiv u_n$ and use \e{3.2} 
to obtain

$$ {d \over dt}\left( {\rho \over B} \right) = {\rho \over B}\left(
   {u_n \over R} - \dl{u_l} \right) \; . \eqno (3.13) $$

\n \e{3.13} represents the ASF form of the Wal\'en equation.
Multiplication of \e{3.11} with \nnu\ and \bu, respectively,
gives the normal and binormal components of the time derivative of the 
tangent vector which describes the change of the flux tube path in time, namely

$$ \eqalignno {
   &{\nnu} \cdot \dt{\,\lu} \; = \; \dl{u_n} + {u_l \over R} 
   + {u_b \over R_t} &(3.14) \cr&&\cr
   &{\bu} \cdot \dt{\,\lu} \; = \; \dl{u_b} - {u_n \over R_t} &(3.15) 
   \cr } $$

\n where we have used the general relation $\nnu = - \lu \times \bu$ and defined
$\ub\cdot\bu\equiv u_b$.

The equation of motion requires somewhat more
consideration.  Let us first write down the Lorentz force on the axis
using \e{3.10}

$$ \eqalignno {
    \ivp ( \nabla \times \Bb ) \times \Bb \; &= \;
    \ivp \Bigl[\,(\nabla B) \times \lu + B\,\nabla \times \lu\, \Bigr] 
    \times B\,\lu & \cr&&\cr
    &= \; \ivp \Bigl[\, B (\nabla B \times \lu ) \times \lu\, + \,
    B^2 ( \nabla \times \lu ) \times \lu\, \Bigr] &(3.16) \cr }
$$

\n Using $(\nabla \times \lu) \times \lu = R^{-1}\,\nnu$ (\eg Smirnov,
1968, Ch.\ts V) and 

$$ ( \nabla B \times \lu ) \times \lu \; = \; 
   - \nabla B + \lu\, ( \lu \cdot \nabla B ) \; \equiv \; - (\nabla B)_\perp
   \eqno (3.17) $$
   
\n where $(\nabla B)_\perp$ denotes the projection of the gradient on the
plane perpendicular to the tangential direction (\ie the cross section of
the tube) we find for the Lorentz force   

$$ {\bf F}_L \; \equiv \; \ivp ( \nabla \times \Bb ) \times \Bb \; = \;
   - \left[ \nabla \left( {B^2 \over 8\pi} \right) \right]_\perp
   + {B^2 \over 4\pi R} \nnu \;. \eqno (3.18) $$

\n The projections of ${\bf F}_L$ on the three directions defined by the
triad of unit vectors are given by

$$ \eqalignno {
   &{\bf F}_L \cdot \lu \; = \; 0 &(3.19{\hbox {a}}) \cr&&\cr
   &{\bf F}_L \cdot \nnu \; = \; - \,{\partial \over \partial n} 
   \left( {B^2 \over 8\pi} \right)\, +\; {B^2 \over 4\pi R} &(3.19{\hbox {b}})
   \cr&&\cr
   &{\bf F}_L \cdot \bu \; = \;  - \,{\partial \over \partial b} 
   \left( {B^2 \over 8\pi} \right) \; . &(3.19{\hbox {c}}) \cr } $$

\n Here we have defined $\nnu \cdot \nabla \equiv \partial/\partial n$ and $\bu
\cdot \nabla \equiv \partial/\partial b$. We recognize the two familiar
constituents of the Lorentz force, \ie the magnetic pressure force in the plane
perpendicular to the axis and the curvature force in the normal direction.
There is no curvature force in the binormal direction and no magnetic force at
all in the tangential direction. Note that the derivatives in the normal and
binormal directions generally do not vanish since, in contrast to the 
ASF for vertical flux tubes, a curved flux tube cannot assumed to be axisymmetric.

Using the expressions derived for the Lorentz force in \e{3.19} we write for
the components of the equation of motion (Eq.\ts3.9)

$$ \eqalignno {
   &\rho\, \dt{\ub} \cdot \lu \; = \; -\, \dl{p} \; + \; \rho\,{\bf g}\cdot\lu 
   &(3.20) \cr&&\cr
   &\rho\, \dt{\ub} \cdot \nnu \; = \; -\, {\partial \over \partial n}
   \bigl( p + {B^2 \over 8\pi} \bigr)\; +\; \rho\,{\bf g}\cdot\nnu\; +\;
   {B^2 \over 4\pi R}\; +\; {\bf F}_D \cdot \nnu &(3.21) \cr&&\cr
   &\rho\, \dt{\ub} \cdot \bu \; = \; -\, {\partial \over \partial b}
   \bigl( p + {B^2 \over 8\pi} \bigr)\; +\; \rho\,{\bf g}\cdot\bu \; + \;
   \; +\; {\bf F}_D \cdot \bu \,. &(3.22) \cr }  $$

\n The component along the flux tube (\e{3.20}) is already in a suitable form
for the ASF. The other components contain derivatives in the normal and
binormal directions which have to be determined by considering
the external fluid. This is achieved by assuming that at the interface between
flux tube and its environment instantaneous pressure equilibrium (more
precisely, continuity of normal stress) is maintained permanently, viz.

$$ p \; + \; {B^2 \over 8\pi} \; = \; p_e\;. \eqno (3.23) $$

\n Since the time required to establish pressure equilibrium is of the order 
of the travel time  of a fast magneto-acoustic wave across the
tube given by $2a/\sqrt{v_A^2+v_S^2}$ ($v_A$: Alfv\'en speed, $v_S$: sound speed)
it can be made arbitrarily small compared to any other dynamical time
scale of the system if the flux tube is sufficiently thin.

Let us first assume
a straight flux tube embedded in a static external fluid of constant pressure
$p_e$, without gravity. We see from \e{3.23} that under these conditions
pressure equilibrium entails $p + B^2/8\pi = \hbox{const.}$ in each cross
section which is identical with the condition for static equilibrium of the
{\it internal\/} fluid since all terms vanish on the r.h.s. of \es{3.21} and
(3.22) except the derivatives of total pressure.

In the case of a curved flux tube, non-vanishing gravity and stratification of
the external fluid, pressure equilibrium at the interface and static
equilibrium of the fluid in the interior of the flux tube generally cannot be
reached simultaneously and a lateral acceleration of the fluid in the tube
results. Since we deal with the zeroth-order ASF, only the terms involving
first derivatives are retained in the Taylor expansion of total pressure
within the cross section which is inserted into \es{3.21} and (3.22).
On the other hand, these derivatives are already fixed by \e{3.23} since a
linear profile of total pressure along the normal and binormal directions is
determined by the values of $p_e$ at the intersections with the boundary of the
flux tube.  We can therefore formally insert \e{3.23} in \es{3.21} and (3.22).
Note that this procedure is legitimate only for linear pressure profiles, \ie
if the requirements for the validity of the ASF are met.  If we take the
external stratification to be hydrostatic (assuming all motions to be far
subsonic), \ie

$$ \nabla p_e \sgs \rho_e {\bf g} \eqno (3.24) $$

\n ($\rho_e$: density of the external fluid), we obtain for the three
components of the ASF form of the equation of motion\fussn{*}
{The transversal force (per unit length along the tube) can also be obtained by
assuming static equilibrium {\it within\/} the flux tube (in the comoving
frame) and integrating the resulting total pressure difference over the
circumference of the cross section. For a thin tube the result is identical to
\es{3.26/27}. In contrast to the procedure described in the text, this method
is as well applicable for non-thin tubes as long as the assumption of internal
hydrostatic equilibrium is valid. It has been used to calculate the buoyancy
force on a circular flux tube of arbitrary radius (Sch\"ussler, 1979).}

$$ \eqalignno {
  &\rho\, \dt{\ub} \cdot \lu \; = \; -\, \dl{p} \; + \; \rho\, {\bf g}\cdot\lu 
  &(3.20)=(3.25) \cr&&\cr
   &\rho\, \dt{\ub} \cdot \nnu \; = \; (\rho - \rho_e) \,
   {\bf g}\cdot\nnu \; + \; {B^2 \over 4\pi R} \; + \; 
   {\bf F}_D\cdot\nnu &(3.26) \cr&&\cr
   &\rho\, \dt{\ub} \cdot \bu \; = \; (\rho - \rho_e) \,
   {\bf g}\cdot\bu \; + \;
   {\bf F}_D\cdot\bu &(3.27) \cr }$$

\n The first terms on the r.h.s.\ts{of} \es{3.26/27} represent the
components of the buoyancy force which are proportional to the density
difference between internal and external fluid. The magnetic field  
enters explicitly only by way of the curvature force in \e{3.26} and in the
pressure balance condition (Eq.\ts3.23).

Let us now discuss the drag term, ${\bf F}_D$, which has been introduced to
describe the dynamical effect of a motion of the flux tube relative to the
surrounding fluid.  Spruit (1981a,b) considered impulsive motions of a flux
tube in an initially static environment in which case this effect can be
described by a larger effective inertia of the tube with respect to perpendicular
motions. This is introduced into the equations by changing 
$ \rho \to \rho + \rho_e $ on the l.h.s.\ts{of} \es{3.26} and
(3.27). In our case we wish to include arbitrary flow fields around the flux
tube (convection, differential rotation, dynamical motion of the tube itself).
Their effect on the tube has to be described explicitly by considering an
aerodynamic drag force (\eg  Parker, 1975a; Sch\"ussler 1977; 1979;
Moreno-Insertis, 1983; 1986; Chou and Fisher, 1989).  Since the drag force has
its origin in pressure differences between the upstream and the downstream
sides of the interface between flux tube and its environment, for the ASF there is
no component along the tube (${\bf F}_D \cdot \lu = 0$). In the simplest
case we may use the expression for the drag force acting on a rigid circular
cylinder of radius $a$ (\cf Batchelor, 1967), viz.

$$ (\pi a^2)\,{\bf F}_D \; = \; C_D\, \rho_e\, a\, v_\perp^2\, {\hat {\bf k}}
   \;. \eqno (3.28) $$

\n $C_D$ is the (dimensionless) drag coefficient and $v_\perp^2$ is given by

$$ v_\perp^2 \; = \; ( {\bf v} \cdot {\hat {\bf k}})^2
          \; = \; \Bigl[ {\bf v} - \lu\, ({\bf v}\cdot\lu ) \Bigr]^2
             \eqno (3.29) $$

\n where ${\bf v} = {\bf v}_e - \ub$ is the relative velocity between the flux
tube and the surrounding fluid moving with velocity ${\bf v}_e$, and 
${\hat {\bf k}}$ is the unit vector in the direction of the component 
of ${\bf v}$ perpendicular to the tube.  \e{3.28} has been derived
for laminar flows in which case wind tunnel measurements give $C_D \approx 1$
for a wide range ($10^2 < Re < 10^5$) of the hydrodynamic Reynolds number $Re =
v_\perp a/\nu$ ($\nu$: kinematic viscosity).  For turbulent flows the effective
Reynolds number of the flow is determined by the turbulent viscosity which can
be orders of magnitude larger than the molecular value. Similar to the
discussion of the turbulent magnetic diffusivity in the preceding chapter, the
turbulent viscosity must be determined taking into account the spatial scale of
the flow to be described.  For $v_\perp$ given by large-scale convection on 
spatial scale $L$ and a flux tube of radius $a$, the relevant
turbulent viscosity is of the order of $0.1\,u(\delta)\,\delta$ where
$u(\delta)$ is the turbulent velocity on a scale $\delta$ which is somewhat
smaller than $a$. For $ \delta = a/10 $ and using \e{2.8} we find for the
Reynolds number calculated with turbulent viscosity: 
$Re = 100 (L/\delta)^{1/3}$. With $a=100$ km
and $L=10^4$ km we finally estimate $Re \approx 10^3$.  Hence, the effective
Reynolds number for turbulent flow is such that we could hope to stay in the
range of validity of \e{3.28} with $C_D \approx 1$ (see also the detailed
discussion by Moreno-Insertis, 1984).

An enhanced inertia for perpendicular motions as introduced by Spruit (1981a,b)
is not used here since it cannot be easily specified for turbulent flows which
we expect in a convection zone.  This may introduce errors if impulsive
perpendicular motions like those connected with transversal tube waves are 
relevant. 

For some applications it is more convenient to write the inertial terms in 
\es{3.25-3.27} in the form of Lagrangian derivatives of the velocity 
components.\fussn{*}
{Note that there is an error in the expressions given by Chou and Fisher
(1989) who treated a plane flux tube ($R_t \to \infty\,,\,u_b = 0$).
The terms involving the derivative of the normal velocity, 
$\partial u_n/\partial l$, are missing in their equations (1) and (2).} 
For example, the longitudinal component of the
inertial term in \e{3.25} can be rewritten using \es{3.14/15} in the form

$$ \eqalignno {
   \rho\, \dt{\ub} \cdot \lu \; &= \; \rho\,\dt{\ub\cdot\lu} \; - \;
          \rho\,\ub\cdot\dt{\,\lu} \; = & \cr&&\cr
   &= \; \rho\,\dt{u_l} \; - \; \rho\left( u_n\,\dl{u_n} + 
          {u_n\,u_l \over R} + {u_n\,u_b \over R_t}\right) \; - \;
          \rho\left( u_b\,\dl{u_b} - {u_b\,u_n \over R_t}\right) \;.
          &(3.30) \cr }
$$

\n Similar expressions can be derived for the other components of the
inertial force.
The set of equations for the ASF derived so far, \ie the equations of
motion, \es{3.25-3.27}, continuity, \e{3.13}, flux tube shape, \es{3.14/15}, 
and instantaneous pressure balance, \e{3.23}, are complemented by the condition 
of magnetic flux conservation along the tube, namely

$$ A \, \cdot B \; = \; \Phi_{mag} \; = \; \hbox{const.}  \eqno (3.31) $$

\n where $A$ is the cross-sectional area. For a flux tube with circular
cross section of radius $a$ we have $A = \pi a^2$.  The form of the
energy equation and the equation of state is determined by the
particular problem to be treated. Since we shall only consider
adiabatic changes we do not specify more complicated forms of the
energy equation here. In most cases the derivation of the appropriate
ASF form is straightforward.

\vfill\eject 

\def\otopline{ 4. Flux tubes in equilibrium }

\vglue 2cm

\def\etopline{ 4.1 Static equilibrium}

\n {\bbbf 4. Flux tubes in equilibrium}
\b

\n The comparatively long lifetime and slow evolution of large solar active
regions after a much shorter phase of flux eruption and dynamical evolution
give rise to the conjecture that the magnetic structures in the convection zone
reach a static or stationary (\ie with a surrounding flow) equilibrium which is
characterized by a time-independent shape of the tube and hydrostatic
equilibrium along its longitudinal direction. Trivial examples of
{\it static\/} equilibria (for constant direction of gravity, \gb) are:
\m
\item {\it a)} a horizontal (${\bf g}\cdot\lu = 0$), straight flux tube with
      constant pressure and density ($\rho = \rho_e$), and
\s
\item {\it b)} a vertical (${\bf g}\cdot\nnu = 0$), straight flux tube 
      in hydrostatic equilibrium \par 
      ($dp/dz = -\rho g\, ;\, z$:
      vertical coordinate).
\m
\n An example for a {\it stationary\/} equilibrium is a straight, horizontal
flux tube with a density difference, $\rho - \rho_e$, such that the resulting
buoyancy force compensates the drag force due to a constant vertical velocity,
${\bf v}_e$, in the exterior. The force balance, \e{3.26}, can be determined
using
\e{3.28} and gives

$$ 1 \; - \; {\rho \over \rho_e} \; = \; {C_D\,v_e^2 \over \pi\,a\,g} \;
             \hbox{sgn(}{\bf g \cdot v}_e) \eqno (4.1) $$

\n with $v_e\equiv\vert\ve\vert$ and $g\equiv\vert\gb\vert$.
 For a downflow, \ie $\hbox{sgn(}{\bf g \cdot v}_e)=+1$, the density of the
fluid in the tube has to be smaller than that of the surroundings and the
resulting upward directed buoyancy force is balanced by the drag force.  In the
case of an upflow the buoyancy force is directed downwards.

In general, we expect more complicated shapes of equilibrium flux tubes in the
solar convection zone. For example, van Ballegooijen (1982a) calculated static
and stationary equilibrium solutions for flux tubes forming loops which are
``anchored'' in a horizontal flux system below the convection zone.  Anton
(1984) determined equilibrium flux tubes which reside completely within the
convection zone for a variety of internal and external temperature
stratifications. We shall not perform detailed calculations here but rather
derive some general properties of equilibrium flux tubes and give a few
illustrative examples.

\b\b

\n {\bbf 4.1 Static equilibrium}
\b

\n The equations describing the static equilibrium of a flux tube in a
hydrostatically stratified environment are obtained by setting the inertial and
the drag terms to zero in \es{3.25-3.27}. In the direction of the binormal the
equilibrium condition reads

$$ (\rho - \rho_e)\, \gb\cdot\bu \; = \; 0 \; . \eqno (4.2) $$

\n Unless we have $\gb\cdot\bu = 0$ this equation can only be satisfied if the
density difference between flux tube and exterior vanishes everywhere along the
flux tube. From the equilibrium condition in the normal direction,

$$(\rho - \rho_e) \, \gb\cdot\nnu \; + \; {B^2 \over 4\pi R} \; = \; 0
   \; ,\eqno (4.3) $$
   
\n we see that in this case the curvature force vanishes, $R \to \infty$, \ie
the tube has to be straight. Hydrostatics of the environment, \e{3.24}, and
along the tube, \e{3.25}, together with $\rho = \rho_e$ entail

$$  {\partial \over \partial l}\,(p - p_e) = 0 \eqno (4.4) $$

\n and from \e{3.23} we find

$$  {\partial \over \partial l}\, {B^2 \over 8\pi} \; = \; 0\;. \eqno (4.5) $$

\n Consequently, the magnetic field strength is constant. Using the perfect
gas law and assuming the molecular weight to be the same inside and outside
the flux tube we find from \e{3.23} for the ratio of internal temperature,
$T$, and external temperature, $T_e$:

$$ {T \over T_e} \; = \; 1 \, - \, {B^2 \over 8\pi p_e} \;. \eqno (4.6) $$ 

\n For a constant direction of gravity, a horizontal flux tube with no variation 
of $p_e$ in its longitudinal direction, and a subadiabatic stratification of the 
external medium such a temperature reduction might be achieved by an
adiabatically expanding, rising flux tube.  For an oblique tube with a
concomitant variation of $p_e$ along its length we can hardly imagine a
thermodynamic process which precisely leads to a temperature variation along
the tube as prescribed by \e{4.6}.

In a real star the direction of \gb\ varies
spatially since gravity is directed towards the center. In this case, a 
longitudinally uniform flux tube assumes circular (toroidal) 
shape with finite $R$\ and therefore we have $\rho \neq \rho_e$.  Thus in
practice straight flux tubes with $\rho = \rho_e$ are irrelevant and we can
conclude from \e{4.2} that $\gb\cdot\bu = 0$ is necessary, \ie static flux
tubes lie in planes which contain
the vector of gravitational acceleration. For the spherical geometry of a
star we may state:

\m {\narrower\narrower
     \it \n In a spherical star with radial gravity  magnetic flux tubes
         in static equilibrium lie in planes which contain the center of the
         star. The singular case of straight tubes without density contrast
         is of no practical importance.\par}
\m

\n For example, a toroidal flux tube in a plane parallel to but outside the
equatorial plane does not fulfill this condition.  It cannot find a static
equilibrium since the component of the buoyancy force perpendicular to the
plane of the tube is not balanced and leads to a poleward drift (Pneuman and
Raadu, 1972; Spruit and van Ballegooijen, 1982).

Another important property of static flux tubes is obtained by rewriting the
Lorentz force in its familiar form

$$\ivp ( \nabla \times \Bb ) \times \Bb \; = \;
       -\,\nabla \left( {B^2 \over 8\pi} \right) \; + \;
       \ivp\,\Bb\cdot\nabla\Bb \;. \eqno (4.7) $$

\n We may use this to obtain the static form of the equation of motion,
\e{3.9}, viz.

$$ -\,\nabla\left( p + {B^2 \over 8\pi} \right) \; + \; \rho\,\gb \; + \;
     \ivp\,\Bb\cdot\nabla\Bb \; = \; 0 \eqno (4.8) $$

\n and with \es{3.23/24} we get

$$ (\rho - \rho_e)\,\gb \; + \; 
     \ivp\,\Bb\cdot\nabla\Bb \; = \; 0 \; . \eqno (4.9) $$

\n We now consider the general case of a spatially varying direction of gravity
and denote by ${\hat \gb}$ the unit vector in the direction of local
gravitational acceleration. We define \hu\ as unit vector perpendicular to
${\hat \gb}$ within the local plane of the flux tube (spanned by \nnu\ and
\lu). Since $\gb\cdot\bu = 0$ for a static tube we have

$$ \hu \; = \; {\hat \gb} \times \bu \; . \eqno (4.10) $$

\n \hu\ defines the {\it local horizontal direction\/}. We multiply 
\e{4.9} by \hu, use \e{3.10}, and find

$$ \hu\,\cdot\,(\lu\cdot\nabla)\,B\,\lu \; = \; 0 \; . \eqno(4.11) $$

\n This may be written as

$$ 0 \; = \; \hu\cdot\dl{B\,\lu} \; = \; {\partial\over\partial l} 
   B(\hu\cdot\lu) \; - \; B\,\lu \cdot \dl{\,\hu} \;. \eqno (4.12) $$

\n Using \es{4.10}, (3.3) and (3.5) we find

$$ \eqalign {
   \dl{\,\hu} \; &= \; \dl{\,{\hat \gb}} \times \bu \; + \; 
   {\hat \gb} \times \dl{\,\bu} \; = \;
   \dl{\,{\hat \gb}} \times (\lu \times \nnu) \; + \;
   R_t^{-1} {\hat \gb}\times \nnu \; = \cr&\cr
   &= \; \lu \left( \dl{\,{\hat \gb}}\cdot\nnu \right) \; - \;
   \nnu \left( \dl{\,{\hat \gb}}\cdot\lu \right) \; + \;
   R_t^{-1} {\hat \gb}\times \nnu \; . \cr } $$

\n Consequently, we have

$$ \lu \cdot \dl{\,\hu} \; = \; \dl{\,{\hat \gb}}\cdot\nnu $$

\n and defining $B_h \equiv B(\hu\cdot\lu) = \Bb\cdot\hu$ we write \e{4.12} as

$$ \dl{B_h} \; = \; B \left(\dl{\,{\hat \gb}}\cdot\nnu\right)\;.\eqno(4.13)$$

\n This equation couples the variation of the strength of the field component
in the local horizontal direction to the variation of the direction of gravity
along the flux tube. In the case of a constant direction of gravitation along
the tube we find from \e{4.13}

$$ \dl{B_h} \; = \; 0 \;, \eqno(4.14) $$

\n \ie the component of the magnetic field perpendicular to gravity is
constant in static equilibrium (see also Parker, 1979, Sec.\ts8.6, and van
Ballegooijen, 1982a).

As we have seen above, static flux tubes in a sphere with radially directed 
gravity lie in planes which contain the center. If we introduce polar coordinates
$(r,\varphi)$ within such a plane, we have ${\hat \gb} = (-1\,,\, 0)$,
$\hu=(0\,,\, 1)$, $\lu=(l_r\,,\, l_\varphi)$, and

$$ \dl{\,{\hat \gb}} \; = \; \lu\cdot\nabla{\hat\gb} \; = \;
   \left( 0\,,\,{-l_\varphi \over r}\right) \; . $$
   
\n Hence, we find from \e{4.13} with $B_h = B(\hu\cdot\lu) = B l_\varphi \equiv
B_\varphi$ :

$$ \dl{B_\varphi} \; = \; -B\, {l_\varphi n_\varphi \over r} \; = \;
  -B_\varphi \, {n_\varphi \over r} \;. \eqno(4.15) $$

\n For $r \to \infty$ this passes over into \e{4.14}.
We may use \e{4.15} to estimate the difference between the
spherical and the plane-parallel case. If we denote by $\delta B_\varphi$ the
change of $B_\varphi$ over a length interval $\delta L$ along the tube an upper
limit for this quantity given by \e{4.15} is

$$ {\delta B_\varphi \over B_\varphi} \; \le \; {\delta L \over r} 
   \; . \eqno(4.16) $$

\n As an example, for a flux tube extending nearly vertically through the whole
depth of the convection zone we have $\delta L = 2\cdot10^5$ km and $r =
5\cdot10^5$ km. Consequently, the change of $B_\varphi$ between top and bottom
of the convection zone is smaller than $0.4\,B_\varphi$. If this flux tube is
anchored in a toroidal flux system near the bottom of the convection zone with
$B_\varphi \approx B_e \approx 10^4$ Gauss, the toroidal field strength at the
top (in the photosphere) cannot be smaller than $6000$ Gauss for a tube in
static equilibrium. Since photospheric magnetic fields and sunspots are
basically vertical with very small net inclination of the magnetic structures
as a whole (the mean horizontal field component is less than 100 Gauss) this
excludes {\it static\/} flux tubes rooted in a toroidal equipartition field
deep in the convection zone as models for sunspots and solar active regions.
Since flux expulsion always tends to establish equipartition field strengths
which are much larger than 100 Gauss the concept of large-scale {\it static\/}
magnetic structures in the convection zone has to be abandoned.

In spite of these pessimistic remarks we shall continue the discussion of static 
equilibrium in the remainder of this section because this 
case is well suited to demonstrate the mathematical methods which are used
to calculate flux tube equilibria in practice. Furthermore, static equilibrium is 
only excluded on a large scale for flux tubes extending through the whole 
convection zone but it might well be {\it locally\/} a reasonable approximation,
for instance for nearly vertical flux tubes in the photosphere. For such 
a small region we neglect the spherical geometry of the star and assume a
constant direction of gravity. The subsequent considerations follow the
approach first proposed by van Ballegooijen (1982a; see also Parker, 1975c).
 
Consider cartesian coordinates $(x,z)$ in a plane which contains the vector
of gravitational acceleration, $\gb = (0\,,\,-g)$. If the external medium is in
plane-parallel hydrostatic equilibrium the path of a static flux tube is given by 
a curve $z=z(x)$ which satisfies \e{4.3}. This situation is sketched in 
Fig.\ts{4}. Hydrostatic equilibrium along the flux tube is described by \e{3.25} 
which gives\fussn{*}
{Since we deal with time-independent quantities we may write non-partial
derivatives.}

$$ -\,\Dl{p} \; + \; \rho\,\gb\cdot\lu \; = \; 0 \; . \eqno (4.17) $$

\n Uniformity of the horizontal field component entails, by virtue of \e{4.14}

$$ B_x \; = \; \hbox{const.} \eqno (4.18) $$

\n Since we may measure the arc length in both directions along the tube, we
remove this ambiguity by requiring $\lu\times{\hat{\bf x}}> 0$ where ${\hat{\bf
x}}$ is the unit vector in $x$-direction. If the shape of the tube is such that
it has a vertical tangent somewhere and turns backwards with respect to $x$
this part of the tube has to be treated separately in the same way as described
below after transforming  $\lu\to -\lu$.  We use the angle $\gamma(l)$ between the
local tangent of the flux tube and the positive $x$-axis defined by

$$ {dz \over dx} \; = \; {B_z \over B_x} \; = \; \tan\gamma 
   \quad ; \quad \Dl{} \; = \; \sin\gamma\,{d \over dz}   \eqno (4.19) $$

\n to write

$$ \eqalignno {
    {\lu} \; &= \; (\cos\gamma , \sin\gamma) &(4.20) \cr&&\cr
    {\nnu} \; &= \; R\, (-\sin\gamma , \cos\gamma) \,\Dl{\gamma} 
       &(4.21) \cr&&\cr
     R^{-1} \; &= \; \left\vert\, \Dl{\gamma}\,\right\vert \; = \;
       \left\vert\, -{d\cos\gamma \over dz} \right\vert \; = \;
       \left( -{d\cos\gamma \over dz} \right)\, \hbox{sgn}
       \left( \Dl{\gamma} \right)\,. &(4.22) \cr&&\cr } $$

\vbox to 8.5cm{\vss
$$\c{\hfill\psfig{figure=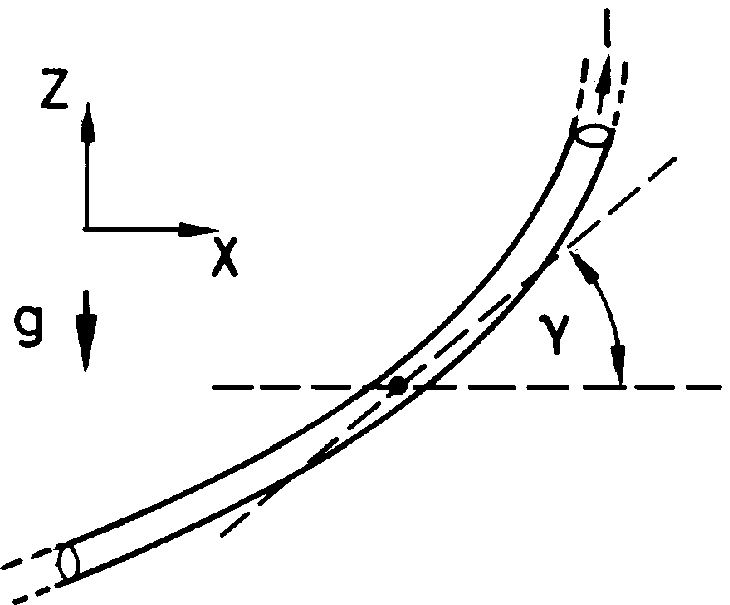,height=7.0cm}\hfill}$$
\vss}

\tenpoint
\baselineskip = 10pt
 
\n {\bf Fig.\ts{4}:} Flux tube in cartesian geometry with gravity directed
downward. $\gamma(l)$ is the angle between the flux tube (direction of
increasing arc length) and the $x$-axis.

\vglue 0.5cm
\elevenpoint

\n Since $\gb\cdot\lu = -g\sin\gamma$ we can rewrite \e{4.17} using \e{4.19}

$$    \dz{p} \; = \; - \rho g \eqno (4.23) $$

\n and in the external medium we have

$$    \dz{p_e} \; = \; - \rho_e g \; . \eqno (4.24) $$

\n Consequently, for given external plane-parallel stratification and given 
internal temperature
profile, $T(z)$, pressure and density {\it within\/} the flux tube depend
only on $z$ and can be determined without prior knowledge of its path.
This applies also to the field strength, viz.

$$    {d\over dz}\left( {B^2\over 8\pi} \right) \; = \;
      {d\over dz} (p_e - p) \; = \; (\rho - \rho_e) g \,. \eqno (4.25) $$

\n With $\gb\cdot\nnu = -g\cos\gamma\,\hbox{sgn}(d\gamma/dl)$ and using
\e{4.25} we find for the sum of the curvature and buoyancy forces which determines
the force balance in normal direction (\cf Eq.\ts{4.3})

$$    0 \; = \; -{d\over dz}\left( {B^2\over 8\pi} \right) \, \cos\gamma
        \; - \; {B^2 \over 4\pi}\, {d\over dz}\cos\gamma \; = \;
        - {B \over 4\pi}{d \over dz}(B\cos\gamma) \; = \; 
      - {B \over 4\pi}\dz{B_x}\,. \eqno (4.26) $$

\n We see that the uniformity of the horizontal component of the magnetic
field is sufficient for static equilibrium if the internal stratification
along the flux tube is hydrostatic. We can use this property to
reduce the calculation of the flux tube path, $z=z(x)$, to a quadrature. Since

$$ 8\pi (p_e - p) \; = \; B^2 \; = \; B_z^2  + B_x^2 \eqno (4.27) $$

\n we can write

$$  {dz \over dx} \sgs {B_z\over B_x} \sgs \left(
    {8\pi(p_e-p) \over B_x^2} \, - 1 \right)^{1/2} . \eqno(4.28) $$
    
\n Since $B_x =$ const. and the pressures are known functions of $z$,
the function $x(z)$ can be determined by integration:

$$  x(z) - x(z_0) \sgs \int_{z_0}^{z} \left(
    {8\pi(p_e-p) \over B_x^2} \, - 1 \right)^{-1/2} d{\tilde z}
    \eqno (4.29) $$

\n The resulting function may be inverted to yield the path, $z=z(x)$.  For
special cases, analytical solutions of \e{4.29} can be obtained (\eg Parker, 1979a,
Sec.\ts8.6).

\b\b

\def\etopline{ 4.2 Stationary equilibrium}

\n {\bbf 4.2 Stationary equilibrium}

\b

\n As we have seen above, flux tubes in static equilibrium do not seem to be
particularly relevant for the description of magnetic structures in the
convection zone. On the other hand, the slow evolution of active regions after
the vigorous dynamical phase of flux eruption indicates some kind of underlying
equilibrium structure. Therefore it seems worthwhile to include the effect of
large scale external velocity fields (convection, differential rotation) and
to consider stationary equilibria of flux tubes. In practice, such an
equilibrium can be determined either by direct (numerical)
integration of a second order ordinary differential equation or by solving a
first order equation and subsequent quadrature. Let us first consider the
latter method (van Ballegooijen, 1982a).

We start from the plane-parallel geometry used in the preceding section and
introduce a relative velocity, \vb, between flux tube and external medium. 
\vb\ is assumed to lie in the $xz$-plane. The unit vector ${\hat{\bf k}}$ 
defined in \es{3.28/29} in this case is given by

$$ {\hat{\bf k}} \sgs \nnu\,\hbox{sgn}(\vb\cdot\nnu)\,. \eqno(4.30) $$

\n We now determine

$$ \vb\cdot\nnu \sgs (-v_x\sin\gamma + v_z\cos\gamma)\,\hbox{sgn}
   \left( \Dl{\gamma} \right) \; \equiv \; v_\perp \,\hbox{sgn}
   \left( \Dl{\gamma} \right)  $$
   
\n and find

$$ \sgn{\vb\cdot\nnu} \sgs \sgn{v_\perp}\,\hbox{sgn}
   \left( \Dl{\gamma} \right)\,. \eqno (4.31) $$    

\n Using \es{3.26} and (3.28) force balance in the normal direction 
leads to

$$(\rho - \rho_e) \, \gb\cdot\nnu \; + \; {B^2 \over 4\pi R} \; + \;
  {C_D \rho_e v_\perp^2 \over \pi\,a}\, {\hat{\bf k}}\cdot\nnu \sgs 0 \;.
   \eqno (4.32) $$

\n In analogy to \e{4.26} we find using \es{4.30/31}

$$ {B \over 4\pi}\,\dz{B_x} \sgs 
   {C_D \rho_e v_\perp^2 \over \pi\,a}\, \sgn{v_\perp} \,. \eqno (4.33) $$

\n In principle, this equation can be used to determine the variation of the
horizontal field component with height which then can be inserted in \e{4.29} to 
determine the path of the flux tube. However, since $v_\perp$ depends on $\gamma$,
the differential equation (4.33) in general is not easily solved, especially if
\vb\ varies spatially.

As an example, let us consider the 
case of a purely horizontal velocity field $\vb = (v(z),\, 0)$ which
may represent a depth-dependent differential rotation.  We have $v_\perp =
-v\sin\gamma$ and assume $v\geq 0$. \e{4.33} now reads

$$ {B \over 4\pi}\,\dz{B_x} \sgs 
   {C_D \rho_e v^2 \over \pi\,a} \sin^2\gamma\,. \eqno (4.34) $$

\n With $\pi a^2 B = \Phi_{mag} =$ const. and $B_x/B = \cos\gamma$
we find

$$ \dz{B_x} \sgs -\, 4 C_D \rho_e v^2 
       \left( \pi\over \Phi_{mag} B \right)^{1/2} 
       \left( 1 - {B_x^2 \over B^2} \right) .
       \eqno (4.35) $$

\n In the special case of a {\it constant magnetic field\/}, $B(z) = B_0$, the
variables can be separated and defining $y \equiv B_x/B_0$ we may write

$$ {dy \over 1 - y^2} \sgs -\,\alpha\,\rho_e v^2\,dz  \eqno (4.36) $$

\n with

$$ \alpha \sgs 4 C_D \left( {\pi\over\Phi_{mag} B_0^3} 
   \right)^{1/2} . \eqno(4.37) $$

\n As we have discussed in the preceding section the case of a constant
magnetic field generally is of not much practical interest because it
requires $\rho=\rho_e$ and, therefore, a very special internal temperature
profile. However, if the scale height is much larger than the height range
covered by the flux tube equilibrium path the variation of density is small
and the assumption of a constant field may be tenable. Such a situation can
be expected near the bottom of the solar convection zone. Integration of \e{4.36}
yields

$$ {1\over2}\,\ln\left( {1\over c_0}\,{y-1\over y+1} \right) \sgs
   \int_0^z \alpha\,\rho_e(\tz) v^2(\tz)\,d\tz \; \equiv \;
   f(z) \eqno(4.38) $$

\n where $c_0$ is a constant of integration which is determined by a boundary
condition at $z=0$. Obviously we must have $y\neq1$ in \e{4.38}. The case $y=1$ is
a singular solution of \e{4.36} and represents a horizontal flux tube
for which all forces vanish individually. Excluding this case we can determine
$y(z)$ from \e{4.38}

$$ y(z) \sgs {1 + c_0 e^{2f(z)} \over 1 - c_0 e^{2f(z)}} \eqno (4.39) $$

\n and determine $c_0$ by specifying $y(0) = y_0 < 1$, \ie

$$ c_0 \sgs {y_0 - 1 \over y_0 + 1}\,. \eqno(4.40) $$

\n For $y_0\in[0,1)$ the value of $c_0$ passes through the interval
$c_0\in [-1,0)$. In the case $y_0 = 0$ (vertical tube at $z=0$) we have
$c_0 = -1$ and \e{4.39} gives

$$  y(z) \sgs -\,\tanh[f(z)]\,. \eqno(4.41) $$

\n Since we have

$$ z'(x) \sgs \tan\gamma \sgs {(B^2-B_x^2)^{1/2} \over B_x} \sgs
         {(1-y^2)^{1/2} \over y} \eqno(4.42) $$

\n integration yields

$$ x(z) \sgs \int_0^z {y \over (1-y^2)^{1/2}} \, d\tz \eqno (4.43) $$

\n where we have assumed $x(0)=0$. Inserting \e{4.39} into \e{4.43} we find

$$ x(z) \sgs \int_0^z {1+c_0 e^{2f(\tz)} \over 2\sqrt{-c_0} e^{f(\tz)}}
   \,d\tz \eqno(4.44) $$

\n and the special case $c_0 = -1$ gives 

$$ x(z) \sgs \int_0^z -\sinh[f(\tz)]\,d\tz\;. \eqno(4.45) $$

\n We give two examples for which \e{4.44} can be directly integrated. First
assume a velocity field with constant kinetic energy density, \ie
$(\rho_ev^2)(z) =$ const. Consequently, we have from \e{4.38} that
$f(z) = \alpha\rho_e v^2 z \equiv \al z$ and thus

$$ x(z) \sgs {1\over 2\al\sqrt{-c_0}}\, \Bigl( c_0 e^{\al z} - e^{-\al z}
        + 1 - c_0 \Bigr) \,. \eqno(4.46) $$

\n For an initially vertical tube, \ie $c_0 = -1$, we have

$$ x(z) \sgs \al^{-1} \Bigl( 1\, - \, \cosh(\al z) \Bigr) \eqno(4.47) $$

\n and normalizing length by $1/\al$, viz. ${\hat z} \equiv \al z$ and
${\hat x} \equiv \al x$, we find 

$$ {\hat x} \sgs 1 \, - \, \cosh {\hat z}\,. \eqno (4.48) $$

\n This solution has earlier been given by Parker (1979d). Another directly 
solvable example is the case of kinetic energy density being proportional to
$z$, viz.

$$ (\rho_e v^2)(z) \sgs (\rho_e v^2)(z_0) \left( {z\over z_0} \right)
   \eqno (4.49) $$

\n with some reference level $z_0$. Inserting \e{4.49} into \e{4.38} we have

$$ f(z) \sgs {\alpha (\rho_e v^2)(z_0) \over 2z_0}\,z^2  \; \equiv \;
       \tal z^2 \eqno (4.50) $$

\n and thus write \e{4.44} as

$$ x(z) \sgs {1\over 2\sqrt{-\tal c_0}} \left( 
        \int_0^{\sqrt{\tal}z} e^{-w^2}\,dw +
        c_0 \int_0^{\sqrt{\tal}z} e^{w^2}\,dw \right)\,. \eqno(4.51) $$

\n The first integral essentially represents the error function
${\bf \Phi}(\sqrt{\tal}z)$ and the second is related to the Dawson integral
${\bf \Psi}$ (\cf Abramowitz and Stegun, 1965)

$$ {\bf \Psi}(u) = e^{-u^2} \int_0^u e^{w^2}\,dw \;. \eqno (4.52) $$

\n For $c_0 = -1$ we have

$$ {\hat x} \sgs {1\over 2} \left( {\sqrt{\pi}\over 2} {\bf\Phi}({\hat z})
   \, - \, e^{{\hat z}^2}{\bf \Psi}({\hat z}) \right) \eqno(4.53) $$

\n with ${\hat z} =\sqrt{\tal}\,z$, ${\hat x} =\sqrt{\tal}\,x$. For both
examples, Fig.\ts{5} shows the resulting flux tube shape.

\vbox to 12.0cm{\vss
$$\c{\hfill\psfig{figure=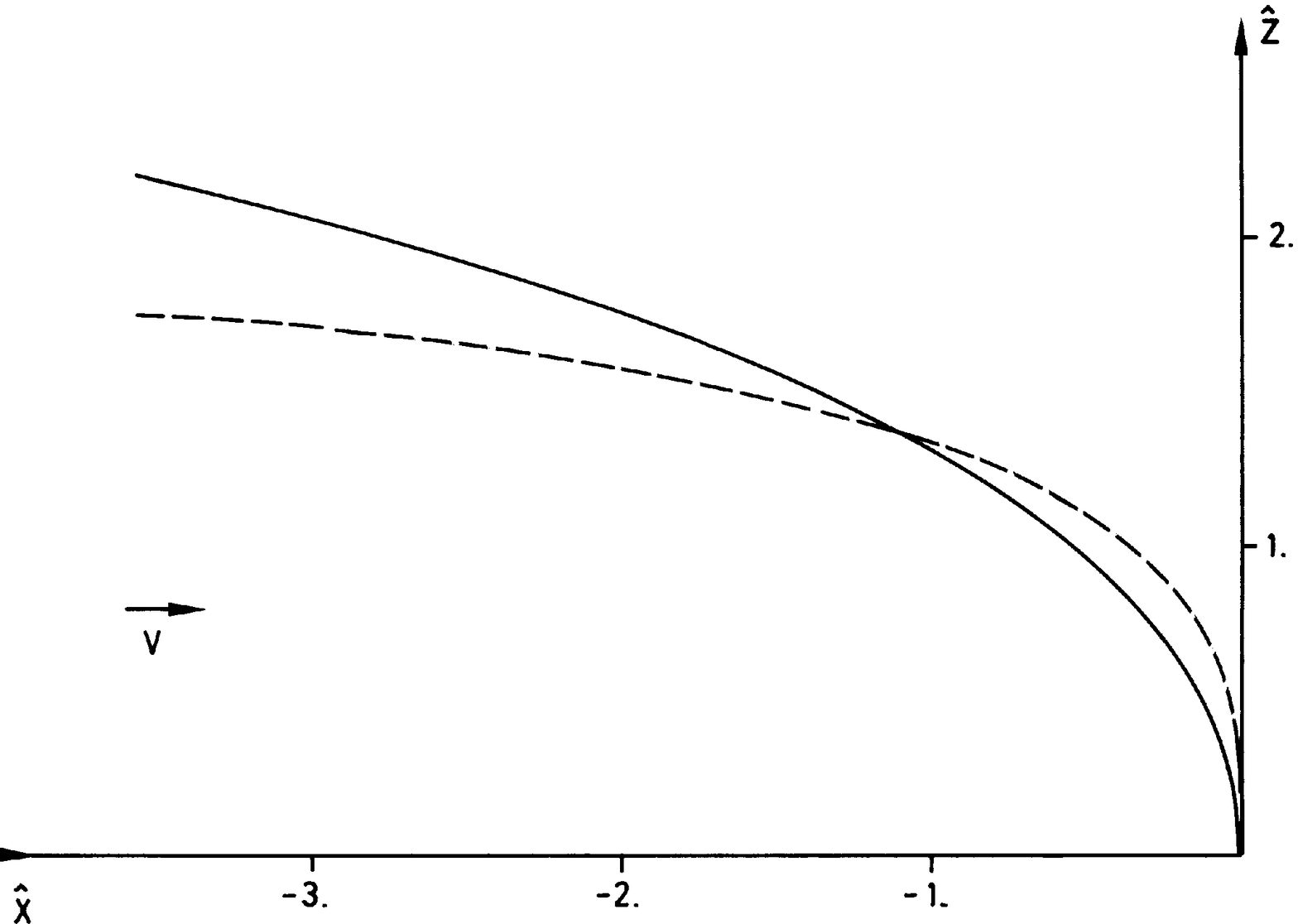,height=10.0cm}\hfill}$$
\vss}

\tenpoint
\baselineskip = 10pt

\n {\bf Fig.\ts{5}:} Stationary equilibrium shape of flux tubes under the
influence of a horizontal velocity field with kinetic energy density $\rho v^2$
constant with height $z$ (full line) and proportional to $z$ (dashed line). The
direction of the flow is from left to right. Note that generally the length
scales are not equal for both examples.

\vglue 0.5cm
\elevenpoint

\n We see that the flux
tube turns {\it towards\/} the flow because only in this way a balance of
forces is possible. This is in contrast to a tree bent by a storm for which
the differential tension force due to bending acts in the opposite direction
of the normal vector. The path given by \e{4.53} and shown by the dashed line
has a smaller curvature for small $\hat z$ than that for constant kinetic energy 
density (Eq.\ts{4.48}, full line) but as $\hat z$ increases it quickly bends
over.  In the case $2 z_0 = 1/\al$ the length scales for both cases are equal
and the curves can be directly compared.  We may use these results to 
estimate the influence of velocity fields on slender flux tubes in the lower
parts of the solar convection zone. Using the values

\m
\hskip 1.5cm $\rho_e = 0.2$ g$\cdot$cm$^{-3}$ \par
\hskip 1.5cm $B_0 = 10^4$ Gauss (equipartition) \par
\hskip 1.5cm $\Phi_{mag} = 10^{18}...10^{20}$ mx ($a = 6\cdot 10^6...6\cdot
             10^7$ cm) \par
\hskip 1.5cm $v_0 = 10^4$ c\ms (convective flow, differential rotation)\par
\hskip 1.5cm $z_0 = 6\cdot10^9$ cm (equal to the pressure scale height) \par
\m

\n we find

\m
\hskip 1.5cm $\al^{-1}\;\;\, \approx 10^7...10^8$ cm \par
\hskip 1.5cm $\tal^{-1/2} \approx 3\cdot10^8...10^9$ cm. \par
\m

\n We see in Fig.\ts{5} that the flux tube paths become almost horizontal at
typical heights ${\hat z} \approx 1...3$ which refers to heights of the order
of $\al^{-1}$ and $\tal^{-1/2}$, respectively, for the two cases. This means that 
in equilibrium an initially vertical flux tube cannot intrude significantly into 
a layer of horizontal flow unless either the field strength is much larger
than the equipartition value, the horizontal flow speed is much smaller than
the typical convective velocities, or the radius of the flux tube is of the
order of the scale height. The dominant r\^ole of external flow fields in
the dynamics of thin flux tubes is an important effect which must be taken into
account when discussing the properties of magnetic structures in
stellar convection zones (see Ch.\ts6).

For other velocity fields $\vb(x,z)$, \e{4.33} in most cases has to be solved by
numerical forward integration in height starting from a suitable initial point. 
The appropriate value of $v_\perp$ for each point is calculated using the
earlier determined angle $\gamma$ and location $x(z)$. The value of $B_x$ which 
results from \e{4.33} can then be used to calculate the values of $\gamma$ and 
$x(z)$ at the next point using \es{4.19} and (4.28/29). Both steps of this 
procedure which correspond to two integrations can be combined in the solution of 
a second order differential equation for the path $z(x)$ (\cf Sch\"ussler, 1980a; 
Parker, 1982c; Anton, 1984).  To this end we use the
relation between radius of curvature and the derivatives, $z'\equiv dz/dx$,
$z''\equiv d^2z/dx^2$, and write for the curvature force

$$ {B^2\over 4\pi R}\, \nnu \sgs {B^2\over 4\pi}\,{z''\over (1+z'^2)^{3/2}} 
   \, \nnu  \,. \eqno (4.54) $$

\n Since we have $\cos\gamma = (1+z'^2)^{-1/2}$ the condition for stationary
equilibrium, \e{4.32}, can be written as

$$(\rho - \rho_e) \, g \; + \; {B^2 \over 4\pi}\,{z''\over 1 + z'^2} \; + \;
  {C_D \rho_e v_\perp^2 \over \pi\,a}\,(1+z'^2)^{1/2}\sgn{v_\perp} \sgs 0
   \eqno (4.55) $$

\n Sometimes it proves useful to rewrite the buoyancy term with aid of 
the relation

$$ {4\pi(\rho_e - \rho)g \over B^2} \sgs - {4\pi\over B^2} {d\over dz}
   \left( {B^2 \over 8\pi} \right) \sgs {1\over 2}{d \over dz} 
   \ln \left( {B^2 \over 4\pi} \right) \,. \eqno (4.56) $$

\n It depends on the properties of the particular problem which of the two
possible ways to calculate the path $z(x)$, \ie \es{4.29/4.33} or \e{4.55}, is
more appropriate. Other examples for the calculation of stationary flux tube
equilibria have been given by Parker (1979d; 1982c,d) for horizontal flows and
for idealized cellular velocity fields, by van Ballegooijen (1982a) for a
constant horizontal drift of flux tubes in the solar convection zone, and by
Anton and Sch\"ussler (unpublished) for giant convective cell patterns (see also
Moreno-Insertis, 1984).

\vfill\eject

\def\otopline{ 5. Stability of flux tubes }

\vglue 2cm

\def\etopline{ 5.1 Previous work }

\n {\bbbf 5. Stability of flux tubes}
\b\m

\n Static or stationary equilibrium configurations of flux tubes can only have 
a practical relevance if they are at least linearly stable, \ie if the
flux tube returns to its equilibrium position and shape after a small
displacement. In Ch.\ts2 we have discussed mechanisms which lead to fragmentation 
of large magnetic structures and to the formation of flux tubes which are much 
smaller than the scale height in the deep parts of a convection zone.
Such tubes are strongly influenced by motions in
their environment like convection, differential rotation, and meridional
circulation.  Even if a flux tube is stable with respect to fragmentation and
reaches a static or stationary equilibrium characterized by a balance of
buoyancy, tension, drag and rotationally induced (Coriolis, centrifugal)
forces, this equilibrium might be unstable due to

\m
\item{\it a)} superadiabaticity of the environment,
\s
\item{\it b)} loop formation with downflows from the crests to the troughs
              with concomitant perturbations of magnetic buoyancy
              (akin to the instability discussed by Parker, 1966),
\s
\item{\it c)} gradients of the drag force exerted by external flows,
\s
\item{\it d)} differential rotation.
\m

\n These instabilities lead to motion and deformation of the flux tube as a
whole and can be treated within the framework of the approximation of slender 
flux tubes.  This allows a considerable simplification of the mathematics and 
simultaneously excludes the fragmentation instabilities (Rayleigh-Taylor, 
Kelvin-Helmholtz) discussed in Ch.\ts2. 

\b\b

\n {\bbf 5.1 Previous work }

\b

\n Vertical flux tubes in the (sub)photosphere of the Sun are liable to
a convective instability (convective collapse) caused by the superadiabatic
stratification of the surrounding fluid (Parker, 1978; Webb and Roberts, 1978;
Spruit and Zweibel, 1979; Unno and Ando, 1979). It is believed that this
process is responsible for the amplification of small-scale solar magnetic
fields far beyond the equipartition field strength (for a more detailed
discussion see Sch\"ussler, 1990).

Apart from preliminary studies 
(\eg Sch\"ussler, 1980a) the first detailed stability analysis of flux tubes 
{\it within\/}  a convection zone has been presented by
Spruit and van Ballegooijen (1982) who analyzed the stability of horizontal,
non-buoyant ($\rho = \rho_e$) flux tubes in cartesian geometry and of toroidal
flux tubes in static equilibrium between buoyancy and tension force in
spherical geometry. They found instabilities which represent a mixture between the
convective and the (Parker type) kink\fussn{*}
{Not to be confused with the ``classical'' kink instability of a plasma pinch
caused by the azimuthal magnetic field component: The kink instability 
discussed here is driven by magnetic buoyancy and superadiabaticity.}
instabilities ({\it a\/} and {\it b\/} above).  The perturbations leading to kink 
instability must have a finite wavelength in the direction along the tube. 
It turns out that
all flux tubes embedded in a superadiabatic environment are unstable and that
kink instability occurs even for slightly subadiabatic stratification.
Moreno-Insertis (1984,\ts1986) performed numerical simulations of the
nonlinear evolution of kink-unstable horizontal flux tubes at the bottom of the
solar convection zone. He found that while the upper part of the unstable loop
rises towards the solar surface, the lower part sinks down and enters the
subadiabatic region below the convection zone where it reaches a stable
equilibrium.  Moreno-Insertis (1984) also investigated the influence of the
drag force exerted by a prescribed giant convective velocity cell on the
evolution of the instability. Recently, Choudhuri (1989; see also Choudhuri and
Gilman, 1987) has included rotation
in a numerical study of kink-unstable flux tubes in the solar convection zone.
He found in most cases that the Coriolis force dominates the dynamics and leads
to a trajectory of the rising loop which is parallel to the axis of rotation.
Consequently, the unstable loops break through the surface
far away from the equatorial regions where solar activity predominantly is
observed. A more radial eruption of unstable loops is only achieved for
quite strong fields (about $10^5$ Gauss) for which buoyancy becomes the
dominating force.

Van Ballegooijen (1983) continued the analytical stability study of toroidal
flux tubes by including (differential) rotation of the external medium and 
a difference between the rotation rate of the gas within and outside of the flux 
tube. Such a difference (a longitudinal flow along the tube with respect 
to a coordinate system which corotates with the external gas) may arise due to
conservation of angular momentum if the flux tube is carried by an equatorward
meridional circulation in the lower part of the convection zone and thus
increases its distance from the axis of rotation (\cf van Ballegooijen, 1982b).
The Coriolis force caused by slower rotation of the internal gas helps
to balance the component of the buoyancy force perpendicular to the axis of
rotation. For flux tubes situated in the equatorial plane,
{\it rigid\/} rotation and longitudinal flow have a stabilizing effect.
However, in the parameter regime relevant for
the lower convection zone of the Sun the stability properties are determined by
{\it differential\/} rotation: The flux tube is stable if
$\partial\Omega_e/\partial r > 0$, i.e. if the angular velocity of the external
medium increases radially outward; it is unstable with respect to
non-axisymmetric disturbances (growing waves along the tube) if $\Omega_e$
decreases.  Similar to the buoyancy-driven kink instability it is the downflow
along the legs of a loop which triggers the instability due to differential 
rotation, in this case by introducing differential Coriolis forces in the 
radial direction.

The more general case of toroidal flux tubes outside the equatorial plane 
including meridional circulation has been treated by van
Ballegooijen and Choudhuri (1988).  In equilibrium, the component of the drag 
force perpendicular to the plane of the tube balances the corresponding
component of the buoyancy force and thus removes the `poleward slip instability'
(Pneuman and Raadu, 1972; Spruit and van Ballegooijen, 1982). The authors found
that an increase of the velocity of meridional circulation in radial direction
has a stabilizing effect on the flux tube. However, their analysis is 
restricted to rigid rotation of the exterior and to axisymmetric perturbations
such that the whole class of kink instabilities induced by buoyancy 
or differential rotation is neglected. The stability properties of toroidal
flux tubes under these conditions remain to be investigated.

For a totally different application, the stability and interaction of jets
from active galactic nuclei, Achterberg (1982, 1988) has derived a
formalism which is similar to these approaches and also to the formalism
developed in this work.

In this chapter we present a formalism which can be used to analyze the
stability of general static and stationary flux tube equilibria in a plane with
constant direction of gravity.  It can be applied to any given plane
equilibrium path $\rb(l)$ along which all quantities may vary.  Such a
formalism is needed in order to determine the stability properties of a number
of non-trivial flux tube equilibria like the examples given in Sec.\ts4.2 or
the loop structures calculated by van Ballegooijen (1982a) and
Anton (1984). We include a velocity field of arbitrary structure but, in order
to limit the complication of the already somewhat involved formalism, we have
refrained from treating the spherical case and also ignored rotation.  However,
this restriction is not fundamental and will be dropped in future work.

\vfill\eject

\def\etopline{ 5.2 Equilibrium}

\n {\bbf 5.2 Equilibrium}
\b

\n We assume a plane flux tube in static or stationary equilibrium described by
the time-independent form of \es{3.25} and (3.26). We take $\gb\cdot\bu=0$ and
assume that the external velocity is restricted to the plane of
the equilibrium tube such that both terms on the \rhs of \e{3.27} (binormal
direction) vanish. We continue to use the notation introduced in Chs.\ts3 and 4.
Denoting all equilibrium quantities by a suffix `0', the
equilibrium path of the flux tube is given by $\rb_0(l_0)$ and the normal,
$\nnun(l_0)$, and tangential, $\lun(l_0)$, unit vectors as well as the radius
of curvature, $R_0$, are defined in the usual way as functions of the
equilibrium path length, $l_0$:

$$ \lun \sgs \dln{\,\rb_0}, \quad \nnun \sgs \Rn\,\dln{\,\lun}, \quad
   \Rn \sgs \left\vert \dln{\,\lun} \right\vert ^{-1}. \eqno (5.1) $$

\n The equilibrium state is characterized by hydrostatic equilibrium in
the longitudinal direction (\cf Eq.\ts{3.25})

$$ 0 \sgs -\dln{p_0} \sps \rho_0\, \gb_0\cdot\lun \eqno(5.2) $$

\n and by a balance of buoyancy, curvature and drag force in the
normal direction (\cf Eq.\ts{3.26})

$$ 0 \sgs (\rho_0-\rho_{e0})g_0\gun \sps {B_0^2 \over 4\pi R_0} \sps
   {C_D\rho_{e0}(\ve_0\cdot\nnun)^2\sgn{\ve_0\cdot\nnun} \over \pi a_0}
   \eqno(5.3) $$

\n $(\gb_0 = g_0{\hat{\gb}}_0)$. \es{5.2/3} are complemented by the conditions
of pressure balance, \e{3.23}, and flux conservation, \e{3.31}, and by 
hydrostatic equilibrium of the external medium, \e{3.24}.
Note that all quantities may depend on $l_0$.
Since the external velocity, \ve, has no component perpendicular to
the plane of the tube the drag force given by \es{3.28/29} can be written
in the simpler form shown in \e{5.3}. We may express the
equilibrium condition in terms of the relative density contrast between
exterior and interior of the flux tube by rewriting \e{5.3} in the form

$$ \beta \left( {\roe \over \ro} -1 \right) \gun  \sgs 
   {2 \hp \over R_0} \sps \beta r \, \sgn{\ve_0\cdot\nnun} \eqno (5.4) $$

\n where $\beta = 8\pi p_0/\bq$, $\hp$ is the internal pressure scale
height defined by $p_0/\hp = \ro g_0$, and $r$ is given by

$$ r \sgs {\roe \over \ro}{C_D(\ve_0\cdot\nnun)^2 \over \pi a_0 g_0}.
   \eqno (5.5) $$

\n The product $\beta r$ can be written in the form

$$ \beta r \sgs \left( {2C_D\over\pi} \right)
                \left( {\roe\over\ro} \right)
                \left( {(\ve_0\cdot\nnun)^2\over v_{A0}^2} \right)
                \left( {\hp\over a_0} \right) \eqno (5.6) $$

\n where $v_{A0} = B_0/\sqrt{4\pi\ro}$ is the Alfv\'en velocity.
For equipartition fields ($B_0 = 10^4 ... 10^5$ Gauss) near the bottom of
the solar convection zone 
($p_0 = 6\cdot 10^{13}\,{\hbox {dyn}}\cdot {\hbox {cm}}^{-2}$) 
we have very large values of $\beta \approx 1.5\cdot 10^5 ... 1.5\cdot 10^7$. 
We shall use this property in Sec.\ts{5.3} to derive a simplified
version of the perturbation equations in the limit $\beta \gg 1$. In most
cases relevant for the deep convection zone, the density contrast in \e{5.4}
is very small such that $\beta (\roe/\ro - 1)$, $\hp/R_0$, and $\beta r$ are all
of order unity or at least have moderate values.

\vfill\eject

\def\etopline{ 5.3 Perturbation equations}

\n {\bbf 5.3 Perturbation equations}
\b

\n In order to determine the linear stability of the equilibrium 
we introduce Lagrangean displacements in the normal and tangential directions
described by the functions $\eps(l_0)$ and $\eta(l_0)$, respectively, and 
write for the perturbed path\fussn{*} 
{The perturbations of all quantities which have a non-vanishing equilibrium value 
are indicated by a suffix `1'. Perturbed quantities (equilibrium value $+$ 
perturbation) are written without suffix.}

$$ \rb \sgs \rb_0 \; + \; \rb_1 \; \equiv \; \rb_0(l_0) \; + \; 
   \eps(l_0)\,\nnun \; + \; \eta(l_0)\,\lun \,. \eqno (5.7) $$

\n Similar to the analyses of Spruit and van Ballegooijen (1982) and of van
Ballegooijen (1983), a displacement in the binormal direction decouples from
the rest of the equations and gives rise to (stable) transversal flux tube
waves which are of no further interest here. We therefore consider only
displacements within the plane of the equilibrium flux tube path.

First we determine the perturbed {\it geometry\/} of the flux tube. The
relation between the arc length of the perturbed tube, $l$, and the
arc length in equilibrium, $l_0$, to first order in the perturbations is given by

$$ \dln{l} \sgs 1 - {\eps \over R_0} + \dln{\eta}\,. \eqno (5.8) $$

\n The perturbed tangent vector, \lu, is

$$ \lu \sgs \dl{\rb} \sgs \dln{\rb}\,\dl{l_0} \sgs
   \lun \sps \nnun \left( {\eta\over R_0} \, + \, \deps \right).
   \eqno (5.9) $$ 

\n Now we can calculate the curvature vector, $\cb \equiv \nnu/R$ :

$$ \cb \sgs \dl{\,\lu} \sgs \dln{\,\lu}\,\dl{\l_0} \sgs 
   {\nnun \over R_0} \left( 1 + {\eps \over R_0} + R_0 {d^2\eps \over
   dl_0^{\,2}} - {\eta \over R_0}\dln{R_0} \right) \; - \; {\lun \over R_0}
   \left( {\eta\over R_0} + \dln{\eps} \right). \eqno (5.10) $$

\n Taking the absolute value of \e{5.10} gives the perturbed radius
of curvature, $R$, and the perturbed normal vector, viz.

$$ R \sgs R_0 \left( 1 + {\eps \over R_0} + R_0 {d^2\eps \over
     dl_0^{\,2}} - {\eta \over R_0}\dln{R_0} \right)^{-1} \eqno(5.11) $$

$$ \nnu \sgs \nnun - \lun \left( {\eta\over R_0} \, + \, \deps \right) .
     \eqno (5.12) $$

\n We now proceed by determining the {\it gravitational acceleration,} \gb,
in the perturbed state. We assume a power law for the dependence of $g_0$ on
height, $z$, and write

$$ \gb_0 \sgs -g_m \left(z_0 \over z_m\right)^s \zu \;\equiv\; -g_0 \zu
       \eqno (5.13) $$

\n (\zu: unit vector in direction of height; $g_m$: gravitational acceleration
at some reference height, $z_m$; $z_0(l_0)$: height of equilibrium flux tube).
With the height displacement, $z_1$, given by

$$ z_1 \;\equiv\; \rb_1\cdot\zu \sgs -\rb_1\cdot\gu
       \sgs -\eps\,\gun - \eta\,\gul \eqno (5.14) $$

\n where $\gu = -\zu$ denotes the unit vector in the direction of gravity we
find, to first order in $z_1$,

$$ \eqalignno{
   \gb\sgs\gb_0 + \gb_1 &\sgs -g_m \left( {z_0 + z_1 \over z_m} \right)^s \zu 
       \sgs -g_m \left( {z_0\over z_m} \right)^s \left( 1 + {z_1\over z_0}
        \right)^s \zu & \cr&&\cr 
       &\sgs \gb_0 \left( 1 - s\, {\eps\,\gun + \eta\,\gul \over z_0}
       \right). &(5.15) \cr}$$

\n Defining

$$ \Delta \; \equiv \; {-z_1 \over z_0} \sgs 
             {\eps\,\gun + \eta\,\gul \over z_0} \eqno (5.16) $$

\n we write for the component of gravity along \lu\ with aid of \es{5.9} and
(5.15)

$$ \gb\cdot\lu \sgs ( 1 - s\Delta) \, \gb_0 \cdot \left[
   \lun \sps \nnun \left( {\eta\over R_0} \, + \, \deps  \right) 
   \right]  \eqno (5.17) $$

\n which gives for the perturbation of longitudinal component of \gb

$$ (\gb\cdot\lu)_1 \sgs g_0 \left[ \gun \left ( {\eta\over R_0} \, + \, \deps 
   \right) \; - \; \gul s \Delta \right]. \eqno (5.18) $$

\n For the perturbation of the normal component of
\gb\ we find using \es{5.12} and (5.15)

$$ (\gb\cdot\nnu)_1 \sgs -g_0 \left[ \gun s \Delta \,+\, 
      \gul \left( {\eta\over R_0} \, + \, \deps  \right)\right]. 
      \eqno (5.19) $$

\n We now consider the {\it longitudinal component of the equation of motion,}
\e{3.25}. To first order in the displacements, the inertial term on its
\lhs can be written using $\ub = d \rb_1/dt \equiv ({\dot\eta},
{\dot\eps})$ in the form

$$ \rho \dt{\ub} \cdot \lu \sgs \rho\dt{\ub\cdot\lu} \sms 
   \rho \ub\cdot\dt{\,\lu} \sgs \rho\left({d\over dt}({\dot\eta}\,\lun + 
   {\dot\eps}\,\nnun)\right)\cdot\lu \sms \rho\ub\cdot \left( \dt{\,\lun} +
   \dt{\,\lu_1} \right) \sgs \ro {\ddot\eta}\,. \eqno (5.20) $$

\n The perturbation of the longitudinal pressure gradient is given by

$$ \left( \dl{p} \right)_1 \sgs
   \dln{p_1} \sps \dln{p_0}\left( {\eps\over R_0} - \deta \right)
   \eqno (5.21) $$

\n and the perturbation of the gravity force reads

$$ \eqalignno {
    (\rho\, \gb\cdot\lu)_1 &\sgs \ro(\gb\cdot\lu)_1 + \rho_1 g_0 \gul &\cr&&\cr
    &\sgs \ro g_0 \left[ \gun \EREL - \gul\sd \right] \sps
     {\rho_1 \over \ro}\dln{p_0}\,. &(5.22) \cr 
    } $$

\n where we have used \es{5.2} and (5.18). Inserting \es{5.20-5.22} into
\e{3.25} we find for the longitudinal equation of motion, to first order in
the perturbations:

$$ \eqalignno {
   {\ddot\eta} \sgs &-{1\over\ro}\dln{p_1} \sms {1\over\ro}\dln{p_0}
    \left( {\eps\over R_0} - \deta - {\rho_1\over\ro} \right) &\cr&&\cr
   &+\; g_0 \left[ \gun \EREL - \gul\sd \right]. &(5.23) \cr
   } $$

\n Similar to \e{5.20} the inertial term in \e{3.26}, the {\it normal component of
the equation of motion,} reduces to $\ro{\ddot\eps}$.\fussn{*}
{An enhanced inertia of the flux tube with respect to transversal motions due
to the acceleration of material in the exterior is not considered here. It would
only affect growth rates (or oscillation frequencies) but does not change the
stability criteria. In the case $\beta = 8\pi p_0/B_0^2 \gg 1$ which we
consider later the enhanced inertia can be crudely taken account of by
multiplying the inertial term of the normal component of the equation of motion
by a factor 2 (\cf Spruit and van Ballegooijen, 1982).}
Using \e{5.19} the perturbation of the buoyancy force on the \rhs of \e{3.26}
is given by

$$ \eqalignno{
   [(\rho-\rho_e)\gb\cdot\nnu]_1 \sgs &(\rho_1-\rho_{e1})g_0\gun &\cr&&\cr
   &\sms (\ro-\roe)g_0\left[ \gun\sd + \gul\EREL \right]. &(5.24) \cr
   } $$

\n The perturbation of the external density, $\rho_{e1}$, in our Lagrangian
approach is

$$ \rho_{e1} \sgs -\roe z_1 H_{\rho e}^{-1} \sgs 
   -\roe z_1 (1 - \nabla) H_{pe}^{-1} \eqno (5.25) $$

\n ($H_{\rho e}$: external density scale height; $\hpe$: external pressure
scale height; $\nabla$: logarithmic temperature gradient in the external medium; 
all these quantities are taken at $z=z_0$). The relation 
$\hpe = (1-\nabla)H_{\rho e}$ is valid if the molecular weight is constant, 
an assumption which is well justified in the lower
parts of the solar convection zone.  With aid of \e{5.11} the perturbation of
the curvature force can be written

$$ \left( {B^2 \over 4\pi R} \right)_1 \sgs
   {B_0 B_1 \over 2\pi R_0} \sps {B_0^2 \over 4\pi R_0}
     \left( {\eps \over R_0} + R_0 {d^2\eps \over
     dl_0^{\,2}} - {\eta \over R_0}\dln{R_0} \right). \eqno (5.26) $$

\n The perturbation of the drag force is slightly more complicated to
determine. We start by noting that for linear analysis we may take
$\sgn{\ve\cdot\nnu} = \sgn{\ve_0\cdot\nnun}$ since the displacement can always
be made sufficiently small. In the case $\ve_0\cdot\nnun = 0$ the perturbation
of the drag term vanishes identically as one can see in \es{5.31/33} below.
The perturbed relative velocity between flux tube and environment is given by

$$ \ve \sgs \ve_0 \sps (\rb_1\cdot\nabla)\ve\vert_{\rb_0} \sms \ub\,.
   \eqno (5.27) $$

\n The third term on the \rhs represents the motion of the tube due
to the displacement while the second term describes the spatial change of
\ve\ which is written in cartesian coordinates $(x,z)$:

$$ {\bf w}_{e1} \; \equiv \; (\rb_1\cdot\nabla)\ve\vert_{\rb_0} \sgs 
    x_1 \left. {\partial\ve\over\partial x}\right\vert_{\rb_0} \sps 
    z_1 \left. {\partial\ve\over\partial z}\right\vert_{\rb_0} \eqno (5.28) $$

\n with

$$ \eqalignno {
  x_1 &\sgs \rb_1\cdot{\hat{\bf x}} \sgs \eps\,({\hat{\bf x}}\cdot\nnun) 
        \sps \eta\,({\hat{\bf x}}\cdot\lun)&\cr
  z_1 &\sgs \rb_1\cdot\zu \sgs  -\eps\,\gun -\eta\,\gul. &(5.29) \cr 
  } $$

\n $\hat{\bf x}$ is the unit vector in direction of the $x$ coordinate, \ie the
horizontal direction. Next we determine the perturbation of $(\ve\cdot\nnu)^2$.
Using \es{5.12} and (5.27) we write

$$ (\ve\cdot\nnu)^2 \sgs \left[(\ve_0 + {\bf w}_{e1} - \ub)\cdot
   \left( \nnun - \lun \left( {\eta\over R_0} \, + \, \deps \right) \right)
   \right]^2 \eqno (5.30) $$

\n which gives

$$ v_{\perp 1} \; \equiv \; (\ve\cdot\nnu)^2_1 \sgs 2(\ve_0\cdot\nnun) \left[ 
   ({\bf w}_{e1}\cdot\nnun) \sms {\dot\eps} \sms (\ve_0\cdot\lun)
   \left( {\eta\over R_0} \, + \, \deps \right) \right]. \eqno (5.31) $$

\n Note that $v_{\perp 1}$ may have positive or negative sign. Using 
\e{5.28/29} we obtain

$$ \eqalignno{
   ({\bf w}_{e1}\cdot\nnun) &\sgs 
   \left[ \eps({\hat{\bf x}}\cdot\nnun)+\eta({\hat{\bf x}}\cdot\lun) \right]
   \Bigl( {\left. {\partial\ve\over\partial x}\right\vert_{\rb_0}}\cdot\nnun
   \Bigr) \sms \left[ \eps\gun + \eta\gul \right] 
   \Bigl( {\left. {\partial\ve\over\partial z}\right\vert_{\rb_0}}\cdot\nnun
   \Bigr) &\cr&&\cr
   &\;\equiv \; 
   \left[ \eps({\hat{\bf x}}\cdot\nnun)+\eta({\hat{\bf x}}\cdot\lun) \right]
   w_x \sms \left[ \eps\gun + \eta\gul \right] w_z\,.
   &(5.32) \cr
   } $$ 

\n Now we can write down the perturbation of the drag term:

$$ \left( {C_D\rho_{e}(\ve\cdot\nnu)^2\sgn{\ve\cdot\nnu} \over \pi a}
   \right)_1 \sgs {C_D\roe(\ve_0\cdot\nnun)^2\sgn{\ve_0\cdot\nnun}\over\pi a_0}
   \left( {v_{\perp 1} \over \ven^2} + {\rho_{e1}\over\roe} -
   {a_1\over a_0} \right). \eqno (5.33) $$

\n The last term on the \rhs can be rewritten using the conservation of
magnetic flux, $Ba^2=$ const., which yields 

$$ {a_1 \over a_0} \sgs - {B_1 \over 2B_0}. \eqno (5.34) $$

\n We combine \es{5.24}, (5.26) and (5.33/34) and obtain for the normal
component of the perturbed equation of motion by inserting into \e{3.26}:

$$ \eqalignno {
   {\ddot\eps} \; = \;\; &{\rho_1-\rho_{e1}\over \ro}\, g_0\gun \sms
   \left( 1 - {\roe\over\ro} \right) g_0\left[ \gun\sd + \gul\EREL \right]
   &\cr&&\cr
   &+ \; {B_0^2 \over 4\pi R_0\ro} \left( {2B_1\over B_0} +{\eps \over R_0} 
      + R_0 {d^2\eps \over dl_0^{\,2}} - {\eta \over R_0}\dln{R_0} \right) 
   &\cr&&\cr
   &+ \; {C_D\roe(\ve_0\cdot\nnun)^2\sgn{\ve_0\cdot\nnun}\over\pi a_0\ro}
      \left( {v_{\perp 1} \over \ven^2} + {\rho_{e1}\over\roe} +
      {B_1\over 2B_0} \right). &(5.35) \cr&&\cr
   } $$

\n Since it is our aim to obtain a set of two coupled equations for the
displacements $\eta$ and $\eps$ alone, we have to eliminate the other
perturbations ($B_1, \rho_1, \rho_{e1}, p_1, ...$) in favor of $\eta$, $\eps$,
and equilibrium quantities. This has already been achieved for $\rho_{e1}$ with
\es{5.25} and (5.14) and for $v_{\perp 1}$ in \es{5.31/32}. 
For the remaining quantities
we use \e{3.13}, the equation of continuity, which can be written to first
order in the perturbations (after a time integration) in the form

$$ {\rho_1 \over \ro} \sms {B_1 \over B_0} \sps \dln{\eta} 
   \sms {\eps \over R_0} \sgs 0\,, \eqno (5.36) $$

\n the equation of state for adiabatic perturbations,

$$ \dt{p} \sgs {\gamma p \over \rho} \dt{\rho}\,, \eqno(5.37) $$

\n which yields (to first order, after time integration)

$$ {p_1 \over p_0} \sgs \gamma {\rho_1 \over \rho_0}\,, \eqno (5.38) $$

\n and the condition of instantaneous pressure equilibrium, \e{3.23}, which
gives

$$ {p_1 \over p_0} \sps {2\over\beta}{B_1 \over B_0} \sgs 
   {p_{e1} \over p_0} \eqno(5.39) $$

\n with $\beta = 8\pi p_0/B_0^2$. Assuming hydrostatic equilibrium in the
environment and using the equilibrium pressure balance condition, 
$p_0 + B_0^2/8\pi = p_{e0}$, we rewrite \e{5.39} in the form

$$ {p_1 \over p_0} \sps {2\over\beta}{B_1 \over B_0} \sgs 
   -\left( 1 + {1\over\beta} \right) \delzh \,. \eqno (5.40) $$

\n \es{5.36} and (5.38/39) are used to obtain

$$ \eqalignno {
   {B_1 \over B_0} &\sgs \left( {\beta\gamma \over \beta\gamma + 2} \right)
   \left( \dln{\eta} - {\eps\over R_0} \right) \sms 
   \left( {\beta + 1 \over \beta\gamma + 2} \right) \delzh &(5.41)\cr&&\cr
   {\rho_1 \over \ro} &\sgs \left( {\beta\gamma\over\beta\gamma + 2}-1 \right)
   \left( \dln{\eta} - {\eps\over R_0} \right) \sms 
   \left( {\beta + 1 \over \beta\gamma + 2} \right) \delzh &(5.42)\cr&&\cr
   {p_1 \over p_0} &\sgs \gamma \left( {\beta\gamma\over\beta\gamma + 2} - 1
   \right)
   \left( \dln{\eta} - {\eps\over R_0} \right) \sms 
   \gamma\left( {\beta + 1 \over \beta\gamma + 2} \right)\delzh\,. &(5.43)\cr
   &&\cr} $$

\n Using the abbreviations

$$ {\beta\gamma\over\beta\gamma+2} \; \equiv \; \alpha_1, \qquad
   {\beta + 1 \over \beta\gamma+2} \; \equiv \; \alpha_2 \eqno(5.44) $$

\n we insert \es{5.42/43} into \e{5.23} and obtain for the equation
which determines the time evolution of the longitudinal displacement $\eta$
(primes denote derivatives with respect to $l_0$):

$$ \eqalignno {
   {\ddot\eta} \sgs \; &\eps \left\{ 
      -(\ai - 1){\gamma p_0\over\ro}{R_0'\over R_0^2} 
      \sps {1\over\ro R_0}\left[ \gamma p_0 (\ai-1) \right]'
      \sps {1\over\ro} \left[ {\aii p_0'\over\hpe} - 
           \left( {\gamma p_0 \aii\over\hpe} \right)' \right] \right. \gun
      &\cr&&\cr &\qquad 
      \left. \sms {\ai p_0' \over \ro R_0} 
      \sps {\gamma p_0 \aii\over \ro \hpe R_0} \gul
      \sms {g_0 s \over z_0}\gul\gun \right\}
      &\cr&&\cr
      &+ \; \eps' \left\{
      (\ai - 1){\gamma p_0\over\ro R_0} + \left( g_0 \, - \, 
      {\gamma p_0 \aii\over\ro \hpe}\right) \gun \right\}
      &\cr&&\cr
      &+ \; \eta  \left\{
      \left[ {\aii p_0' \over \ro\hpe} - {1\over\ro}\left(
             {\gamma p_0 \aii \over \hpe}\right)' \right] \gul \right.
      &\cr&&\cr &\qquad
      \left. \sps \left( {g_0\over R_0} \, - \, {\gamma p_0 \aii \over
                  \ro \hpe R_0} \right) \gun - {g_0 s \over z_0}\gul^2
      \right\}
      &\cr&&\cr
      &+ \; \eta' \left\{ -{1\over\ro}\Bigl[ \gamma p_0 (\ai - 1) \Bigr]'
            \sps {p_0'\ai\over\ro} \sms {\gamma p_0\aii\over\ro\hpe}\gul
      \right\}
      &\cr&&\cr
      &+ \; \eta'' \left\{ -(\ai - 1){\gamma p_0\over\ro} \right\}.
      &(5.45) \cr } $$

\n In order to derive \e{5.45} we have used \es{5.14}, (5.16) and the relation

$$ \left[ \eps\,\gun + \eta\,\gul \right]' \sgs
   \left( \eta' \,-\, {\eps\over R_0} \right) \gul \sps
   \left( \eps' \,+\, {\eta\over R_0} \right) \gun. \eqno (5.46) $$

\n In a similar way we obtain the equation which determines the time
evolution of $\eps$, the displacement in normal direction, by inserting
into \e{5.35}:

$$ \eqalignno {
   {\ddot\eps} \sgs \; &\eta \left\{ 
      - {2 p_0 R_0'\over \ro \beta R_0^2}
        \sps{4 p_0 \aii \over \ro \beta R_0 \hpe} \gul
        \sps {g_0\over\hpe}\left[ \aii \,+\, {\roe\over\ro}(\nabla-1) \right]
        \gul\gun \right. 
      &\cr&&\cr &\qquad 
        \sms \left( 1 - {\roe\over\ro} \right) \left[
        {g_0 s \over z_0}\gul\gun \,+\, {g_0\over R_0}\gul \right]   
      &\cr&&\cr &\qquad
        \sps {C_D\roe\sgn{\ve_0\cdot\nnun}(\ve_0\cdot\nnun)^2
        \over\pi a_0\ro} \biggl[
        {2\over (\ve_0\cdot\nnun)} \left( ({\hat{\bf x}}\cdot\lun) w_x
        - \gul w_z \right) 
      &\cr&&\cr &\qquad
        \left. \left. - \;{2 (\ve_0\cdot\lun) \over R_0(\ve_0\cdot\nnun)} \sms
        {1\over\hpe} \left( \nabla \, - \, 1
        \,-\, {\aii\over 2} \right) \gul \right] \right\}
      &\cr&&\cr
      &+ \; \eta' \left\{
        {4 p_0 \ai \over \ro \beta R_0} \sps g_0 (\ai -1)\gun \sps
        {C_D\roe(\ve_0\cdot\nnun)^2\sgn{\ve_0\cdot\nnun}\ai\over 2\pi a_0\ro}
        \right\}
      &\cr&&\cr
      &+ \; \eps  \left\{
        -{4 p_0 \ai \over \ro \beta R_0^2} \sps
        {4 p_0 \aii \over \ro \beta R_0 \hpe} \gun \sps
        {2 p_0 \over \ro R_0^2 \beta} \sms
        {g_0 s \over z_0} \left( 1-{\roe\over\ro} \right) \gun^2 \right.
      &\cr&&\cr &\qquad
        \sms \left[ {g_0\over R_0}(\ai - 1) \, - \, {g_0\over\hpe}
             \left( \aii + {\roe\over\ro}(\nabla-1) \right)\gun \right] \gun
      &\cr&&\cr &\qquad
        \sps {C_D\roe\sgn{\ve_0\cdot\nnun}(\ve_0\cdot\nnun)^2 
        \over\pi a_0\ro} \biggl[
        {2\over(\ve_0\cdot\nnun)} \left( ({\hat{\bf x}}\cdot\nnun) w_x
        - \gun w_z \right) 
      &\cr&&\cr &\qquad
        \left. \left. \sms
        {1\over\hpe} \left( \nabla \, - \, 1
        \,-\, {\aii\over 2} \right) \gun \sms
        {\ai\over 2R_0} \right] \right\}
      &\cr&&\cr
      &+ \; \eps' \left\{ -g_0 \left( 1-{\roe\over\ro} \right) \gul \sms
         {2 C_D\roe\sgn{\ve_0\cdot\nnun}\over\pi a_0\ro}(\ve_0\cdot\nnun)
         (\ve_0\cdot\lun) \right\}
      &\cr&&\cr
      &+ \; \eps'' \left\{ {2 p_0\over \ro\beta} \right\}
      &\cr&&\cr
      &- \; {\dot\eps} \left\{
        {2 C_D\roe(\ve_0\cdot\nnun)\sgn{\ve_0\cdot\nnun}\over \pi a_0\ro}
        \right\}.
      &(5.47) \cr } $$

\n In the case of a non-constant molecular weight, the term $(1-\nabla)/\hpe$
has to be replaced by $1/H_{\rho e}$ (\cf Eq.\ts5.25). A form of
\es{5.45/5.47} which is more convenient for our purposes is obtained below
by considering the limit $\beta\gg 1$ and a suitable non-dimensionalization.

\b\b

\def\etopline{ 5.4 Non-dimensionalization and the case $\beta\gg 1$}

\twelvepoint
\n {\bf 5.4 Non-dimensionalization and the case $\beta\gg 1$}
\elevenpoint

\b

\n As we have discussed in Sec.\ts{5.2}, the value of $\beta$ for
equipartition flux tubes in the deep layers of the solar convection zone is
very large. Having this application in mind, it is convenient to rewrite and
simplify \es{5.45/47} in the limit $\beta\gg 1$ and 
to introduce non-dimensional quantities. As units for all quantities we
use their values in the interior of the flux tube at the reference level,
$z=z_m$, which has already been defined in \e{5.13}.  As length scale we
choose the internal scale height, $H_{pm} \equiv H_{p0}(z_m)$, and the units of
magnetic field, pressure, density etc. are defined by $B_m\equiv B_0(z_m)$,
$p_m\equiv p_0(z_m)$, $\rho_m\equiv \ro(z_m)$ etc., respectively.  As time unit,
$\tau$, we take, similar to Spruit and van Ballegooijen (1982)

$$ \tau \sgs \left( {\beta_m H_{pm} \over g_m} \right)^{1/2} \eqno (5.48) $$

\m

\n with $\beta_m=8\pi p_m/B_m^{\,2}$. It is easy to see that $\tau$ is
$\sqrt 2$ times the Alfv\'en travel time across one scale height.
For values of $\beta_m$ of the order
$10^5...10^7$, $H_{pm}=10^9$ cm, $g_m=$\dex{6}{4} cm$^2\cdot$s$^{-1}$ we find
$\tau = 10^5...10^6$s $\approx 1...2$ days. The velocity unit is defined as
$V = H_{pm}/\tau = v_A/\sqrt{2}$ where $v_A = B_m/\sqrt{4\pi\rho_m}$ is the 
Alfv\'en speed at the reference height. Since \es{5.45/47} are linear and
homogeneous, we can separate the time dependence by writing 
$f \propto \exp(i\omega t)$ for all quantities where $\omega$ is a (generally
complex) frequency. This leads to

$$ ({\ddot \eta}, {\ddot\eps}) \sgs -\omega^2 (\eta, \eps)\,. \eqno(5.49) $$

\m

\n We insert \e{5.49} into \es{5.45/47} and take the limit $\beta\gg 1$ such
that only terms of order unity and order $\beta^{-1}$ are retained. It turns
out that the terms of order unity cancel. Consequently, non-dimensionalization by
multiplication of the resulting equations with $\tau^2 = \beta_m H_{pm}/g_m$
leads to terms of order unity and to terms of the form $\beta\delta$, $\beta r$,
and $\beta(1-\roe/\ro)$ which are also of order unity ($\delta$ is the
superadiabaticity of the external gas and $r$ is defined in Eq.\ts{5.5}). For
example, a flux tube in temperature equilibrium with its surroundings
($T_0 = T_{e0}$) has $\beta(1-\roe/\ro) = -1$ and, using a standard mixing-length 
model, Spruit and van Ballegooijen (1982) found a value of $\beta\delta=3.6$ for 
an equipartition flux tube in the deep parts of the solar convection zone. Thus
for $\beta\delta\ll 1$ we have fields strong compared to the equipartition
value, for $\beta\delta\gg 1$ we have weak fields.  

We give some examples of how the transformation of the coefficients 
in \es{5.45/47} is carried out but avoid a presentation of the whole
calculation. The algebra is straightforward although lengthy and tedious, filling
dozens of pages.
Let us first write down the quantities $\ai$ and $\aii$ up to first order
in $\beta^{-1}$ (\cf Eq.\ts5.44), viz.

\b

$$ \eqalignno {
   \ai &\sgs  1 \sms {2\over\beta\gamma} \sps O(\beta^{-2})
   &\cr&&\cr
   \aii &\sgs {1\over\gamma} \sps {1\over\beta\gamma}
         \left( 1 - {2\over\gamma} \right) \sps O(\beta^{-2})\,.
   &(5.50) \cr
   } $$

\n It is easy to show that for $\ai$ and $\aii$ the operations of taking the
limit $\beta\gg 1$ and taking the derivative with respect to $l_0$ can be 
interchanged.
This simplifies the calculation of the coefficients which contain derivatives
of these quantities. We now consider the first coefficient on the
\rhs of \e{5.45} and use \e{5.50} to obtain the limit for large $\beta$:

$$ -(\ai - 1){\gamma p_0\over\ro}{R_0'\over R_0^2} \sgs
   {\bq\over 4\pi\ro}\,{R_0'\over \rq} \sps O\left(\beta^{-2}\right).
   \eqno (5.51) $$

\n The non-dimensional form is found by multiplication with 
$\beta_m H_{pm}/g_m$ which yields

$$    {\bq\over 4\pi\ro}\,{R_0'\over \rq} \cdot {8\pi p_m H_{pm} \over
      B_m^{\,2} g_m} \sgs {2 {\tilde{B}}_0^{\,2} \over {\tilde{\rho}}_0}
      \,{{\tilde R}_0'\over {\tilde R}_0^{\,2}}\,. \eqno (5.52) $$

\n The tilde denotes dimensionless quantities 
(\eg ${\tilde{\rho}}_0 = \rho_0/\rho_m$) and we have used the
hydrostatic relation $p_m/g_m = \rho_m H_{pm}$. A similar procedure is carried
out for all other coefficients. Sometimes it is helpful to use the relation
between the derivative along the flux tube and the derivative with respect to
$z$ for quantities which only depend on height such as $\ro$, $p_0$ and $B_0$:

$$ f' \sgs \lun\cdot\nabla f \sgs -\gul {df\over dz}. \eqno (5.53) $$

\n An often needed quantity is the $z$-derivative of a scale height 
(external or internal) which can be written for the case of constant 
mean molecular weight, ${\bar\mu}$, in the form

$$ {dH_p\over dz} \sgs {d\over dz} \left( {{\cal R}T\over{\bar\mu}g} \right)
   \sgs {{\cal R} \over{\bar\mu}g}\,{dT\over dz} \, + \, 
   {{\cal R}T \over{\bar\mu}}\,{d\over dz} \left( {1\over g} \right)
   \sgs -\nabla \, - \, {s H_p\over z} \eqno (5.54) $$

\n where we have used the equation of state for a perfect gas. $\cal R$ is the
gas constant and $T$ denotes the temperature. Another helpful relation is given
by the condition of internal hydrostatic equilibrium, \ie $p_0/\hp = \ro g_0$.
We give another example for the transformation of a coefficient which is not 
quite straightforward. Consider the large $\beta$ limit for the third term on the
\rhs of \e{5.45}:

$$  \eqalignno {
    {1\over\ro} &\left[ {\aii p_0'\over\hpe} \, - \,
       \left( {\gamma p_0 \aii\over\hpe} \right)' \right] \;\; \to \;\;
       {1\over\ro} \left\{ \left[ {1\over\gamma} + {1\over\beta\gamma}
       \left (1 - {2\over\gamma} \right) \right] {p_0'\over\hpe} \, - \,
       \left[ {\gamma p_0\over\hpe} \left( {1\over\gamma}
       + {1\over\beta\gamma}\left( 1 - {2\over\gamma} \right) \right) \right]'
       \right\} 
    &\cr&&\cr
       &\qquad = \; {1\over\ro} \left\{ {p_0'\over\gamma\hpe} \,+\,
       {p_0'\over\beta\gamma\hpe} \left( 1 - {2\over\gamma} \right) \,-\,
       \left( {p_0\over\hpe} \right)' \,-\, \left[ {p_0\over\beta\hpe}
       \left( 1 - {2\over\gamma} \right) \right]' \right\}. &(5.55). \cr
    } $$

\n We denote the superadiabaticity of the external gas by
$\delta = \nabla - \nabla_{ad}$, take $\nabla_{ad} = (\gamma-1)/\gamma$,
such that $1/\gamma = 1 - \nabla + \delta$ and write using \es{5.53/54}

$$ {p_0'\over\gamma\hpe} \,-\, \left( {p_0\over\hpe} \right)' \sgs
   -\gul \left[ {\delta\over\hpe}{dp_0\over dz} \,+\, \nabla {p_0\over\hpe^2}
        \left( {\hpe\over\hp} - 1 \right) \,-\, {p_0 s \over \hpe z_0}
   \right]. \eqno (5.56) $$

\n Inserting \e{5.56} and using $dp_0/dz = -p_0/H_{p0} = -\rho_0 g_0$ we find
for the \rhs of \e{5.55}:

$$ \eqalignno {
   ... \; &= \; -\gul {1\over\ro} \left\{
          -{\delta p_0\over \hpe\hp} \,+\, \nabla {p_0\over \hpe^2}
           \left( {\hpe\over\hp} - 1 \right) \,-\,  {p_0 s\over \hpe z_0}
           \right. &\cr&&\cr
     &\qquad\left. -\left( 1-{2\over\gamma}\right) {p_0\over\beta\gamma\hpe\hp}
         \,-\, \left( 1 - {2\over\gamma} \right) \left[ 
         {1\over\hpe} {d\over dz} \left( {p_0\over\beta} \right) +
         {p_0\over\beta} {d\over dz} \left( {1\over\hpe} \right) \right] 
         \right\} &\cr&&\cr
      &= \; \gul \left\{
          {g_0\over \hpe} \left[ \delta \,-\, \nabla
           \left( 1 - {\hp\over\hpe} \right) \,+\, {s\hp \over z_0} \right]
           \right. &\cr&&\cr &\qquad \left.
         \,+\, \left( 1-{2\over\gamma} \right) {1\over \ro\hpe} \left[
           {d\over dz} \left( {\bq\over 8\pi} \right) + {\bq\over 8\pi\hpe}
           \left( {\hpe\over\gamma\hp} + \nabla + {s \hpe \over z_0} \right) 
           \right] \right\}. &(5.57) \cr&&\cr
      } $$

\n By multiplication with $\beta_m H_{pm}/g_m$ which can be rewritten as

$$  {\beta_m H_{pm} \over g_m} \sgs {\beta_m H_{pm} \over g_m\beta} \beta
    \sgs {H_{pm}^{\,2}\rho_m\bq\over g_0\hp\ro B_m^{\,2}} \beta 
    \eqno (5.58)$$

\n and

$$ {\beta_m H_{pm} \over g_m} \sgs {8\pi H_{pm}^{\,2} \rho_m\over B_m^{\,2}},
   \eqno (5.59) $$

\n \e{5.57} is transferred to non-dimensional form:

$$ \eqalignno {
   ... \; &\to \; \gul {{\tilde{B}}_0^{\,2}\over{\tilde\rho}_0 {\tilde H}_{pe}}
          \left\{ {1\over{\tilde H}_{p0}} \left[ \beta\delta \,-\, \nabla\beta
           \left( 1 - {{\tilde H}_{po}\over{\tilde H}_{pe}} \right) \,+\, 
           {\beta s {\tilde H}_{p0} \over {\tilde z}_0} \right]
           \right. &\cr&&\cr &\qquad \left.
         \,+\, \left( 1-{2\over\gamma} \right)\left[ {1\over{\tilde B}_0^{\,2}}
           {d {\tilde B}_0^{\,2} \over d{\tilde z}} + {1\over {\tilde H}_{pe}}
           \left( {{\tilde H}_{pe}\over\gamma{\tilde H}_{p0}} + \nabla +
           {s {\tilde H}_{pe} \over {\tilde z}_0} \right) 
           \right] \right\}. &(5.60) \cr&&\cr
      } $$
 
\n In what follows we shall omit the tildes and tacitly assume that all
quantities are non-di\-men\-sio\-nal unless the contrary is explicitly stated.
The procedures described above have been applied to each term in 
\es{5.45/47}. 
For equipartition flux tubes in the deep solar convection zone we have
$\hpe/\hp-1 = O(\beta^{-1}) ,\, 
\roe/\ro-1 = O(\beta^{-1})$ and we can take $\hpe/\hp \approx 1, \,
\roe/\ro \approx 1$ unless such a term is multiplied by $\beta$.  Under these
conditions we have

$$ \beta \left( 1 - {\hpe\over\hp} \right) \,+\, 1 \sgs
   \beta \left( {\roe\over\ro} - 1 \right ) \sps O(\beta^{-2}). \eqno(5.61) $$

\n Another useful relation is

$$ {1\over\bq} {d\bq\over dz} \sgs  
   {8\pi\over\bq} {d\over dz}(p_{e0} - p_0) \sgs
    {\beta\over\hp} \left( 1 - {\roe\over\ro} \right ). \eqno (5.62) $$

\n After some tedious algebra, \es{5.45/5.47} in their final form are
given by

$$ \eqalignno {
   -\omega^2 {\ro\over\bq} \eta \; &= \; \eps \left\{ \gul \left[
      {1\over R_0\hp}\betro - {2\over\rq}{dR_0\over dz} \right] \right.
  &\cr&&\cr
   &\qquad \left. + \; \gul\gun \left\lgroup -{s\over z_0\hp} \left[
      \betro + \gamiz \right] \, + \, {1\over H_{p0}^{\,2}} \left[
      \beta\delta - {1\over\gamma} - {\beta\over\gamma} 
      \left( 1 - {\roe\over\ro} \right) \right] \right\rgroup \right\}
  &\cr&&\cr
   &\; + \; \eps' \left\{ -{2\over R_0} \, + \, \gun {1\over\hp} \left[ 
      \betro + \gamiz \right] \right\}
  &\cr&&\cr
   &\; + \; \eta \left\{ \gul^2 \left\lgroup -{s\over z_0\hp} \left[
      \betro + \gamiz \right] \, + \, {1\over H_{p0}^{\,2}} \left[
      \beta\delta - {1\over\gamma} - {\beta\over\gamma} 
      \left( 1 - {\roe\over\ro} \right) \right] \right\rgroup \right.
  &\cr&&\cr
   &\qquad \left. + \; \gun {1\over R_0\hp} \left[ \betro + \gamiz \right]
      \right\}
  &\cr&&\cr
   &\; + \; \eta' \left\{ -\gul {1\over\hp} \betro \right\} \; + \; 2\eta''
  &(5.63)\cr&&\cr
   } $$

\n for the longitudinal displacement and

$$ \eqalignno {
   -\omega^2 {\ro\over\bq} \eps \; &= \; \eta \left\{ \gul \left[
      {2\over\rq}{dR_0\over dz} + {2\over\gamma R_0 \hp} - {1\over R_0\hp}
      \betro \right] \right.
  &\cr&&\cr
   &\qquad \left. + \; \gul\gun {1\over H_{p0}^{\,2}} \left[ \beta\delta
      \, - \, {s\hp\over z_0} \, + \, 
      {1\over\gamma} \left( 1 - \gamiz \right) \right] \right. 
  &\cr&&\cr
    &\qquad \left. + \; {\beta r m\over\hp} \left\lgroup 
       {2\over\ve_0\cdot\nnun} \left[ ({\hat{\bf x}}\cdot\lun) w_x
        - \gul w_z - {\ve_0\cdot\lun \over R_0} \right] \, + \,
       {1\over 2\gamma\hp}\gul \right\rgroup \right\}
  &\cr&&\cr
   &\; + \; \eta' \left\{ {4\over R_0} \, - \, \gun {2\over\gamma\hp}
       \, + \, {\beta r m\over 2 \hp} \right\}
  &\cr&&\cr
   &\; + \; \eps \left\{ -{2\over\rq} \, + \, \gun {4\over\gamma R_0 \hp} 
       \, + \, \gun^2 {1\over H_{p0}^{\,2}} \left[ \beta\delta 
       + {1\over\gamma} \left( 1 - \gamiz \right) - {s H_{p0}\over z_0}
       \betro \right] \right.
  &\cr&&\cr
   &\qquad \left. + \; {\beta r m\over\hp} \left\lgroup 
      {2\over\ve_0\cdot\nnun} \left[ ({\hat{\bf x}}\cdot\nnun) w_x
      - \gun w_z \right] \, + \,
      {1\over 2\gamma\hp}\gun \, - \, {1\over 2R_0} \right\rgroup \right\}
  &\cr&&\cr
   &\; + \; \eps' \left\{ -\gul {1\over\hp} \betro \, - \,
      {2\beta r m (\ve_0\cdot\lun)\over\hp(\ve_0\cdot\nnun)} \right\}
  &\cr&&\cr
   &\; + \; 2\eps'' \sms i\omega\eps 
       {2 \beta r m \over \hp (\ve_0\cdot\nnun)} 
  &(5.64)\cr&&\cr
   } $$

\n for the displacement in normal direction. In \e{5.64} we have abbreviated
$\sgn{\ve_0\cdot\nnun}\equiv m$ and used the equilibrium condition, \e{5.4},
in some places. In the special case of a horizontal flux tube in static
equilibrium and uniform gravity ($R_0\to\infty$, $\gul=0$, $\gun=-1$, 
$s=0$, $\ve=0$, $\ro=\roe$, $\bq=\ro=1$) \es{5.63/64} take the form

$$ \eqalignno {
    -\omega^2 \eta \; &= \; 2\eta'' \sms \eps' \gamiz &\cr&&\cr
    -2\omega^2 \eps \; &= \; 2\eps'' \sps \eta' \gamiz  \sps
     \eps \left( \beta\delta + {1\over\gamma} - {2\over\gamma^2} \right).
    &(5.65)
    } $$

\n In order to compare with the result of Spruit and van Ballegooijen (1982)
we have multiplied the inertial term in the equation for $\eps$ by a factor 2
intending to describe the enhanced inertia of the flux tube with respect
to transversal motions in the same way as done by these authors. 
Since the coefficients are constant we are permitted to write
$(\eta, \eps) \propto \exp(ikl_0)$ with wavenumber $k$ in the longitudinal
(here: horizontal) direction. Inserting this into \e{5.65} we obtain the
dispersion relation

$$ \omega^4 \sps \omega^2 \left[ -3k^2 + {\beta\delta\over 2} +
   {1\over 2\gamma} \left( 1 - \gamiz \right) \right] \sps
   2k^2 \left( k^2 - {\beta\delta\over 2} - {1\over 2\gamma} \right) \sgs 0
   \eqno (5.66) $$
    
\n which is identical to the relation found by Spruit and van Ballegooijen
(1982, their Eq.\ts39).

\vfill\eject
\b\b

\def\etopline{ 5.5 Symmetric form for static equilibrium}

\n {\bbf 5.5 Symmetric form for static equilibrium}

\b

\n If we ignore the drag force there is no dissipation in the system and the 
force operator is self-adjoint (Spruit, 1981b). Hence, in the case
$\ve_0 = 0$ (static equilibrium) the perturbation equations (5.63/64) 
must lead to a self-adjoint eigenvalue problem with real eigenvalues.\fussn{*}
{Since the drag force is quadratic in the relative velocity between flux tube
and environment, its perturbation vanishes if $\ve_0 = 0$.}
Symbolically, we can write \es{5.64/65} in the form

$$  - \omega^2 \left ( \eqalign { &\eta \cr
                                   &\eps \cr} \right )  \sgs
    {\bf F}\left ( \eqalign { &\eta \cr
                                   &\eps \cr} \right )  \eqno (5.67) $$

\n where the operator {\bf F} is defined by

$$  {\bf F} \left ( \eqalign { &\eta \cr
                                   &\eps \cr} \right )  \sgs
    \left ( \eqalign { &A\eps + B\eps' + C\eta + D\eta' + E\eta'' \cr
                       &F\eta + G\eta' + H\eps + D\eps' + E\eps'' \cr}\right ).
    \eqno (5.68) $$

\n The functions $A(l_0),...,H(l_0)$ can be obtained from the \rhs of
\es{5.64/65} by multiplication with $\bq/\ro$ and setting $\ve_0 = 0$.
Using the equilibrium condition given by \e{5.4} is readily shown that 
in this (static) case we have $B = -G$. However, the operator {\bf F} is
not yet written in a form in which its self-adjointness becomes apparent.
Such a form is necessary for application of variational methods like the
energy principle (Bernstein et al., 1958). It is also useful for 
numerical calculations since it leads to symmetric matrices which are much
easier to deal with. A symmetric form can be obtained by transforming the
eigenvector as

$$ \left ( \eqalign { &\eta \cr
                                   &\eps \cr} \right )  \sgs
    \left ( \eqalign { &ax \cr
                                   &by \cr} \right )  \eqno (5.69) $$

\n where $x$ and $y$ are the new variables and the functions $a(l_0)$ and 
$b(l_0)$ are determined such that the eigenvalue problem given by 
\es{5.67/68} attains the form

$$ - \omega^2 \left ( \eqalign { &x \cr
                                   &y \cr} \right )  \sgs
    \left ( \eqalign { &Ky \,+\, {1\over 2}\left [ (By)' + By' \right ] 
            \,+\, Mx \,+\, {1\over 2}\left [ (Ex)'' + Ex'' \right ] \cr
            &Kx \,-\, {1\over 2}\left [ (Bx)' + Bx' \right ] 
            \,+\, Ny \,+\, {1\over 2}\left [ (Ey)'' + Ey'' \right ]  \cr} \right )
    \eqno (5.70) $$

\n which possesses the same eigenvalues as \es{5.67/68}. Inserting \e{5.69}
into \e{5.67} leads to the following pair of equations:

$$ \eqalignno {
       -\omega^2 x \; = \; y\,&{b\over a} \left ( A + B {b'\over b} - {1\over 2}B'
             \right) \sps {b\over 2a} \left[ (By)' + By' \right] \sps
             x \left( C + D {a'\over a} + E {a''\over a} - {1\over 2}
             E'' \right ) & \cr&&\cr
       &\quad + \; x' \left ( D + 2E {a'\over a} - E' \right )               
              \sps  {1\over 2}\left [ (Ex)'' + Ex'' \right ] & (5.71) \cr
              } $$

$$ \eqalignno {
       -\omega^2 y \; = \; x\,&{a\over b} \left ( F - B {a'\over a} + {1\over 2}B'
             \right) \sms {a\over 2b} \left[ (Bx)' + Bx' \right] \sps
             y \left( H + D {b'\over b} + E {b''\over b} - {1\over 2}
             E'' \right ) & \cr&&\cr
       &\quad + \; y' \left ( D + 2E {b'\over b} - E' \right )               
              \sps  {1\over 2}\left [ (Ey)'' + Ey'' \right ]. & (5.72) \cr
               } $$

\n By comparison with the desired form given by \e{5.70} we find that the
functions $a(l_0)$ and $b(l_0)$ are determined by the requirement that the
terms which multiply $x'$ in \e{5.71} and $y'$ in \e{5.72}, respectively, must
vanish:

$$ \eqalignno {
                D \sps 2E\, {a'\over a} - E' & \; = \; 0 &\cr&&\cr
                D \sps 2E\, {b'\over b} - E' & \; = \; 0 \,.  & (5.73) \cr
              } $$

\n Consequently, we have $a'/a = b'/b$ and since $a$ and $b$ may be multiplied
by any constant number we can take $a\equiv b$ without loss of generality.
\e{5.73} may be integrated to give

$$ \ln a \sgs {1\over 2}\ln E \sms \int {D\over 2E}\,dl_0 \,. \eqno (5.74) $$

\n The coefficient functions $M$ and $N$ in \e{5.70} are determined using
\e{5.73} which leads to the condition

$$ \eqalignno { M \; &= \; C \sps {(E')^2 - D^2 \over 4E} \sps
              E \left( {E' - D \over 2E} \right)' - {E''\over 2} &\cr&&\cr
                N \; &= \; H \sps {(E')^2 - D^2 \over 4E} \sps
              E \left( {E' - D \over 2E} \right)' - {E''\over 2}. &(5.75) \cr
              } $$             

\n The requirement that the function which multiplies $y$ in the first term
on the \rhs of \e{5.71} is identical to that which multiplies $x$ in the first
term on the \rhs of \e{5.72} in order to conform with \e{5.70} leads to 

$$ A \sps B {b'\over b} \sms {B'\over 2} \sgs K \sgs 
   F \sms B {a'\over a} \sps {B'\over 2}\,. \eqno (5.76) $$

\n This implies the compatibility relation

$$ F \sms A \sgs {B ( E' - D) \over E} \sms B' \eqno (5.77) $$

\n which must be fulfilled if the system is self-adjoint. Hence, we may use
\e{5.77} in order to check the
correctness of the algebraic manipulations and the consistency of the 
approximation which led to \es{5.63/64}.
It turns out, after some lengthy algebra, that \e{5.77} indeed is valid
for our problem, \ie the self-adjointness of the problem is reflected
in the symmetric structure of the resulting system of equations. Defining

$$ {\bf \xi}\; \equiv \; \left ( \eqalign { &x \cr  &y \cr} \right )  $$

\n it is easy to show that \e{5.70} indeed is selfadjoint, \ie the equality

$$ \int {\tilde{\bf\xi}}\cdot{\bf G}({\bf\xi})\,dV \sgs
   \int {\bf\xi}\cdot{\bf G}({\tilde{\bf\xi}})\,dV \eqno (5.78) $$

\n holds. {\bf G} denotes the operator on the \rhs of \e{5.70},
the tilde indicates the complex conjugate, and
the integration is taken over an appropriate volume. Since integrations
by parts are necessary to demonstrate the validity of \e{5.78} the volume of
integration must be chosen large enough such that the displacements can be
assumed to vanish at the boundaries.

\b\b

\def\etopline{ 5.6 Horizontal tubes with vertical external flow}

\n {\bbf 5.6 Horizontal tubes with vertical external flow}

\b

\n The preceding sections give a basis for the determination of the
stability properties of rather general flux tube equilibrium structures.
For realistic convection zone models and flow patterns an equilibrium
can only be determined by numerical means (\eg van Ballegooijen, 1982a;
Anton, 1984; Anton and Sch\"ussler, unpublished) and thus the perturbation
equations, \es{5.63/64}, have to be transformed into a numerically tractable
matrix eigenvalue problem. Such an undertaking is outside the scope of the
present investigation and has to be deferred to future work. In what follows
we shall consider a couple of analytically tractable cases instead to which
we nevertheless attribute some general relevance.

As a first simple application of the formalism we consider
the case $R_0 \to \infty$, $\gul = 0$, $\gun = -1$, $s = 0$, $\ve = v(z)
{\hat {\bf z}}$, \ie a horizontal, straight flux tube whose equilibrium is
determined by a balance between the buoyancy and drag forces (\cf Eq.\ts 5.4/5):

$$ \betro \sgs \beta r m\,.     \eqno (5.79) $$

\n Here we have $m = \sgn{\ve_0\cdot\nnun} = \sgn{v(z_0)}$, \ie $m=1$ describes
an upflow, $m=-1$ a downflow. Defining

$$ {C_D v_0^2 \over \pi a_0 g_0} \; \equiv \; q  \eqno (5.80) $$

\n with $v_0 = \vert v(z_0) \vert$ 
we obtain the equilibrium density contrast as function of the positive number 
$q$ as

$$ 1 - {\roe\over\ro} \sgs {q\over m + q}\,. \eqno (5.81) $$

\n It is clear from \e{5.81} that $q$ is limited to $q<1$ in the case of a
downflow ($m=-1$) since the tube becomes completely evacuated for $q\to 1$
and the buoyancy force cannot be increased beyond that limit. On the other
hand, for an upflow ($m=1$) the range of values for $q$ is not restricted
because the internal
density can be made large enough to balance any upward directed drag force.

If we insert \e{5.79} into the perturbation equations, \es{5.63/64}, and
have regard to the properties of the equilibrium as described in the beginning
of this section we obtain the following pair of equations:

$$ \eqalignno {
    -\omega^2 \eta \; &= \; - \eps' \left( \beta r m + \gamiz \right)
                   \sps 2\eta''  & (5.82) \cr&&\cr
    -\omega^2 \eps \; &= \; \eta' \left( {\beta r m \over 2} + \gamiz \right) \sps
     \eps \left[ \beta\delta + {1\over\gamma} \left ( 1 - {2\over\gamma} \right)
     + \beta r m \left( {2\over v_0} \dz{v_0} - {1\over 2\gamma} \right) \right]
     &\cr&&\cr
     &\qquad \sps 2\eps'' \sms i\omega\eps {2\beta r\over v_0} 
   &(5.83) \cr
    } $$

\n Here we have written $\vert v(z_0) \vert \equiv v_0$ and the notation
$dv_0/dz$ has been used to abbreviate $d\vert v(z) \vert/dz$ taken at $z=z_0$.
Let us first consider displacements which do not depend on the horizontal 
coordinate
$l_0 \equiv x$. All derivatives vanish in this case and we find $\eta \equiv 0$
from \e{5.82}. The equation for $\eps$ is transformed into the dispersion
relation

$$ \omega^2 \sms i\omega {2 \beta r\over v_0} \sps
        \left[ \beta\delta + {1\over\gamma} \left ( 1 - {2\over\gamma} \right)
     + \beta r m \left( {2\over v_0} \dz{v_0} - {1\over 2\gamma} \right) \right]
   \sgs 0\,.  \eqno (5.84) $$

\n Writing 

$$  {\beta r\over v_0} \; \equiv \; S \, \ge 0  \eqno (5.85) $$

\n and

$$  \beta\delta + {1\over\gamma} \left ( 1 - {2\over\gamma} \right)
     + \beta r m \left( {2\over v_0} \dz{v_0} - {1\over 2\gamma} \right) 
     \; \equiv \; T \eqno (5.86) $$

\n we find from \e{5.84} 

$$ \omega_{\pm} \sgs iS \; \pm \; \left( -S^2 - T \right)^{1/2} \eqno (5.87) $$

\n and by multiplication with the imaginary unit we obtain

$$ i\omega_{\pm} \sgs -S \; \mp \; \left( S^2 + T \right)^{1/2}. \eqno (5.88) $$

\n We now have to consider two cases. Firstly, if $S^2 + T < 0$, the square root
in \e{5.88} is imaginary and the growth rate (real part of $i\omega_{\pm}$) 
becomes
equal to $-S$. Consequently, we have a damped oscillation in this case and the
equilibrium is stable. In the second case, $S^2 + T > 0$, both the square root
and $i\omega_{\pm}$ are real. Depending on the sign of the latter,
the perturbation will grow or decay monotonically. It is easy to see that
the stability in this case depends solely on the sign of $T$: If $T<0$ we have 
$i\omega_{\pm}<0$
and the displacement decays, if $T>0$ we have $i\omega_{-}>0$ and a monotonic
growth of the perturbation ensues. Combining the results for both cases we find

$$ T \; \cases
            {\; > 0\; : {\hbox{ monotonic growth}}& \cr &\cr
             \; <0 \; : \cases { S^2 + T < 0 \; : &damped oscillation \cr &\cr
                                S^2 + T > 0 \; : &monotonic decay \cr
                              }&\cr &\cr
            } $$

\n Hence, the stability of the flux tube depends only on the sign of $T$
while the imaginary term in \e{5.83} influences the way in which stable
perturbations decay, monotonically or oscillatory. This behavior is plausible
from the fact that this term, being proportional to $\dot\eps$, 
describes the drag force caused by the perturbation itself and therefore only
has a damping effect. Similar to a damped oscillator the case of 
`creeping motion' is achieved if the damping rate (described by $S$) exceeds 
a critical value. The equilibrium is unstable if $T>0$, \ie if the 
superadiabaticity satisfies the inequality

$$ \beta\delta \; > \; {1\over\gamma} \left ( {2\over\gamma} - 1 \right)
     + \beta r m \left(  {1\over 2\gamma} - {2\over v_0} \dz{v_0} \right)\,.
   \eqno (5.89) $$

\n Consequently, positive terms on the \rhs of \e{5.89} have a stabilizing
influence. For constant velocity, \ie $dv_0/dz = 0$, we find that an upflow
($m=1$) has a stabilizing effect while a downflow ($m=-1$) is destabilizing.
This behavior is caused by the changes in flux tube radius and external density.
An upward displacement leads to an expansion of the tube and a decrease of
external density. Consequently, the drag force decreases (\cf Eq.~5.33).
Similarly, for a downward displacement we find an increase of the drag force.
If we now have a flux tube which is in equilibrium with a downflow, 
for both upward
and downward displacements the perturbation of the (downward) drag force 
tends to increase the displacement and thus favors instability while the reverse 
is true for an upflow.

For $dv_0/dz \ne 0$, the situation is more complicated. Whether the change
of velocity with height acts stabilizing or destabilizing depends on its sign
and on the direction of the flow.
For example, an upflow ($m=1$) whose velocity increases
with height ($dv_0/dz > 0$) leads to a perturbation of the drag force which
tends to increase an initial displacement and thus favors instability.
The influence of the velocity term on the stability properties in the case
of purely vertical displacements is summarized Tab.\ts1:

$$ 
\offinterlineskip \tabskip=0pt
\vbox{
 \halign to 0.90\hsize
  { \strut \vrule # \tabskip=0pt plus 100pt
                  & \hfil # \quad & \vrule#& \quad # \hfil &
                  \vrule#& \quad # \hfil & \vrule#& \quad # \hfil \tabskip=0pt&
                  \vrule# \cr
           \noalign{\hrule}{\vrule height 11.5pt width 0pt}
                           {\vrule depth   6.5pt width 0pt}
 & $m$  &&  $dv_0/dz$  &&  $1/ 2\gamma$  &&  $(1/2\gamma)-(2/v_0)(dv_0/dz)$ &\cr
           \noalign{\hrule}{\vrule height 11.5pt width 0pt}
 & 1 && $> 0 \; \to \;$ destabilizing && stabilizing && ? & \cr
 & 1 && $< 0 \; \to \;$ stabilizing && stabilizing && stabilizing & \cr
 & -1 && $> 0 \; \to \;$ stabilizing && destabilizing && ? & \cr
 {\vrule depth 6.5pt width 0pt}
 & -1 && $< 0 \; \to \;$ destabilizing && destabilizing && destabilizing & \cr
      \noalign{\hrule}
}}$$
 
\tenpoint
\baselineskip = 10pt
 
\n {\bf Tab.\ts{1}:} Influence of the velocity-related terms in the
stability criterion (Eq.\ts5.89) for horizontal magnetic flux tubes and
purely vertical displacement (infinite longitudinal wavelength). 

\vglue 0.5cm
\elevenpoint

\n The combined effect of both terms in \e{5.89} which are affected by the 
external velocity is
indicated in the last column. A question mark indicates that 
one term is stabilizing and the other destabilizing such that it depends
on their relative sizes which one dominates. In dimensionalized form, the
term involving the velocity derivative can be written as

$$ {2\over{\tilde v}_0}{d {\tilde v}_0\over d{\tilde z}} \; \to \;
   {2H_{p0}\over v_0}\dz{v_0} \; \equiv \; {2H_{p0}\over H_v} \eqno (5.90) $$

\n where $H_v$ is the scale height of the external velocity. If we assume a flow 
with constant mass flux density, $\rho v$, and assume further that the temperature
varies much more slowly with height than the density, we have $p_0v_0 =$ const. 
and thus $H_{p0} \approx H_v$. The relation between the two terms in the second
bracket on the \rhs of \e{5.89} is then given by

$$ {(1/2\gamma) \over (2/v_0)(dv_0/dz)} \;\approx\; {1\over 4\gamma} 
   \sgs {3\over 20}  \eqno (5.91) $$

\n for $\gamma = 5/3$. We see that for a flow with constant mass flux density 
the stability properties are mainly determined by the velocity gradient. 
Since density
decreases with height, the flow velocity increases and we see from the table
above that an upflow has a destabilizing influence, while a downflow stabilizes.

If we now consider all terms in 
\e{5.89} we find that a flow can stabilize/destabilize a convectively
unstable/stable flux tube. For a flux tube in temperature equilibrium 
($T_0 = T_{e0}$, 
$\beta r m = -1$) and a downflow with constant mass flux density instability
requires

$$ \beta\delta \; > \; {2\over\gamma^2} - {3\over 2\gamma} + 2 \sgs 1.82
   \eqno (5.92) $$

\n Since equipartition flux tubes in the deep convection zone 
of the Sun have $\beta\delta\approx 3.6$ (Spruit and van Ballegooijen, 1982)
they cannot be stabilized by a flow of this kind. On the other hand, \e{5.92}
predicts stability for flux tubes located in a downflow within an overshoot 
region which have $\beta\delta\le 0$.

The case of purely vertical displacements which do not depend on the horizontal
coordinate treated so far excludes an important destabilizing
mechanism, the flow from the crests to the troughs of a wavelike disturbance
of the flux tube which is the mechanism for the so-called Parker instability
(Parker, 1966). If we allow for a dependence of the displacements $\eta$ and
$\eps$ on the horizontal coordinate, $x$, we may write

$$ \eta\,,\eps \; \propto \; e^{ikx} \eqno (5.93) $$

\n with (real) wavenumber $k$ since the coefficients in \es{5.82/83} are
constant. Using the symbols $\eta$ and $\eps$ again for the (constant) 
{\it amplitudes\/} of the perturbations we obtain by inserting \e{5.93}
into \es{5.82/83}:

$$ \eqalignno {
    -\omega^2 \eta \; &= \; - ik\eps \left( \beta r m + \gamiz \right)
                   \sms 2k^2\eta  & (5.94) \cr&&\cr
    -\omega^2 \eps \; &= \; ik\eta\left( {\beta r m \over 2} + \gamiz \right) \sps
     \eps \left[ \beta\delta + {1\over\gamma} \left ( 1 - {2\over\gamma} \right)
     + \beta r m \left( {2\over v_0} \dz{v_0} - {1\over 2\gamma} \right) \right]
     &\cr&&\cr
     &\qquad \sms 2k^2\eps \sms i\omega\eps {2\beta r\over v_0} .
   &(5.95) \cr
    } $$

\n This linear, homogeneous system of equations has non-trivial solutions
for eigenvalues $\omega$ which satisfy the dispersion relation

$$ \eqalignno {
   \left( \omega^2-2k^2 \right)& \left[ \omega^2 -i\omega {2\beta r\over v_0}
        -2k^2 + \beta\delta + {1\over\gamma} \left ( 1 - {2\over\gamma} \right)
     + \beta r m \left( {2\over v_0} \dz{v_0} - {1\over 2\gamma} \right) \right]
   &\cr&&\cr
   &- \; k^2 \left( \gamiz + {\beta r m\over 2} \right)
             \left( \gamiz + \beta r m \right) \sgs 0 \,. & (5.96) \cr
  } $$

\n \e{5.96} can be written as a fourth order polynomial in
$i\omega\equiv{\hat\omega}$ with real coefficients:

$$ \homeg^4 \sps A\homeg^3 \sps B\homeg^2 \sps C\homeg \sps D \sgs 0
   \eqno (5.97) $$

\n with

$$ \eqalign {
        A \; &= \; {2\beta r\over v_0}  \cr
        B \; &= \; 4k^2 - \beta\delta - {1\over\gamma} 
                                        \left( 1 - {2\over\gamma} \right)
       - \beta r m \left( {2\over v_0} \dz{v_0} - {1\over 2\gamma} \right) \cr
        C \; &= \; {4k^2\beta r\over v_0} \cr
        D \; &= \; 2k^2 \left[ 2k^2 - \beta\delta - {1\over\gamma}
           - \beta r m \left( {2\over v_0} \dz{v_0} + {1\over\gamma} \right) 
           - {(\beta r)^2\over 4} \right] \cr
    } $$

\n Since the coefficients are real, the roots of this polynomial can be real
numbers and pairs of complex conjugates. A positive, real root means monotonic
instability, i.e.~exponential growth of the perturbation, while a complex
root with positive real part represents overstability, \ie
oscillations or waves with exponentially growing amplitude. 
In order to establish a sufficient condition for monotonic instability
we may utilize Descartes' sign rule: 

\smallskip
\n {\it The number of positive, real roots
of a fourth order polynomial with real coefficients is smaller than or equal
to the number of sign changes in the sequence of its coefficients. The difference
is an even number.} 
\smallskip

\n Since in our case we have $A\ge 0$ and $C\ge 0$ the number of sign changes
is determined by the signs of $B$ and $D$ as indicated in Tab.\ts2:

$$ 
\offinterlineskip \tabskip=0pt
\vbox{
 \halign to 0.90\hsize
  { \strut \vrule # \tabskip=0pt plus 100pt
                  & \hfil # \hfil & \vrule#& \hfil # \hfil &
                  \vrule#& \hfil # \hfil & \vrule#& \hfil # \hfil & 
                  \vrule#& \hfil # \hfil \tabskip=0pt&
                  \vrule# \cr
           \noalign{\hrule}{\vrule height 11.5pt width 0pt}
                           {\vrule depth   6.5pt width 0pt}
 & Case  &&  $B$  &&  $D$  &&  Sign changes && Positive, real roots &\cr
           \noalign{\hrule}{\vrule height 11.5pt width 0pt}
 & 1 && $> 0$ && $<0$ && 1 && 1 & \cr
 & 2 && $> 0$ && $>0$ && 0 && 0 & \cr
 & 3 && $< 0$ && $<0$ && 3 && 1 or 3 & \cr
 {\vrule depth 6.5pt width 0pt}
 & 4 && $< 0$ && $>0$ && 2 && 0 or 2 & \cr
      \noalign{\hrule}
}}$$

\tenpoint
\baselineskip = 10pt
 
\n {\bf Tab.\ts{2}:} Estimate of the number of positive, real roots (\ie 
monotonically unstable modes) 
of \e{5.97} using the number of sign changes in the sequence of its coefficients
(Descartes' rule). Since $A$ and $C$ are positive, this number depends 
only on the signs of $B$ and $D$.

\vglue 0.5cm
\elevenpoint

\n In the cases 1 and 3 we have at least one positive root. Consequently, 
$D<0$ {\it is a sufficient condition for monotonic instability.}
In case 2 there is no monotonic instability (but possibly overstability)  while
for case 4 no definite statement can be made at this stage.
The condition $D<0$ can be transformed into a condition for the wavenumber, $k$:

$$ k^2 \; < \; k_0^2 \; \equiv \; {1\over 2} \left[ \beta\delta + {1\over\gamma}
           + \beta r m \left( {2\over v_0} \dz{v_0} + {1\over\gamma} \right) 
           + {(\beta r)^2\over 4} \right] \eqno (5.98) $$

\n Since any value of $k$ is allowed in our cartesian model (in contrast
to a spherical system where the wavelength cannot be larger than the 
circumference), the flux tube is monotonically unstable if only $k_0^2>0$, \ie

$$ \beta\delta \; > \; -{1\over\gamma}
           + \beta r m \left( -{1\over\gamma}  - {2\over v_0} \dz{v_0} \right) 
           - {(\beta r)^2\over 4}\,. \eqno (5.99) $$

\n We immediately see that a flow with constant velocity 
($dv_0/dz=0$) cannot stabilize
a tube within a superadiabatic environment: For an upflow ($m=+1$) both
remaining velocity terms are negative while it is easy to see that for
a downflow ($m=-1$) the \rhs of \e{5.99} is always negative since $\gamma>1$.
If the flow speed depends on depth, stabilization is possible.
A downflow whose speed decreases with depth such that the mass flux density
stays constant [$m=-1$, $(2/v_0)(dv_0/dz)=2$] leads to positive values of
the \rhs of \e{5.99} provided that $0.23<\beta r<10.16$, \ie within a certain
range of values for the parameter $\beta r$. 

Due to the last term in \e{5.99} a flux tube is always monotonically unstable
for sufficiently large values of $\beta r$, irrespective of the flow direction.
This is caused by the change of the flux tube radius in the course of a
wave-like displacement, a mechanism which can be understood with aid of the
heuristic approach followed in Sec.\ts{5.7.3}.

For the example discussed in conjunction with the criterion \e{5.92}
(temperature equilibrium, $T_0 = T_{e0}$, and downflow with 
constant mass flux density, $\beta r m = -1$) we find from \e{5.99} 
the criterion $\beta\delta > 1.75$ for monotonic instability, \ie a slightly 
smaller critical value than that obtained for the case of purely vertical 
displacements. Generally we find that the condition given in \e{5.99}
leads to instability for smaller values of the $\beta\delta$, \ie monotonic
instability is easier to excite for wave-like
perturbations (even if $k\to 0$) than for those with
$k\equiv 0$. This is similar to the case without 
external flow treated by Spruit and
van Ballegooijen (1982) whose results may be recovered by setting $\beta r = 0$.
However, in our case flux tubes embedded in a subadiabatic region ($\beta\delta\le
0$) are stable with respect to wave-like perturbations while in the case without
external flow instability results if $\beta\delta > -1/\gamma$.

The general properties of the roots of \e{5.97} unfortunately are not
as easily obtained as the sufficient condition for positive, real roots.
However, since the coefficients $A$ and $C$ result from the imaginary term
in \e{5.83} which mainly has a damping effect and does not affect the
stability conditions for the case $k=0$, we may conjecture that this is the
case for $k \ne 0$ as well. We therefore consider a reduced dispersion
relation by setting $A=C=0$ in \e{5.97}. The conclusions drawn from
the discussion of this equation have been verified by (numerical)
determination of the roots of the full equation (5.97).\fussn{*}
{The general algorithm for the determination of roots of quartic equations in
closed form is not used here because, given the complicated structure of the
coefficients, it leads to lengthy algebraic expressions which give no direct
insight in the properties of the stability problem.}
For $A=C=0$ \e{5.97} is transformed into a biquadratic equation which is
readily solved, viz.

$$ \homeg^2 \sgs -{B\over 2} \pm \left( {B^2\over 4} - D \right)^{1/2}.
   \eqno (5.100) $$

\n We see immediately that we recover the sufficient condition for monotonic
instability, $D<0$, since in this case always one solution $\homeg^2>0$ exists.
A second possibility for monotonic instability is the case $B<0$ and 
$0\le D \le B^2/4$ while the case $B>0$, $0\le D \le B^2/4$ leads to
stable solutions ($\homeg^2<0$). Complex roots which signify oscillatory 
instability (overstability)
appear for $B^2/4 - D < 0$, irrespective of the sign of $B$. Since we have

$$ {B^2\over 4} - D \sgs k^2 \left[ {4\over \gamma^2} + {(\beta r)^2\over 2}
   + {3\over \gamma}\beta r m \right] \sps {1\over 4}\left[ \beta\delta
   + {1\over\gamma} \left( 1 - {2\over\gamma} \right)
   + \beta r m \left( {2\over v_0} \dz{v_0} - {1\over 2\gamma} \right) \right]^2
   \eqno (5.101) $$

\n complex roots exist only if the first term on the \rhs is negative.
This requires $m=-1$ (downflow) and 

$$ {2\over\gamma} \; < \; \beta r \; < \; {4\over\gamma} \,. \eqno (5.102) $$

\n For $\gamma=5/3$ this range is given by $1.2<\beta r<2.4$. If both conditions
are satisfied we see from \e{5.101} that the wavenumber can always be made large 
enough to give $B^2/4 - D < 0$. Hence, the condition for overstability is 
$k^2>k_1^2$ with

$$ k_1^2\sgs {1\over 4}\left[ -{4\over \gamma^2} - {(\beta r)^2\over 2}
   - {3\over \gamma}\beta r m \right]^{-1} 
   \left[ \beta\delta +{1\over\gamma} \left( 1-{2\over\gamma} \right)
 + \beta r m \left( {2\over v_0} \dz{v_0} - {1\over 2\gamma} \right) \right]^2.
 \eqno (5.103) $$

\n Note that in order to observe the limits set by the approximation
of slender flux tubes the radius of the flux tube has to be much smaller
than the perturbation wavelength. This can be a severe restriction if
overstability appears only for very small wavelength (very large wavenumber).
If we have $k^2>k_1^2$ the four solutions of \e{5.100} are given by

$$ \homeg \sgs \pm {1\over 2} \left( 2D^{1/2} - B \right)^{1/2} 
               \; \pm \; {1\over 2} i\left( 2D^{1/2} + B \right)^{1/2} .
               \eqno (5.104) $$

\n We see that if the conditions for complex roots are satisfied there
are always unstable solutions, \ie roots with a positive real part. 
The overstability is caused by the drag force which is added to the restoring
force due to magnetic tension such that an oscillation with growing amplitude
results. The conditions for overstability depend only on the sign and the
amplitude of the velocity (\cf Eq.~5.102). In particular, they are 
independent of the velocity gradient, a fact which is due to the linear approach.

Numerical solutions of the exact dispersion relation, \e{5.97}, basically 
confirm the
criteria derived on the basis of the discussion of the reduced form, \e{5.100}.
In most cases, finite values of $A$ and $C$ only affect the growth or decay
rates or change oscillatory into monotonic decay. A notable exception is
the result that finite values of $A$ and $C$ lead to positive growth rates
of oscillatory modes whenever the conditions $m=-1$ and \e{5.102} are fulfilled, 
even for wave numbers which give $B^2/4 - D > 0$. The second possibility for 
monotonic instability mentioned above, $B < 0$ and $0 \le D \le B^2/4$, does
not appear in practice since it requires that the conditions $m=-1$ and
\e{5.102} are fulfilled which immediately lead to oscillatory instability.
Hence, we can summarize the stability properties as follows:

\m {\narrower\narrower
    \it \n A horizontal flux tube whose equilibrium is determined by a
    balance between buoyancy and drag force exerted by an external, vertical flow
    is monotonically unstable if the condition $D<0$ (\cf Eq.\ts{5.99}) is
    satisfied. In particular, flux tube equilibria with a sufficiently large 
    value of $\beta r$ are always unstable, irrespective of the flow direction.
    A flux tube may exhibit oscillatory instability (overstability) if the
    external flow is a downflow and the value of $\beta r$ is within a
    certain range (\cf Eq.\ts{5.102}). \par}
\m

\n Oscillatory instability can be excited whenever the conditions
$m=-1$ and $2/\gamma < \beta r < 4/\gamma$ are fulfilled. This excludes
isothermal flux tubes which require $\brm = -1$. Overstability
cannot be stabilized by the stratification, however subadiabatic it may be.
Consequently, it is potentially relevant for all flux tubes, 
irrespective of their location in convection zones, overshoot layers or regions
in radiative equilibrium. Since $\beta r$ depends on the field strength and
radius of the tube as well as on the flow  velocity, for given values of two
of these parameters \e{5.102} defines a range of the third parameter which 
leads to overstability. The tube radius, however, must always be small enough
to stay consistent with the utilization of the approximation of slender flux
tubes.

While the convective and Parker-type instabilities are most easily excited for
large wavelength along the flux tube, overstable modes have their largest growth
rates for small wavelengths. Consequently, we expect that overstability
is not only relevant for horizontal flux tubes but also for more general forms
of equilibrium. If we restrict ourselves to perturbation wavelengths which are
much smaller than the wavelength of the equilibrium path of a flux tube
a kind of local analysis is possible. Overstable modes can be revealed by this
kind of treatment while the convective and Parker-type instabilities are
suppressed since they only appear for large enough wavelengths. An example of
such a local analysis is given in Sec.\ts{5.7.1} below.

Let us finally consider the influence of the coefficients
$A$ and $C$ on the growth rates of unstable modes.
Tab.\ts3 below illustrates the dependence of $\homeg$ 
on the value of $v_0$ which determines the size of $A$ and $C$ for given values
of $\beta r$ (\cf Eq.\ts{5.97}). We consider two examples: Monotonic
instability of equipartition flux tubes ($\beta\delta = 3.6$) in temperature
equilibrium with the environment ($\beta r m = -1$) and overstability of
equipartition tubes with $\beta r m = -1.8$.\fussn{ }{\textindent{*}
Strictly spoken, $v_0\to\infty$ is inconsistent with the assumption of a 
finite value for $\beta r$ unless the flux tube diameter becomes infinite, too.
We nevertheless consider this case (which gives $A=C=0$) for the purpose of
comparison with the results for finite values of $v_0$.}

$$ 
\offinterlineskip \tabskip=0pt
\vbox{
 \halign to 0.90\hsize
  { \strut \vrule # \tabskip=0pt plus 100pt
                  & \quad # \hfil & \vrule#& \quad # \hfil &
                  \vrule#& \quad # \hfil \tabskip=0pt&
                  \vrule# \cr
           \noalign{\hrule}{\vrule height 11.5pt width 0pt}
                           {\vrule depth   6.5pt width 0pt}
 & $v_0$  &&  $\beta r m = -1.$, $k=0.$ (monotonic)  &&  
$\beta r m = -1.8$, $k=1.$ (overstable) &\cr
           \noalign{\hrule}{\vrule height 11.5pt width 0pt}
 &  && \quad Re($\homeg$)\hfil Im($\homeg$) && 
       \quad Re($\homeg$)\hfil Im($\homeg$) & \cr
           \noalign{\hrule}{\vrule height 11.5pt width 0pt}
 & $\infty^*$  && \quad 1.334\hfil 0. && \quad  0.137\hfil 1.345 & \cr
 & 10.    &&\quad  1.238\hfil 0. && \quad 0.077\hfil 1.348 & \cr
 & 1.     && \quad 0.667\hfil 0. && \quad 0.012\hfil 1.413 & \cr
 {\vrule depth 6.5pt width 0pt}
 & 0.1    && \quad 0.089\hfil 0. && \quad 0.001\hfil 1.414 & \cr
      \noalign{\hrule}
}}$$
 
\tenpoint
\baselineskip = 10pt
 
\n {\bf Tab.\ts{3}:} Dependence of growth rate, Re($\homeg$), and
oscillation frequency, Im($\homeg$), on the the vertical velocity $v_0$
which determines the size of the coefficients of the damping terms $A$ and $C$.
Two cases are considered: Monotonic 
instability of equipartition flux tubes ($\beta\delta = 3.6$) in temperature 
equilibrium with the environment ($\beta r m = -1$) and overstability of 
equipartition tubes with $\beta r m = -1.8$. While Im($\homeg$) is barely
affected, the growth rate decreases drastically for small values of $v_0$ and
fixed $\beta r$. 

\vglue 0.5cm
\elevenpoint

\n As explained in Sec.\ts{5.4}, the velocity $v_0$ is measured in
units of $v_A/\sqrt 2$ and the time unit is
of the order of one day for the deep layers of the solar convection zone.
Smaller values of $v_0$ (for fixed values of $\beta r$, see discussion
below) entail larger values of $A$ and $C$ which lead to smaller growth
rates of both monotonic and oscillatory instabilities. The overstable modes
which typically have much smaller growth rates than the monotonic modes
are more strongly affected by this damping mechanism. 
The {\it growth time} $2\pi/{\hbox {Re}}(\homeg)$ of the overstable mode in our 
example increases from about 80 days for $v_0 = 10.$ to 17 years for $v_0 = 0.1$
while for the monotonic mode the numbers are 7 days and 70 days, respectively.
On the other hand, the oscillation frequencies are hardly affected.

We may estimate the value of $v_0$ associated with a given value of $\beta r$
with aid of \e{5.6} which may be simplified
by assuming $\beta\gg 1$ and $2C_D/\pi\approx 1$. Written in non-dimensional
quantities we obtain

$$ \beta r \; \approx \; {v_0^2\over a_0}. \eqno (5.105) $$

\n Here $a_0$ is the flux tube radius in units of the external pressure
scale height. For the interesting range $\beta r = O(1)$ we have

$$ v_0 \; \approx \; {\sqrt a_0} \eqno(5.106) $$

\n We see that for {\it fixed\/} $\beta r$ the quantity $v_0$ effectively is 
a measure of the flux tube radius. Tab.\ts3 shows that small flux tubes 
are more strongly influenced by the
drag forces and suffer from a stronger damping than those with larger radius.
For $\beta r = 1$ and $H_{p0}$ = \dex{5}{4} km we find the range 
$v_0 \approx 0.03...0.3$
for flux tube radii between 50 km and 5000 km. Consequently, we have to expect
a strong effect of the damping terms on the growth rates,
especially in the case of overstable modes.

\b\b

\def\etopline{ 5.7 Symmetric loops with vertical external flow}

\n {\bbf 5.7 Symmetric loops with vertical external flow}

\b\m

\n As an example of a flux tube equilibrium with a curved path we investigate
a planar loop which is symmetric with respect to a vertical line through its
maximum or minimum. Since we shall consider the stability only in the vicinity
of the point of extremum, the results are relevant also for flux tubes
which wind like a serpentine line, \ie consist of a sequence of (symmetric) 
loops with alternating maxima and minima. Such configurations are of practical 
interest
since one could imagine that the kink instability of a horizontal (or toroidal)
flux tube in a convection zone leads to `sea serpent' structures, \ie a 
series of erupted active regions connected by loops with minima in
the convection zone. Another interesting question is whether `dived sea serpents',
a series of minimum and maximum loops fully within the convection zone 
represent a stable alternative to the unstable horizontal tubes.

As in the preceding section we
assume a purely vertical external velocity field which does not depend on $x$, 
the coordinate in the horizontal direction, and continue to use the 
notation of the velocity terms introduced there.
The equilibrium is determined by a balance of buoyancy, curvature and drag force
which is described by \e{5.4}. For the point of extremum with
$\gun = -1$ in the case of a minimum and $\gun = +1$ in the case of a maximum
we have (in non-dimensional form)

$$ \mp \betro \sgs {2\over\Rn} \sps \brm\,. \eqno (5.107) $$

\n The upper sign on the \lhs applies for a maximum, the lower sign for a minimum.
In both cases we have $\gul = 0$, \ie the tangent vector has a horizontal
direction at the point of extremum.

\b
\n {\bf 5.7.1 Local analysis}
\m

\n Overstable modes of a symmetric loop can be treated analytically by way of a
local stability analysis assuming 
perturbations with small wavelength (large wavenumber) in the direction along 
the flux tube. We consider the extremum point (maximum or minimum) of a
symmetric loop and investigate its stability with respect to growing oscillations
of short wavelength in the neighborhood of this point.

We assume wave-like perturbations, \ie $\eps,\, \eta \propto \exp(ikx)$ whose 
wavelength $\lambda = 2\pi/k$ is small enough such that we can take all quantities
describing the equilibrium flux tube ($\Rn,\hp,\ro,\gul,\gun...)$ to be constant 
within a wavelength. On the other hand, the approximation of slender flux tubes
demands $\lambda\gg a_0$ where $a_0$ is the radius of the tube.
Consequently, we require $a_0\ll\lambda\ll\Rn$ and also 
$\lambda\ll (8\Rn\hp)^{1/2}$ which results from the requirement that the 
height increment of the path of the equilibrium flux tube within one wavelength
must be small compared to the scale height. Both conditions lead to

$$ {2\pi\over a_0} \; \gg \; k \; \gg \; 
   {2\pi\over \min(\Rn,\,8\Rn\hp)^{1/2}}. \eqno (5.108) $$

\n Since overstability often requires that the wavenumber exceeds some
critical value one has to keep in mind that the following results are only
applicable within the limits of the approximation of slender flux tubes
if the tube radius satisfies the left part of the above relation.
For wavenumbers which satisfy \e{5.108} we may use \e{5.107} to rewrite 
the general perturbation equations, \es{5.63/64},  
taking the extremum as reference point, $z_m$, for the 
non-dimensionalization of all quantities and obtain

$$ \eqalignno { 
    -\omega^2 \eta \; &= \;  ik\eps \left( -{4\over R_0} - \brm \pm \gamiz \right) 
              \sps \eta {1\over\Rn} \left( -{2\over\Rn} - \brm \pm \gamiz \right)
              \sms 2k^2\eta  & (5.109) \cr&&\cr
    -\omega^2 \eps \; &= \; ik\eta\left({4\over\Rn}+{\brm\over 2}+\gamiz \right)
                            \sms 2k^2\eps \sms i\omega\eps {2\beta r\over v_0}
     &\cr&&\cr
     &\quad + \; \eps \left[ -{2\over\rq} \pm {4\over\gamma\Rn} +
      \beta\delta + {1\over\gamma} \left ( 1 - {2\over\gamma} \right)
     + \brm \left(\mp {2\over v_0} \dz{v_0} \pm {1\over 2\gamma}
     - {1\over 2\Rn} \right) \right].  &(5.110) \cr
    } $$
   
\n Note that for the lower signs ($\gun=-1$) and
in the limit $\Rn\to\infty$ these equations pass over to 
\es{5.94/95} for a horizontal flux tube.
We follow the same procedure as in the preceding section and determine
from \es{5.109/110} a dispersion relation for $i\omega\equiv\homeg$, again
of a fourth order polynomial with real coefficients:

$$ \homeg^4 \sps A\homeg^3 \sps B\homeg^2 \sps C\homeg \sps D \sgs 0
   \eqno (5.111) $$
%
%
$$ \eqalign {
       &A \; = \; {2\beta r\over v_0}  \cr&\cr
       &B \; = \; 4k^2 + {2\over\Rn} \left( {2\over\Rn}\mp{3\over\gamma}
                \right) - \beta\delta - {1\over\gamma} 
                                        \left( 1 - {2\over\gamma} \right)
                + \brm \left( \pm {2\over v_0} \dz{v_0} \mp {1\over 2\gamma}
                              + {3\over 2\Rn} \right) \cr&\cr
       &C \; = \; {2\beta r\over v_0} \left[ 2k^2 + {1\over\Rn} \left(
                   {2\over\Rn} + \brm \mp \gamiz \right) \right]  \cr&\cr
       &D \; = \; 2k^2 \left[ 2k^2 - \beta\delta - {1\over\gamma}
           + {2\over\Rn} \left( \pm {1\over\gamma} - {2\over\Rn} \right)
           + \brm \left( \pm {2\over v_0} \dz{v_0} \pm {1\over\gamma} 
           - {3\over 2\Rn} \right) - {(\beta r)^2\over 4} \right] \cr&\cr
             &\qquad + \; {1\over\Rn} \left\{ {4\over\rq} \left( {1\over\Rn}
           \mp {3\over\gamma} \right) + {2\over\gamma\Rn} \left( {6\over\gamma}
           -1 \right) \pm {2\over\gamma^2} \left( 1 - \gamiz \right) 
           + \beta\delta \left( -{2\over\Rn} - \brm\pm\gamiz\right)\right.\cr&\cr
             &\qquad + \; \left. \brm \left[ {2\over v_0}\dz{v_0} 
             \left( \gamiz \mp {2\over\Rn}
             \right) + {1\over\gamma} \left( {3\over\gamma} - {6\over R_0} - 1
             \right) + {3\over\rq} \right] + (\beta r)^2 \left( 
             \mp {2\over v_0}\dz{v_0} \mp {1\over 2\gamma} + {1\over 2\Rn} \right)
    \right \} \cr
    } $$

\n In analogy to the treatment in Sec.\ts{5.6} we consider
a reduced dispersion relation obtained by setting $A=C=0$ in \e{5.111}
which leads to 

$$ \homeg^2 \sgs -{B\over 2} \pm \left( {B^2\over 4} - D \right)^{1/2}.
   \eqno (5.112) $$

\n We have obtained a number of numerical solutions of the full dispersion 
relation, \e{5.111}, which confirm the 
criteria derived below on the basis of \e{5.112}.
Finite values of $A$ and $C$ only affect the growth or decay rates.
Monotonic instability for large values of $k$ requires extreme values for
$\beta\delta$ or $\beta r$ which are unrealistic for a convection zone.
Oscillatory instability, on the other hand, which sets in if $B^2/4-D<0$
can be well described by local analysis. This expression is given by

$$ \eqalignno {
   {B^2\over 4} - D &\sgs k^2 \left[ {4\over \gamma^2} + {(\beta r)^2\over 2}
   + 3\brm \left( {2\over\Rn} \mp {1\over \gamma} \right) + {16\over\Rn} 
   \left( {1\over\Rn} \mp {1\over\gamma} \right) \right] & \cr && \cr
   & \sps {1\over 4}\left[ -{2\over\Rn} \left( {2\over\Rn}\mp{3\over\gamma} 
     \right) + \beta\delta + {1\over\gamma} \left( 1 - {2\over\gamma} \right) 
     - \brm \left( \pm {2\over v_0} \dz{v_0} \mp {1\over 2\gamma}
     + {3\over 2\Rn} \right) \right]^2 &\cr &&\cr
   & \sms {1\over\Rn} \left\{ {4\over\rq} \left( {1\over\Rn}
           \mp {3\over\gamma} \right) + \; ... \; + (\beta r)^2 \left( 
             \mp {2\over v_0}\dz{v_0} \mp {1\over 2\gamma} + {1\over 2\Rn} \right)
    \right \} &(5.113) \cr
    } $$

\n where the term in braces is the same as the last term in the above 
definition of $D$. Since local analysis demands large values of $k$ 
we conclude that the sign of $B^2/4-D$ is determined by the sign of the 
term multiplied by $k^2$ in \e{5.113}. Note that the superadiabaticity does
not enter into this term, \ie the excitation condition for oscillatory
instability is independent of the stability of the external stratification
which may only affect the wavelength range of overstable modes. This applies
also to horizontal flux tubes (\cf Eqs.\ts{5.101/103}). A simple calculation shows
that the term under consideration is negative if

$$ (\beta r)_{\min} \; < \; \beta r \; < \; (\beta r)_{\max} \eqno (5.114) $$

\n where

$$ 
   (\beta r)_{\min} \; = \; \min \left[ (\beta r)_1, (\beta r)_2 \right],
   \qquad
   (\beta r)_{\max} \; = \; \max \left[ (\beta r)_1, (\beta r)_2 \right] 
   \eqno (5.115) $$

\n with

$$ \eqalignno {
   (\beta r)_1 \; &= \; (3m - 1)\left( \pm {1\over\gamma}-{2\over\Rn}\right) &
   \cr&&\cr
   (\beta r)_2 \; &= \; (3m + 1)\left( \pm {1\over\gamma}-{2\over\Rn}\right). 
   &(5.116) 
   }
$$

\n Since $\beta r$ is a positive quantity
and $(\beta r)_1$ and $(\beta r)_2$ always have the same sign, overstability
is only possible if both are positive, too. We now have to distinguish 4 cases, 
\ie maximum/minimum and upflow/downflow, respectively. Note that since 
$m = \sgn{\ve_0\cdot\nnun}$ the value $m=1$ signifies
an upflow for a minimum and a downflow for a maximum while $m=-1$ corresponds
to a downflow for a minimum and an upflow for a maximum. Tab.\ts4 
summarizes the stability properties for the four possible cases. The fourth column
gives necessary conditions on $R_0$ for overstability while the last column
shows the range of values of $\beta r$ which lead to overstability. Cases 3 and 4
with $\Rn\to\infty$ pass over to the case of a horizontal flux tube treated
in the preceding section (\cf Eq.$\,$5.102).

$$  
\offinterlineskip \tabskip=0pt 
\vbox{ 
 \halign to 1.0\hsize 
  { \strut \vrule # \tabskip=0pt plus 100pt 
                  & \hfil # \hfil & \vrule#& \enskip # \hfil &
                  \vrule#& \quad # \hfil & \vrule#& \hfill # \hfill & 
                  \vrule#& \hfil # \hfil \tabskip=0pt&
                  \vrule# \cr
           \noalign{\hrule}{\vrule height 11.5pt width 0pt}
                           {\vrule depth   6.5pt width 0pt}
 & Case  &&  Extremum  && \hfil $m$ \hfil &&  Condition && 
   Oscillatory instability for &\cr
           \noalign{\hrule}{\vrule height 11.5pt width 0pt}
                           {\vrule depth   8.5pt width 0pt}
 & 1 && maximum && $-1$ (up) && $\Rn<2\gamma$ && 
 $2 (2/\Rn - 1/\gamma) < \beta r < 4 (2/\Rn - 1/\gamma)$ & \cr
                           {\vrule depth   8.5pt width 0pt}
 & 2 && maximum && $+1$ (down) && $\Rn>2\gamma$ && 
 $2 (1/\gamma - 2/\Rn ) < \beta r < 4 (1/\gamma - 2/\Rn )$ & \cr
                           {\vrule depth   8.5pt width 0pt}
 & 3 && minimum && $-1$ (down) &&   && 
 $2 (1/\gamma + 2/\Rn ) < \beta r < 4 (1/\gamma + 2/\Rn )$ & \cr
                             {\vrule depth 6.5pt width 0pt}
 & 4 && minimum && $+1$ (up) &&   &&  no overstability & \cr
      \noalign{\hrule}
}}$$

\tenpoint
\baselineskip = 10pt
  
\n {\bf Tab.\ts{4}:} Range of values for $\beta r$ which lead to overstability 
of symmetric loops with fixed radius of curvature, $\Rn$. The four possible cases 
(maximum/minimum, upflow/downflow) are given. For the first pair of cases 
necessary conditions for overstability exist which are indicated in the fourth 
column.

\vglue 0.5cm
\elevenpoint

\n We must keep in mind, however, that the quantities $\Rn$ and $\beta r$ 
generally cannot be chosen independently since they are related through the 
equilibrium condition given by \e{5.107}. If the relation between internal and 
external temperature and the value of $\brm$ are given, the radius of curvature 
is fixed. Hence, \e{5.116} and Tab.\ts4 are only applicable if the internal
temperature and $\beta r$ are chosen in such a way that $\Rn$ stays constant.
For more realistic cases we must prescribe the relation between external and
internal temperature and determine $\Rn$ according to \e{5.107} for any given
value of $\beta r$.

In what follows we discuss the case of an {\it isothermal\/} flux tube, \ie 
internal equal to external equilibrium temperature. In this case the equilibrium 
(written in dimensionless form) is given by 

$$ {2\over\Rn} \sps \brm \sgs \pm 1\,. \eqno (5.117) $$

\n We notice that no equilibrium is possible for a minimum with upflow. This is 
due to the fact that an isothermal flux tube has an upward
directed buoyancy force and the curvature force in case of a minimum has
the same direction. Consequently, force balance can only be achieved by a
downflow. In other cases there are restrictions on $\beta r\,$: For a minimum
with downflow we must have $\beta r \ge 1$ while a maximum with downflow requires
$\beta r \le 1$ and $\Rn\ge 1$. In the case of a maximum with 
upflow any value of $\beta r$ is permitted. We insert \e{5.117} into
\e{5.113} and determine the range of oscillatory instability by the same procedure
which led to \es{5.114-116}. We find

$$ 
   (\beta r)_1 \; = \; {(\pm 5m - 1)\over 3}\left( 1 - {1\over\gamma}\right),
   \qquad
   (\beta r)_2 \; = \; {(\pm 5m + 1)\over 3}\left( 1 - {1\over\gamma}\right).
   \eqno(5.118) 
$$

\n Since we certainly have $\gamma \ge 1$ the sign of $(\beta r)_{1,2}$ for a
given equilibrium is determined only by $m$. The stability criteria for the 
isothermal case as they follow from \es{5.114/118} are summarized in Tab.\ts5:

$$  
\offinterlineskip \tabskip=0pt 
\vbox{ 
 \halign to 1.0\hsize 
  { \strut \vrule # \tabskip=0pt plus 100pt 
                  & \hfil # \hfil & \vrule#& \enskip # \hfil &
                  \vrule#& \quad # \hfil &
                  \vrule#& \hfil # \hfil \tabskip=0pt&
                  \vrule# \cr
           \noalign{\hrule}{\vrule height 11.5pt width 0pt}
                           {\vrule depth   6.5pt width 0pt}
 & Case  &&  Extremum  && \hfil $m$ \hfil &&
   Oscillatory instability for &\cr
           \noalign{\hrule}{\vrule height 11.5pt width 0pt}
                           {\vrule depth   8.5pt width 0pt}
 & 1 && maximum && $-1$ (up) &&  no overstability & \cr
                           {\vrule depth   8.5pt width 0pt}
 & 2 && maximum && $+1$ (down) && 
 $(4/3) (1 - 1/\gamma) < \beta r < 2 (1 - 1/\gamma)$ & \cr
                           {\vrule depth   8.5pt width 0pt}
 & 3 && minimum && $-1$ (down) && 
 $(4/3) (1 - 1/\gamma) < \beta r < 2 (1 - 1/\gamma)$ & \cr
                             {\vrule depth 6.5pt width 0pt}
 & 4 && minimum && $+1$ (up) &&  no equilibrium & \cr
      \noalign{\hrule}
}}$$
\nobreak
\tenpoint
\baselineskip = 10pt
\n {\bf Tab.\ts{5}:} Range of values for $\beta r$ which lead to overstability 
of {\it isothermal,} symmetric loops. 
Only configurations with an external downflow are liable to overstability.
 
\elevenpoint

\n For $\gamma=5/3$ the range of values
of $\beta r$ which lead to overstability in cases 2 and 3 is given by

$$ 0.53 \; < \; \beta r \; < \; 0.8 \eqno (5.119) $$

\n Since the equilibrium condition given by \e{5.117} requires $\beta r\ge 1$ for
case 3 (minimum with downflow) there is no possibility for overstability in 
this case if the flux tube is isothermal. Consequently, {\it the minimum region
of a symmetric loop formed by a flux tube which is in temperature equilibrium
with the external gas is always stable with respect to growing oscillations and
a maximum with a downflow is the only configuration which can lead to
overstability of an isothermal loop.} For $\Rn\to\infty$ in all cases we 
have $\beta r\to 1$ and we thus recover the result of the preceding section,
namely that an isothermal, horizontal flux tube does not show overstability.

Another interesting special case is the {\it neutrally buoyant flux tube}, 
$\ro = \roe$, whose equilibrium is determined by a balance between curvature
force and drag force. \e{5.107} then leads to (non-dimensional form)

$$ {2\over\Rn} \sgs -\brm \eqno (5.120) $$

\n and it is clear that this kind of equilibrium requires $m=-1$, \ie either
a minimum with a downflow or a maximum with an upflow. Inserting \e{5.120}
into \e{5.113} with $m=-1$ we now obtain

$$ \eqalignno {
   (\beta r)_1 \; &= \; {\pm 5 - 1\over 3\gamma} &
   \cr&&\cr
   (\beta r)_2 \; &= \; {\pm 5 + 1\over 3\gamma} \,.
   & \cr
   }
$$

\n Consequently, the minimum region of a neutrally buoyant flux loop is always
stable with respect to growing oscillations while a maximum may show
overstability. For a maximum loop and $\gamma=5/3$ we find $(\beta r)_1=0.8$ and 
$(\beta r)_2=1.2$. Consequently, overstability of a neutrally buoyant loop can 
occur if it has a maximum with a radius of curvature of about 2 scale heights. 

\b
\n {\bf 5.7.2 Constant vertical displacement}
\m

\n If we assume that the perturbation wavelength along the equilibrium
flux tube is infinite and consider purely
vertical displacements (in $z$-direction) the perturbation equations can be
simplified considerably and are analytically solvable in the case
of symmetric loops. In this way we can investigate {\it monotonic instability.}
As in the preceding section we consider an extremum
point (local minimum or maximum) of a flux tube whose path has the form of a
symmetric loop. We assume that the whole structure is displaced
vertically by a constant amount, $z_1$. Since the tube is not stretched
by this operation we have $l = l_0$ and conclude from \e {5.8} that

$$ \eta' \sgs {\eps \over R_0}\,.   \eqno (5.121) $$

\n At the point of extremum we have for reasons of symmetry

$$ \eta \sgs \eta'' \sgs \eps' \sgs 0 \eqno (5.122) $$

\n and since the radius of curvature is unaffected by this kind of displacement
$(R = R_0)$ we find from \e{5.11}

$$  \eps'' \sgs - {\eps\over\rq}\,. \eqno (5.123) $$ 

\n Furthermore, we have $\gul=0$ and $\gun=\pm 1$ at an extremum point. 
Using this together
with \es{5.121-123} we find that both sides of \e{5.63} vanish and 
\e{5.64} has non-trivial solutions provided that

$$ \omega^2 \sms i\omega {2 \beta r\over v_0} \sps
        \left[ \beta\delta + {1\over\gamma} \left ( 1 - {2\over\gamma} \right)
             \pm {2\over\gamma\Rn} 
   + \beta r j \left( {2\over v_0} \dz{v_0} - {1\over 2\gamma} \right) \right]
   \sgs 0  \eqno (5.124) $$

\n where (as before) the upper ($+$) sign applies for a maximum 
and the lower ($-$) sign for a minimum. In this equation we have introduced 
$j\equiv\sgn{\vb_{e0}\cdot\zu}=\mp m$ such 
that for both cases (minimum and maximum) $j=+1$ indicates an upflow and $j=-1$
a downflow. The quantities have been non-dimensionalized with respect to their
values at the point of extremum. As in the preceding sections we have 
assumed a constant gravitational acceleration ($s=0$) in order to simplify
the discussion. The influence of a variation of \gb\ on the stability
properties is marginal, however. Note that in the case of a horizontal tube
 $(\Rn\to\infty)$ \e{5.124} transforms into \e{5.84}, the dispersion relation 
for vertical displacements of a horizontal tube.\fussn{*}
{In the derivation of \e{5.84} we have assumed $\gun=-1$ such that we have
$j=m$ in this case.}
The same line of arguments as in that case shows that a necessary and sufficient 
criterion for instability is that the expression in square brackets 
in \e{5.124} is positive, viz.

$$ \beta\delta + {1\over\gamma} \left ( 1 - {2\over\gamma} \right)
             \pm {2\over\gamma\Rn} 
   + \beta r j \left( {2\over v_0} \dz{v_0} - {1\over 2\gamma} \right) 
   \; > \; 0\,. \eqno (5.125) $$

\n The effect of the terms which depend on the external velocity is identical
to the case of a horizontal tube, \ie the curvature of the tube does not
change the effect of the flow on the stability of the point of extremum.
This is not obvious from the original equations (5.63/64) since we find
terms there which depend on both velocity {\it and\/} curvature. In the special
case of purely vertical displacements these terms cancel, however. 

For the heuristic argumentation presented in the following subsection it is 
important to note here that the stability criterion given by \e{5.125} can be also
obtained by considering the perturbations of the buoyancy, curvature and
drag forces brought about by a vertical displacement of the point of extremum.
The sign of the resulting perturbation
of the total force then determines the stability properties of the equilibrium.
In the present case we assume that a purely vertical displacement does
not lead to a flow of matter along the tube $(\eta\equiv 0)$ 
such that the ratio $B/\rho$
is constant and the perturbations (indicated by an index {\it 1}) are related
to the equilibrium quantities as $B_1/B_0=\rho_1/\ro$. Together with the
adiabaticity of the perturbations (Eq.\ts{5.38}) and the condition
of pressure equilibrium (Eq.\ts{5.39}) this leads to the following relation
for the magnetic field perturbation as function of the displacement $z_1$:

$$ {B_1 \over B_0} \sgs 
   - \left( {\beta + 1 \over \beta\gamma + 2} \right) \delzh \,. \eqno (5.126) $$

\n Consequently, for $\beta\gg 1$ (ignoring terms of order $\beta^{-1}$) we
find for the perturbation of the curvature force (with $R_1 = 0$):

$$ F_{C1} \sgs \mp {B_0 B_1 \over 2\pi\Rn} \sgs 
               \pm {\bq z_1\over 2\pi\gamma\Rn\hp}\,. \eqno (5.127) $$

\n In a similar way we determine the perturbations of the buoyancy and
drag forces as functions of $z_1$ and the equilibrium quantities. We add
all force perturbations together, take the limit $\beta\gg 1$, and 
non-dimensionalize with respect to the values of the quantities at the
point of extremum. Taking then $z_1 \propto \exp(i\omega t)$ we obtain 
{\it exactly} the stability criterion given by \es{5.124/125}. This result lends 
some support to the treatment presented in the subsequent section where a
similar heuristic approach is used for displacements with large but finite
wavelength. 

The perturbation of the curvature force given by \e{5.127} is helpful for
understanding the effect of curvature on the stability of the loop which is
expressed by the term $\pm 2/\gamma\Rn$ in \e{5.124}. Since the magnetic field
decreases for an upward displacement and increases for a downward 
displacement (\cf Eq.\ts{5.126}), the absolute value of the curvature force
always decreases for an upward displacement and increases for a downward
displacement. Now consider a {\it minimum}: The curvature force is directed
upward (positive) and is reduced by an upward displacement and increased
by a downward displacement; the result is a restoring force which acts against
the displacement -- this stabilizing effect of curvature shows up
in the negative sign of the term $2/\gamma\Rn$ in the case of a minimum.
An analogous consideration shows that in the case of a {\it maximum} the
effect of curvature is destabilizing. Tab.\ts6 below summarizes the
effect of the different terms in \e{5.125} on the stability properties
of the loop. Regarding the stratification of the convection zone we
assume that the flow speed decreases with depth.

$$  
\offinterlineskip \tabskip=0pt 
\vbox{ 
 \halign to 1.0\hsize 
  { \strut \vrule # \tabskip=0pt plus 100pt 
                  & \hfil # \hfil & \vrule#& \enskip # \hfil &
                  \vrule#& \quad # \hfil & \vrule#& \qquad # \hfill & 
                  \vrule#& \quad # \hfil \tabskip=0pt&
                  \vrule# \cr
           \noalign{\hrule}{\vrule height 11.5pt width 0pt}
                           {\vrule depth   6.5pt width 0pt}
 & Extremum  &&  Flow  && \hfil $\pm 2/(\gamma\Rn)$ \hfil && 
   \hfil $(2/v_0)(dv_0/dz)>0$ \hfil && \hfil $-1/(2\gamma)<0$ \hfil  &\cr
           \noalign{\hrule}{\vrule height 11.5pt width 0pt}
                           {\vrule depth   8.5pt width 0pt}
 & minimum && up && stabilizing && destabilizing && stabilizing & \cr
                           {\vrule depth   8.5pt width 0pt}
 & maximum && up && destabilizing && destabilizing && stabilizing & \cr
                           {\vrule depth   8.5pt width 0pt}
 & minimum && down && stabilizing && stabilizing && destabilizing & \cr
                             {\vrule depth 6.5pt width 0pt}
 & maximum && down && destabilizing && stabilizing && destabilizing & \cr
      \noalign{\hrule}
}}$$
  
\tenpoint
\baselineskip = 10pt
  
\n {\bf Tab.\ts{6}:} Influence of individual terms in the
stability criterion (Eq.\ts5.125) for various configurations of a symmetric 
loop in the case of purely vertical displacements. The term given in the third
column describes the direct effect of curvature while the terms in the
fourth and fifth column represent the effect of the external velocity.

\vglue 0.5cm
\elevenpoint

\n According to \e{5.91} we expect that the velocity gradient term dominates
over the last term in \e{5.125} such that in particular minimum loops with
a downflow are possible examples of stable configurations.

In order to estimate the quantitative effect of the curvature term
we consider the case of a flux tube in thermal equilibrium with the
environment, \ie $T_0 = T_{e0}$ and $\beta (\roe/\ro-1) = 1$, such that 
the equilibrium condition, \e{5.107}, can be written in non-dimensional form as

$$ 1 \, + \, \brj \sgs \pm {2\over\Rn}\,. \eqno (5.128) $$

\n We see that for a minimum (lower sign) we must have a downflow $(j=-1)$ 
of sufficient strength since in temperature equilibrium the internal density
is always smaller than the external density leading to an upward directed
buoyancy force. Inserting \e{5.128} into the instability criterion \e{5.125}
and assuming a constant flow velocity ($dv_0/dz=0$) we find instability for

$$ \beta\delta \sps {2\over\gamma} \left ( 1 - {1\over\gamma} \right)
   \sps {\beta r j \over 2\gamma}  \; > \; 0\,. \eqno (5.129) $$

\n The fact that in this case a downflow always stabilizes while an upflow
destabilizes can be understood by considering \e{5.128}: With the exception 
of the small range
$-1\le\brj\le 0$ a downflow leads to a minimum and an upflow to a
a maximum. In both cases the effect of the flow is to decrease $R_0$, the radius
of curvature. Thus the stabilizing effects of curvature for a minimum (downflow)
and its destabilizing effects for a maximum (upflow) are amplified by the flow.

In the absence of a flow we only can have a maximum and
the criterion for instability reads

$$ \beta\delta \; > \;-\, {2\over\gamma} \left ( 1 - {1\over\gamma} \right)
               \sgs -0.48 \eqno(5.130) $$

\n for $\gamma=5/3$. If we compare this with the criterion for a horizontal
tube without flow, \ie $\beta\delta > +0.12$ (\cf Eq.\ts{5.89}) 
we see that the effect of
curvature is quite significant: While for the horizontal tube a positive
superadiabaticity was necessary in the case of purely vertical perturbations, 
the static, isothermal loop is unstable even in a moderately 
subadiabatic environment. On the other hand, an external downflow of 
sufficient amplitude leads to stable
local minimum configurations for any value of the superadiabaticity. Take
for example the value $\beta\delta=3.6$ for equipartition flux tubes in the
deep solar convection zone. With $j=-1$ we find from \e{5.129} that such
an equilibrium is stable if $\beta r > 13.6$. From \e{5.105} we see that
this condition is realized for flux tube radii smaller than about 
$\hp/13.6\approx 4000$ km if the external velocity is of the order of the 
typical convective velocities obtained from mixing length theory.

\b
\n {\bf 5.7.3 Heuristic approach for perturbations with large wavelength}
\m

\n As far as monotonic instability is concerned, it has already been shown
by Spruit and van Ballegooijen (1982) that for horizontal flux tubes without 
external flow the most unstable perturbations are those which lead to wave-like 
displacements of very large, {\it but finite,} wavelength. In this case a 
Parker-type/convective  instability with a flow along the tube can be excited
which is largely undisturbed by curvature forces. We have generalized this result
to horizontal tubes with external flow and we suspect that this kind of 
perturbation is also decisive for the monotonic instability of symmetric loops 
for which we until now have only considered very small and infinite wavelength
(subsections 5.7.1 and 5.7.2, respectively).
Unfortunately, finite wavelength perturbations for non-horizontal flux tube
equilibria in most cases do not permit analytical treatment since
the coefficients in the perturbation equations, \es{5.63/64}, become variable.
One then has to resort to a numerical solution of the eigenvalue problem 
for any given equilibrium tube. However, for symmetric loops and 
displacements with very large, but finite, wavelength which turned out to be most
unstable kind of perturbation for horizontal tubes we can extend the heuristic
approach sketched in the preceding subsection. Since in contrast to the treatment
there we cannot check against the exact result,
no definite proof can be given that the approach described below is correct.
However, we will show that in the limit of horizontal tubes ($R_0\to\infty$) 
the result becomes identical to the exact criterion for monotonic instability
and $k\to 0$ (but finite, \cf Eqs.\ts{5.98/99}). Together with the success
of the method for purely vertical displacements demonstrated in the preceding 
subsection this gives some confidence in its validity. 

We consider a symmetric loop in a vertical flow and determine the perturbations 
of the various forces (buoyancy, curvature, drag) at the point of extremum which
are brought about by a vertical displacement. In contrast to the previous
treatment we now assume a perturbation with finite
wavelength such that the flux tube is not displaced as a whole --
parts of the equilibrium tube are lifted while other parts are displaced
downward.
Since in our cartesian model the wavelength of the perturbation can be made
arbitrarily large, it can in particular be chosen large enough such that
the perturbations of the tube geometry (arc length, radius of curvature etc.)
can be neglected in comparison with the relative perturbations of the other 
quantities.
For example, for a displacement $z_1$ with wavenumber $k$, the perturbation 
of the radius of curvature at the extremum point of a symmetric loop
is given by

$$ \vert R_1 \vert \; = \; k^2 \rq \vert z_1 \vert  \eqno (5.131) $$

\n where we have used \es{5.11/14} and the symmetry properties. We see that
for any given displacement
we can make $R_1$ as small as we want by decreasing $k$ sufficiently, \ie by
increasing the wavelength of the perturbation. 

The important difference to the treatment in the preceding subsection lies
in the determination of the perturbations of internal density and pressure. Since
the wavelength of the displacement is now finite (albeit very large) 
parts of the equilibrium tube are lifted while other parts are displaced
downward such that an internal flow along the tube sets in which tends to 
establish hydrostatic equilibrium along the magnetic field lines 
according to the principle of communicating tubes. We therefore determine 
the perturbed internal density and pressure by assuming that
the flow along the tube has already restored hydrostatic equilibrium
at the point of extremum, viz.

$$ p_1 \sgs - {p_0\over\hp} z_1 \eqno (5.132) $$

\n which entails for adiabatic perturbations (\cf Eq.\ts{5.38})

$$ \rho_1 \sgs - {\ro\over\gamma\hp} z_1 \,. \eqno (5.133) $$

\n By assuming that hydrostatic equilibrium is reestablished we 
probably loose information about growth rates and we also
cannot reproduce the overstable modes but we conjecture that the stability
criteria for monotonic instabilities are correctly described by this approach.
We shall prove this below for the special case of horizontal tubes
for which the exact solution is available. 

We continue by determining $F_{B1}$, the perturbation of the buoyancy force,
which is given by

$$ F_{B1} \sgs ( \rho_{e1} - \rho_1 ) g_0 \eqno (5.134) $$

\n where we have assumed that the gravitational acceleration is constant ($s=0$).
With $\nabla_{ad} = (\gamma - 1)/\gamma$, $\delta = \nabla - \nabla_{ad}$  
and using \es{5.25} and (5.133) we find, after some algebra

$$ F_{B1} \sgs {\bq\over 8\pi\hp^{\,2}} \left\{ (1 - \nabla )
          \left[ \betro \, + \, {\roe\over\ro}
               {\hp\over\hpe}\,\,\beta \left( {\hpe\over\hp} - 1 \right) \right]
          \, + \, \beta\delta \right\} z_1\,. \eqno (5.135) $$

\n We now perform the same procedure as in Sec.\ts{5.4} and take
the limit $\beta\gg 1$ such that all terms of order $\beta^{-1}$ are
neglected unless they are multiplied by $\beta$. In this limit we have

$$ \beta \left( {\hpe\over\hp} - 1 \right) = 1 \sps \betro \sps O(\beta^{-1}) 
   \eqno (5.136) $$

\n as well as $1-\nabla = 1/\gamma + O(\beta^{-1})$, 
$\roe/\ro = 1 + O(\beta^{-1})$,
$\hp/\hpe = 1 + O(\beta^{-1})$, and \e{5.135} is transformed into

$$ F_{B1} \sgs {\bq\over 8\pi\hp^{\,2}} \left[ \beta\delta \, + \, {1\over\gamma}
          \, + \, \gamiz \, \betro \right] z_1\,. \eqno (5.137) $$

\n The perturbation of the magnetic field is determined by the
condition of pressure balance, \e{5.39}, which yields together with \es{5.25} 
and (5.132/233)

$$ {B_1\over B_0} \sgs \betro {z_1\over 2\hp}\,. \eqno (5.138) $$

\n Inserting into the first (general) part of \e{5.127} we obtain the 
perturbation of the curvature force

$$ F_{C1} \sgs \mp {\bq\over 4\pi\hp\Rn}\,\, \betro z_1\,. \eqno (5.139) $$

\n Note that we have assumed a sufficiently large wavelength of the 
displacement such
that the perturbation of the radius of curvature can be neglected.
Finally we determine $F_{D1}$, the perturbation of the drag force. As in the
preceding sections we assume a purely vertical external velocity field
and continue to use the notation introduced in \es{5.83} and (5.124).
Since our present approach cannot adequately describe impulsive motion
of the flux tube we omit the contribution to $F_{D1}$ due to the motion
of the tube itself which gives rise to the last term in \e{5.64}. We
have discussed at some length in Sec.\ts{5.6} that this term does only
affect the growth rates but {\it not} the stability criteria. A derivation along
the lines of \es{5.27-5.33} applied to the special case discussed here
yields

$$ F_{D1} \sgs {C_D\roe v_0^{\,2} j\over\pi a_0}
      \left( {v_{\perp 1} \over v_0^{\,2}} \, + \, {\rho_{e1}\over\roe} \, +
      \, {B_1\over 2B_0} \right). \eqno (5.140) $$

\n Using \es{5.25}, (5.90) and (5.138) and taking 
the limit $\beta\gg 1$ we obtain, after some algebra

$$ F_{D1} \sgs {\bq\over 8\pi\hp^{\,2}} \, \brj \left[ {2\hp\over H_{v0}} 
    \, - \, {1\over\gamma} \, + \, {1\over 4} \, \betro \right] z_1 
    \eqno (5.141) $$

\n where $H_{v0}\equiv v_0(dv_0/dz)^{-1}$ denotes the scale height 
of the external velocity.
The total force perturbation is obtained by adding \es{5.137}, (5.139),
(5.141), and using the equilibrium condition, \e{5.4}, which gives

$$ \eqalignno {
       F_{B1} \sbs F_{C1} \sbs F_{D1} \sgs &{\bq\over 8\pi\hp^{\,2}}
         \left[ \beta\delta \, + \, {1\over\gamma}  \, + \,
         {4\hp\over\Rn} \left( {\hp\over\Rn} \mp {1\over\gamma} \right)
         \, + \, {(\beta r)^2\over 4} \right.
         & \cr && \cr
         &\qquad \left. + \, \brj \left( {2\hp\over H_{v0}} \sbs {1\over\gamma} 
         \, \pm \, {5\over 2}{\hp\over\Rn} \right) \right] z_1 .
         & (5.142) \cr && \cr 
              } $$

\n The equilibrium is unstable if the expression within square brackets
is positive which means that the perturbation of total force has the same
sign as the displacement and thus tends to increase the latter.
The exact result for a horizontal tube obtained in Sec.\ts{5.6} is recovered
by taking $R_0 \to \infty$ which yields as condition for monotonic instability

$$       \beta\delta \, + \, {1\over\gamma}  \, + \,
         {(\beta r)^2\over 4} \, + \,
         \brj \left( {2\hp\over H_{v0}} + {1\over\gamma} 
         \right) \; > \; 0 \,. \eqno (5.143) $$

\n Regarding \e{5.90} we find that this is {\it identical} to \e{5.99},\fussn{*}
{Since we have taken $\gun = -1$ in Sec.\ts{5.6} we have $j = m$.}
the exact criterion in the limit of very large wavelength.
For $\beta r = 0$ the result of Spruit and van Ballegooijen (1982) is
recovered. For finite radius of curvature the general criterion for monotonic
instability is

$$       \beta\delta \, + \, {1\over\gamma}  \, + \,
         {4\hp\over\Rn} \left( {\hp\over\Rn} \mp {1\over\gamma} \right)
         \, + \, {(\beta r)^2\over 4}
         \, + \, \brj \left( {2\hp\over H_{v0}} + {1\over\gamma} 
         \pm {5\over 2}{\hp\over\Rn} \right) \; > \; 0 . \eqno (5.144) $$

\n This could easily be transformed to a non-dimensional form by formally
taking $\hp\equiv 1$ but we do not change the notation here. 
Let us now discuss the influence of a finite curvature of the equilibrium
flux tube on its stability which is expressed in the terms containing the
ratio $\hp / \Rn$. The third term on the \lhs of \e{5.144} is always
positive and therefore destabilizing for a minimum (lower sign) while for
a maximum (upper sign) it is stabilizing if $\hp/\Rn < 1/\gamma$. However,
since $\gamma\ge 1$ 
it is easy to show that the sum of the second and the third term on the \lhs
of \e{5.144} is always positive, i.e. 

$$ {1\over\gamma}  \sps {4\hp\over\Rn}
     \left( {\hp\over\Rn} \mp {1\over\gamma} \right) \; > \; 0 \eqno (5.145) $$

\n such that {\it without external velocity curvature cannot stabilize a
flux tube in a superadiabatic environment\/} ($\beta\delta > 0$). 
For a minimum the effect of curvature always is destabilizing, in contrast
to the case of a constant vertical displacement (cf. Eq.\ts{5.125}). This
is caused by the perturbation of the curvature force. If we take
$\beta r = 0$ the equilibrium condition (Eq.\ts{5.4}) for a minimum reads

$$ \betro \sgs \mp {2\hp\over\Rn}  \eqno (5.146) $$

\n and we find from \e{5.139}

$$ F_{C1} \sgs +{\bq z_1 \over 2\pi\rq} \eqno (5.147) $$

\n such that the perturbation of the curvature always tends to increase
the displacement. 

We summarize
the influence of the velocity terms in \e{5.144} on the stability properties
of a loop in Tab.\ts7 below, assuming that the flow speed decreases
with depth ($\hv > 0$).

$$  
\offinterlineskip \tabskip=0pt 
\vbox{ 
 \halign to 1.0\hsize 
  { \strut \vrule # \tabskip=0pt plus 100pt 
                  & \hfil # \hfil & \vrule#& \enskip # \hfil &
                  \vrule#& \quad # \hfil &
                  \vrule#& \quad # \hfil \tabskip=0pt&
                  \vrule# \cr
           \noalign{\hrule}{\vrule height 11.5pt width 0pt}
                           {\vrule depth   6.5pt width 0pt}
 & Extremum  &&  Flow  && 
   \hfil $2\hp/\hv+1/\gamma>0$ \hfil && \hfil $\pm (5/2)(\hp/\hv)$ \hfil  &\cr
           \noalign{\hrule}{\vrule height 11.5pt width 0pt}
                           {\vrule depth   8.5pt width 0pt}
 & minimum && up && destabilizing && stabilizing & \cr
                           {\vrule depth   8.5pt width 0pt}
 & maximum && up && destabilizing && destabilizing & \cr
                           {\vrule depth   8.5pt width 0pt}
 & minimum && down && stabilizing && destabilizing & \cr
                             {\vrule depth 6.5pt width 0pt}
 & maximum && down && stabilizing && stabilizing & \cr
      \noalign{\hrule}
}}$$

\tenpoint
\baselineskip = 10pt
  
\n {\bf Tab.\ts{7}:} Influence of velocity-related terms in the
criterion for monotonic instability (Eq.\ts5.144) for various configurations 
of a symmetric loop in the case of displacements with large wavelength. 
The fourth term in the criterion, $(\beta r)^2/4$, is always positive and
destabilizing.

\vglue 0.5cm
\elevenpoint

\n The sum of the second, third and fourth term in \e{5.144} is always
positive (destabilizing) such that in a superadiabatic environment
a maximum loop with an upflow cannot be stabilized by a flow with $\hv > 0$.
The other three configurations can be stabilized provided that certain conditions
are fulfilled. 

As example let us consider the {\it isothermal\/} case, $T_0 = T_{e0}$,
for which the equilibrium is determined by \e{5.128}. If we insert this
condition into the criterion given by \e{5.144} we find

$$ \beta\delta \sbs 1 \, - \, {1\over\gamma} \sbs {5\over 2}(\beta r)^2 \sbs
   \brj \left( {2\hp\over H_{v0}} \, - \, {1\over\gamma} \sbs {13\over 4}\right ) 
   \; > \; 0 \,. \eqno (5.148) $$

\n In the static case ($\beta r = 0$) which leads to a loop with a maximum 
(\cf Eq.\ts{5.128}) we have

$$ \beta\delta \; > \; \gami \sms 1 \sgs -0.4 \eqno (5.149) $$

\n for $\gamma = 5/3$ (this value will also be used in the following numerical 
examples). Consequently, a static isothermal loop is unstable
in both superadiabatic and slightly subadiabatic regions. If we include the
velocity terms and assume a constant velocity ($H_{v0}\to\infty$) we find,
similar to the case of displacements with infinite wavelength (Eq.\ts{5.129}), 
that an upflow ($j=+1$) always destabilizes while a downflow ($j=-1$) may exert a
stabilizing influence provided that

$$ {5\over 2}(\beta r)^2 \sms \beta r \left( {13\over 4} - \gami \right)
   \; < \; 0 \eqno (5.150) $$

\n which means

$$  0 \; < \; \beta r \; < \; 1.06 \eqno (5.151) $$

\n For $\beta r>1$ the loop form changes from a maximum to a minimum.
The smallest value that the expression on the \lhs of \e{5.150} can reach
is $-0.7$ (for $\beta r = 0.53$) which leads to the criterion
$ \beta\delta>-0.4+0.7=0.3$. Although some stabilization has
been achieved, a constant downflow cannot stabilize an isothermal equipartition
loop in the deep convection zone of the Sun where we have $\beta\delta = 3.6$.

In which way does a velocity gradient affect the stability of an isothermal loop?
From \e{5.148} we see that a velocity which increases with height ($H_{v0} > 0$)
has a stabilizing effect in the case of a downflow and
a destabilizing effect for an upflow (and vice versa). In the case of a
flow with constant mass flux density we have seen in Sec.\ts{5.6}
that $H_{v0} \approx \hp$ such that we find the criterion

$$ \beta\delta \, + \, 0.4 \,+\, {5\over 2}(\beta r)^2
               \, + \, 4.65 (\brj) \; > \; 0\,. \eqno (5.152) $$

\n Consequently, a downflow has a stabilizing effect for $0 < \beta r < 1.86$. 
The velocity terms 
attain their most negative value for $\beta r = 0.93$ which gives
as condition for instability

$$ \beta\delta \;>\; 1.76 \eqno (5.153) $$

\n Thus an equipartition tube with $\beta\delta = 3.6$ is still unstable. 
In principle we may expect larger velocity gradients
for convective downflows near the bottom of the convection zone or within an
overshoot region where the flows are strongly decelerated
due to the strong subadiabaticity of the radiative region below. Take for example
a large equipartition flux tube ($a_0 = 10^4$ km) near the bottom of the 
solar convection zone ($\hp = 5\cdot 10^4$ km, $v_0 = v_{A0} = 100$ m/s
which gives $\beta r = 2.5$, $\beta\delta = 3.6$). 
We find from \e{5.148} that a minimum loop formed by such a tube
is stable in a downflow provided that
$H_{v0} / \hp < 0.4$ or $H_{v0} < 2\cdot 10^4$ km, a value which does not
appear unrealistic. On the other hand, a smaller tube with $a_0 = 10^3$ km 
and $\beta r = 25$ already requires a value of $H_{v0} < 160$ km in
order to be stabilized which is clearly unrealistic. We conclude that minimum
loops formed by relatively large, isothermal flux tubes can possibly be 
stabilized by a strongly decelerating downflow near the bottom of 
the solar convection zone. 

The considerations above were for the isothermal case, \ie a flux tube which
is in temperature equilibrium with its environment. Although there is a
natural tendency towards this state due to radiative energy
exchange the relevant time scale becomes very large in the deep layers of
a convection zone.
If a loop has evolved out of an initially horizontal tube and hydrostatic 
equilibrium along the field lines has been established adiabatically, 
a temperature difference with respect to a superadiabatic environment 
is the consequence. A loop with a minimum would be somewhat cooler and
a loop with a local maximum somewhat hotter than its surroundings. In order
to assess the influence of such a temperature difference on the stability of
a loop we define the parameter

$$ \alpha \; \equiv \; \beta \left( {\roe\over\ro} - 1 \right) \eqno (5.154) $$

\n which expresses the relation between external and internal
temperature: If we assume that the mean molecular weight is the same inside and
outside the tube we find from \e{5.136}:

$$ \beta \left( {\roe\over\ro} - 1 \right) \sgs 1 \sms
   \beta \left( {T_{e0}\over T_0} - 1 \right) \eqno (5.155) $$

\n where $T_{e0}$ and $T_0$ denote external and internal temperature,
respectively. Consequently, temperature equilibrium entails $\alpha = 1$ 
while for a cooler interior we find $\alpha<1$ and for a hotter interior
we have $\alpha>1$. If the temperature difference $\Delta T\equiv T_0-T_{e0}$
is small compared to the external temperature we can approximate \e{5.155} 
as

$$  \alpha \; \approx \; 1 \sps \beta \, {\Delta T\over T_{e0}} \eqno (5.156) $$

\n and for equipartition flux tubes near the bottom of the solar convection 
zone ($T_{e0} = 2\cdot 10^6$ K, $\beta = 10^6$) we have 

$$ \alpha \; \approx \; 1 \sps {\Delta T\over 2} \eqno (5.157) $$

\n where $\Delta T$ is assumed to be given in degrees Kelvin.
Using \e{5.154} the equilibrium condition (Eq.\ts{5.4}) at the
point of extremum is written as

$$ \alpha \, + \, \brj \sgs \pm {2\hp\over\Rn}\,. \eqno (5.158) $$

\n If we insert \e{5.158} into \e{5.144} we obtain the following condition
for instability:

$$ \beta\delta \sbs \alpha^2 \sbs \gami (1-2\alpha) \sbs {5\over 2}(\beta r)^2
\sbs \brj \left( {2\hp\over H_{v0}} - {1\over\gamma} + {13\over 4}\alpha \right )
   \; > \; 0 \,. \eqno (5.159) $$

\n For given values of 
$\beta\delta$ and $\brj$ we can determine a range of values
of $\alpha$ for which the \lhs of \e{5.159} is {\it negative\/} describing
a stable loop. For the case $H_{v0}=\hp$ and $\gamma=5/3$ this is achieved 
by solving the quadratic inequality

$$ \alpha^2 \sps \alpha P \sps Q \; < \; 0  \eqno (5.160) $$

\n where 

$$ \eqalign {
       &P \sgs {13\over 4}(\brj) \, - \, {6\over 5} \cr 
       &Q \sgs \beta\delta \sbs {5\over 2}(\beta r)^2 
          \sbs {7\over 5}(\brj) \sbs {3\over 5}  \cr
            } $$

\n This inequality has a range of real solutions provided that 
$W\equiv P^2/4 - Q \ge 0$ which leads to the condition

$$ {9\over 64}(\beta r)^2 \,- \, {67\over 20}(\brj) \, - \,  {6\over 25} 
   \, - \, \beta\delta  \; \ge \; 0\,. \eqno (5.161) $$

\n For an equipartition flux tube in an external downflow ($j=-1$)  
within the deep layers of the convection zone ($\beta\delta = 3.6$)
we find that \e{5.161} is satisfied for $\beta r > 1.1$. For example,
if we take $\brj = -2$, \e{5.160} is fulfilled for $2. < \alpha < 5.7$, \ie
if the tube is slightly hotter than its environment. The equilibrium
condition (Eq.\ts{5.158}) shows that this case refers to a loop with
a maximum. Minima can be stabilized by an upflow ($j=+1$) in which case
\e{5.161} is satisfied for $\beta r > 24.9$. For example, if we take
$\beta r j = 25.$ we find that \e{5.160} is fulfilled for 
$-40.57 < \alpha < -39.48$, \ie a cool loop which forms a minimum.
Generally we can conclude from \es{5.160/161} and (5.158) that in the 
case $H_{v0}=\hp$ and $\beta\delta=3.6$ only {\it cool
minima in an upflow and hot maxima in a downflow represent a stable
configuration\/} provided that the temperature difference and the flow velocity
correspond to the relationship expressed by the above inequalities.

From the bottom to the middle parts of the solar convection zone the
integrated temperature difference between adiabatic and actual stratification
amounts to only $\Delta T\approx 1$K (Parker, 1987) such that \e{5.157} gives
rather small values for the parameter $\alpha$. Consequently, the estimates
discussed above suggest that unless more efficient cooling or heating takes place
(\eg if a loop sinks down from the top layers of the convection zone) 
thermal effects in conjunction with drag forces cannot effectively stabilize  
flux tubes with or without loops in a stellar convection zone.

\b\b

\def\etopline{ 5.8 Summary of the stability properties}

\n {\bbf 5.8 Summary of the stability properties}

\b

\n The formalism derived in Secs.\ts{5.3/4} provides a tool which can
be used to investigate the stability of a wide class of
flux tube  equilibrium structures. In most cases, however, a numerical
treatment of the resulting eigenvalue problem is necessary. Such an undertaking
is intended for the future but outside the scope of the work presented here. 
However, we have been able to determine analytically the stability properties in 
a number of cases which are not without general relevance. We have considered 
in particular

\m
\item{-} horizontal flux tubes with purely vertical and wave-like displacements
         (Sec.\ts{5.6}),
\s
\item{-} symmetric loops with perturbations of small wavelength
         (Sec.\ts{5.7.1}),
\s
\item{-} symmetric loops with purely vertical displacements (infinite wavelength)
         (Sec.\ts{5.7.2}), and
\s  
\item{-} symmetric loops with displacements of large wavelength  
         (Sec.\ts{5.7.3}).
\m

\n For {\it horizontal flux tubes\/} we have 
generalized the results of Spruit and van Ballegooijen (1982) to the case
of a vertical external flow. The monotonic instability found by these 
authors can be stabilized if the flow speed has a large gradient,
for example by a downflow whose velocity strongly decreases with depth.
If perturbations with finite wavelength along the flux tube are considered
the value of the quantity $\beta r$ has to be within a specific range
for this stabilization to become effective. Irrespective of the direction
or gradient of the flow, sufficiently large values of $\beta r$ (due
to small field strength, small radius, or large velocity) provoke monotonic 
instability caused by the radius change of the tube during its displacement.
An estimate based on the properties of the deep layers of the solar convection
zone shows that the effect of a downflow with constant mass flux
density is insufficient to prevent isothermal equipartition flux tubes
from monotonic instability due to the superadiabatic stratification.

The monotonic mode has been investigated also for the case of a {\it symmetric
loop\/} structure with a horizontal tangent vector at the point of extremum 
(maximum or minimum). We have derived the
exact solution for constant vertical displacements (infinite longitudinal
wavelength) and used it as a guideline and test for a heuristic approach
which allowed to treat also the case of large (but finite) wavelength. This
represents the most unstable perturbation for horizontal tubes and it turned out
that this is true also for symmetric loop structures.
In the absence of an external flow, all loops (maximum or minimum) 
are monotonically unstable in a superadiabatic or slightly subadiabatic
environment. A flow may exert a stabilizing influence: For an isothermal
tube, a downflow with values of $\beta r$ within a certain range stabilizes.
In particular, a strongly decelerating downflow leads to a stable 
minimum loop even if $\beta\delta=3.6$. If the tube is
non-isothermal, cool minima in an upflow or hot maxima in a downflow
may be stabilized in a superadiabatic region provided that the 
temperature difference and, again, the value of $\beta r$ are within specific 
intervals. As for horizontal tubes, sufficiently large values of $\beta r$
always lead to instability.

While monotonic instability preferentially evolves for displacements with
large (but finite) wavenumber in the longitudinal direction, another mode
of instability preferentially  appears for large wavenumbers, 
\ie {\it overstable transversal oscillations.} In the case of overstability, 
the drag force conspires with the magnetic tension 
force such that oscillations with growing amplitude result.
For a horizontal tube this requires a downflow and is restricted to
a certain interval of values for $\beta r$. The fact that overstability
occurs preferentially for large wavenumbers suggests a local analysis 
for non-horizontal equilibrium tubes. We have carried out such an analysis for 
symmetric loops and found that
from the four possible combinations of loop geometry (minimum/maximum) and
flow (up/down) overstability is excluded for the minimum loop with an 
upflow. For the other cases overstability may appear within specific intervals
for $\beta r$. In the case of an {\it isothermal\/} flux tube, however,
overstability is restricted to maximum loops with downflow while a {\it
neutrally buoyant\/} loop may only become overstable if it represents a
maximum with an upflow. 
Since the approximation of slender flux tubes demands that the
perturbation wavelength is large compared to the flux tube radius while 
overstability requires small wavelengths, the applicability of the present results
is restricted to tubes of sufficiently small radius. 
Apart from the effects discussed so far, the introduction of external flows
and drag forces leads to a decrease of the growth rates of
unstable perturbations which is most significant for oscillatory instability.

The excitation of the overstable mode depends only on the direction
of the flow and the value of $\beta r$; in particular, it cannot be stabilized 
by the stratification and therefore may appear also in convectively stable layers
like overshoot zones or regions in radiative equilibrium. Excited locally,
for instance in a loop formed by a downflow or an upflow, such oscillations
could propagate as transversal tube waves. Under the influence of rotationally
induced Coriolis forces these waves may even exhibit helicity and contribute
to the field-regeneration mechanism which is necessary for the operation of
a dynamo. In order to investigate this conjecture we have to extend the
present formalism by moving to spherical geometry and including
a (differential) rotation.

Convective flows in the deep parts of the solar convection zone have a 
typical time scale of the order of a month. Even a stable, stationary
flux tube configuration within the convection zone cannot be expected to
exist for a significantly longer time. This is in general accordance with
the lifetime of large active regions. We have found that 
stable flux tube equilibria within a superadiabatic region require 
fine-tuned relations between the various parameters which determine the
equilibrium such that they probably are not of great practical importance.
A realistic convection zone, of course, is much more complicated than can be
expressed by the simple analytical examples treated here. For example, the 
superadiabaticity probably shows 
significant spatial variations, be it between upflow and downflow regions
or related to differential rotation (Durney, 1989). Hence, the estimates given
in the preceding sections must not be taken too serious; however, they strongly
indicate that, in the long run, the unstable stratification of the 
convection zone itself cannot be overcome.

\vfill\eject

\def\otopline{ 6. Dynamics of flux tubes in a convection zone }

\vglue 2cm

\def\etopline{ 6.1 Size distribution }

\n {\bbbf 6. Dynamics of flux tubes in a convection zone }

\b\m

\n In the preceding chapters we have investigated in some detail certain
aspects of the structure and dynamics of concentrated fields.
We may have obtained some pieces of a yet unfinished jigsaw
puzzle in this way but we certainly are not in a position to present a
full theory. In the present chapter we shall nevertheless try to sketch
a tentative picture of the magnetic field dynamics in stellar convection zones,
based upon our own results and on the work of other researchers.
This picture is largely based on heuristic arguments, sometimes supported
by more solid results. It is not thought as a comprehensive model but more as
an orientation and stimulus for further work.

\b\b

\n {\bbf 6.1 Size distribution}

\b

\n We have already discussed in Ch.\ts{2} that the
magnetic Rayleigh-Taylor instability and other fragmentation processes
tend to produce magnetic structures with sizes of less than 100 km within
the convection zone. Fragmentation proceeds until the fragments are so small
that they merge by magnetic diffusion as fast as they are formed, \ie until
the diffusive time scale becomes equal to the growth time of the instability
considered. For the case of the interchange instability this minimum fragment
size, $d_i$, is given by \e{2.12} which we repeat here:

$$ d_i \sgs \left( {2\, R\,\eta^2\over v_A^{\,2}} \right)^{1/3} 
            \eqno (2.12)=(6.1)$$

\n ($R$: radius of curvature, $\eta$: magnetic diffusivity, $v_A$: Alfv\'en
velocity). Taking equipartition fields, \ie Alfv\'en velocity equal to $v_c$, the
typical velocity of convective flows, and $R$ equal to $L$, the length scale 
of the convective flows, we can determine $d_i$ for different depths in the 
solar convection zone using the model of Spruit (1977b). For this rough estimate
we use the depth as typical size of the dominant convective cell. 
The result is given in the following table:

$$   
\offinterlineskip \tabskip=0pt 
\vbox{ 
 \halign to 1.0\hsize 
  { \strut \vrule # \tabskip=0pt plus 100pt 
                  & \enskip # \hfil & \vrule#& \enskip # \hfil &
                  \vrule#& \enskip # \hfil & \vrule#& \enskip # \hfil & 
                  \vrule#& \enskip # \hfil & \vrule#& \enskip # \hfil &
                  \vrule#& \enskip # \hfil & \tabskip=0pt&
                  \vrule# \cr
           \noalign{\hrule}{\vrule height 11.5pt width 0pt}
                           {\vrule depth   6.5pt width 0pt}
 & Depth (cm) && $\eta$ (cm$^2\cdot$s$^{-1}$) && $v_c$ (cm$\cdot$s$^{-1}$) &&
   $d_i$ (cm) && $d_r$ (cm) && $\tau_r$ (s) && $\tau_s$ (s) &\cr
           \noalign{\hrule}{\vrule height 11.5pt width 0pt}
                           {\vrule depth   8.5pt width 0pt}
 & \dex{1.0}{8} && \dex{1.5}{6} && \dex{1.3}{5} && \dex{3.0}{3} && 
   \dex{3.4}{4} && \dex{7.7}{2} && \dex{3.0}{-1} & \cr
                           {\vrule depth   8.5pt width 0pt}
 & \dex{1.0}{9} && \dex{1.0}{5} && \dex{3.2}{4} && \dex{2.7}{3} && 
   \dex{5.6}{4} && \dex{3.1}{4} && \dex{9.1}{2}  & \cr
                           {\vrule depth   8.5pt width 0pt}
 & \dex{1.0}{10}&& \dex{7.8}{3} && \dex{9.1}{3} && \dex{2.5}{3} && 
   \dex{9.3}{4} && \dex{1.1}{6} && 2.1  & \cr
                             {\vrule depth 6.5pt width 0pt}
 & \dex{1.8}{10}&& \dex{7.5}{3} && \dex{4.9}{3} && \dex{4.4}{3} && 
   \dex{1.7}{5} && \dex{3.7}{6} && 2.3  & \cr
        \noalign{\hrule}
}}$$

\tenpoint
\baselineskip = 10pt

\n {\bf Tab.\ts8}: Properties of magnetic filaments as a function of depth
in the solar convection zone. $\eta$: molecular magnetic diffusivity;
$v_c$: convective velocity; $d_i$: diffusive scale for the interchange 
instability; $d_r$: resistive boundary layer thickness; $\tau_r=d_r^2/\eta$:
resistive diffusion time; $\tau_s$: radiative diffusion time on spatial scale 
$d_i$.

\vglue 0.5cm
\vfill\eject

\elevenpoint

\n The table also gives $d_r$, the thickness of the resistive boundary layer:

$$ d_r \sgs \left( {L\, \eta \over v_c} \right)^{1/2}. \eqno (6.2) $$

\n This quantity is determined by the balance of magnetic diffusion and
advection by convective flows and represents the scale of structures
formed by kinematical magnetic flux expulsion. The relevant time scale for
this process is the eddy turnover time, $L/v_c$, which is equal to the
resistive diffusion time on the spatial scale $d_r$, viz.

$$ \tau_r \sgs {d_r^2 \over \eta}. \eqno (6.3) $$

\n In the dynamical case \e{6.2} still gives a lower limit: Structures 
smaller than $d_r$ diffuse too rapidly to be held together by the convective 
cell and therefore cannot be sustained individually. On the other hand,
we see from Table 6.1 that always $d_i < d_r$ which means that the growth
time for the interchange instability for a structure of size $d_r$ is shorter
than its diffusion time, \ie the time scale of flux expulsion. Consequently,
even if flux expulsion produces structures with sizes larger than $d_r$,
these are actually bundles of smaller filaments with a typical size
$d_i$. It is the very influence of the collecting flow itself
which leads to fragmentation: It provokes interchanging by deforming the magnetic
structures as well as by exerting a destabilizing pressure gradient at the
interface between magnetic and non-magnetic gas (see the discussion in
Sch\"ussler, 1984b).

The last column in
Table 6.1 gives the radiative diffusion time, the time scale which determines
the heat exchange between a fragment of size $d_i$ and its environment,

$$ \tau_s \sgs {d_i^2 \over \eta_s}\, , \eqno (6.4) $$

\n with the radiative diffusivity $\eta_s$ given by

$$ \eta_s \sgs { 16\,\sigma\, T^3\over 3\, \rho^2 c_V \kappa_R} \eqno (6.5) $$

\n ($\sigma$: Stefan's radiation constant; $T$: temperature; $c_V$: specific
heat capacity; $\kappa_r$: Rosseland mean opacity).
For both resistive and thermal diffusion we have assumed that turbulence
on length scales smaller than the filament size which could give rise to turbulent
diffusivities is suppressed by the magnetic field. If the field is in
equipartition with flows on the dominant convective scale it will always
be stronger than the respective equipartition field on any other
scale of the convective/turbulent velocity field. 

We find from Table 6.1 that $\tau_s$ is very small compared to any relevant
dynamical time scale throughout the whole convection zone. This statement remains
valid if the larger spatial scale $d_r$ is used in \e{6.4}. Furthermore, since
we always have $d_i<d_r$, the diffusion time on the scale of the filaments, $d_i$,
is shorter than the dynamical time scale $L/v_c=\tau_r$. 
Consequently, for magnetic structures which have been
fragmented down to the resistive diffusion limit, $d_i < 100$~m, as well as for
structures of the size of the resistive boundary layer, $d_r \approx 1$ km, which
are formed by flux expulsion, two
statements can be made which become important in the following sections:

\medskip
\item{-} The magnetic diffusion time is small enough to allow {\it mass
         exchange\/} between magnetic structures and their surroundings
         within a dynamical time scale (\eg while being stretched 
         by differential rotation), and

\smallskip

\item{-} the thermal diffusion time is sufficiently small to ensure
         {\it temperature equilibrium\/}  between interior and exterior 
         of a filament during its dynamical evolution.

\medskip

\n As we have seen above, convective flows accumulate magnetic flux
and form larger structures which appear in the form of 
bundles of small filaments, not monolithic flux tubes. For spatial scales 
which are much larger than the filament size a mean field treatment can
be employed (\cf Parker, 1982b) and appropriate turbulent diffusivities have
to be used. Knobloch (1981; see also Knobloch and Rosner, 1981, and
references therein) has taken this approach and used the scaling laws
given by Galloway et al. (1978) for nonlinear Boussinesq magnetoconvection
to estimate a size spectrum of magnetic structures in a turbulent fluid
for which he assumed a Kolmogorov spectrum with a (viscous) cut-off. Even
without taking into account fragmentation processes he found that
most of the structures have sizes at or below the length scale defined by the
cut-off. He comes to the conclusion that larger structures can only be formed 
in the form of flux tube bundles whose size distribution
is difficult to obtain.

This may not be the whole story though: A mechanism not mentioned so far is
{\it coalescence\/} of two parallel twisted flux tubes with the same
sense of twist (Parker, 1982e, 1983a,b; Bogdan, 1984; Choudhuri, 1988). 
If the tubes collide
a neutral sheet forms at their interface, the azimuthal field reconnects
and builds a binding sheath of magnetic flux about both tubes. An azimuthal
field component at the same time exerts a stabilizing influence with
respect to the interchange and Kelvin-Helmholtz instabilities. Consequently,
besides advection by convective flows and fragmentation,  coalescence may be
another important factor which determines the size spectrum of
flux tubes in the convection zone. First attempts to include this effect
in a consistent treatment of the size distribution have been made by
Bogdan and Lerche (1985) and by Bogdan (1985) who found that, in principle,
sunspot-size structures can be produced within the convection zone. However,
the models used so far are very idealized and it is not clear whether the results
hold under more realistic conditions, \ie non-straight flux tubes, 
inclusion of fragmentation
processes, convective and Parker-type instability, advection by large scale
flows, magnetic buoyancy. All these effects tend to either fragment large
structures or to quickly remove them from the deeper layers of the convection 
zone. It seems doubtful that under these conditions sunspot-size structures 
can be formed but this claim can be substantiated or disproved only by more
detailed models which include also the more subtle effects of flux tubes and
flux tube arrays in a convectively unstable medium like `convective counterflow'
(Parker, 1985a,b) and `convective propulsion' (Parker, 1979e).

How can we reconcile our picture of the magnetic field in the deep convection 
zone as an ensemble of very thin flux tubes with
observations of the surface fields where a whole spectrum of structures 
from large spots to small magnetic elements is present? The key to an answer
seems to be the fact that except for the first phase of flux emergence in an 
active region only {\it fragmentation\/} of large
structures into smaller structures is observed, never the opposite process:
Old active regions do not `rejuvenate' and again form pores or spots unless
new magnetic flux erupts. The lognormal distribution of sunspot umbral areas
found by Bogdan et al.\ts\ts(1988) also is consistent with their origin in a
fragmentation sequence. 

We can understand the observation of large sunspots in view of a
convection zone which continually shreds and fragments magnetic structures
if the magnetic flux does not originate there but is injected in large portions
from below (where it resides in a non-fragmented form) and rises rapidly
to the surface. Indeed, flux emergence and the appearance of large structures
at the surface always take place within the very first days of the life
of an active region. The simulations of Moreno-Insertis (1984, 1986) show
that a kink-unstable large flux tube traverses the convection zone and breaks out
at the surface within a few days. Even within this short time fragmentation
apparently has occurred since a whole spectrum of structures appears at the
photosphere and large sunspots always form out of a number of pieces
(McIntosh, 1981) which seem to know where to go (the ``rising magnetic tree''
of  Zwaan, 1978): Initially the rise and emergence of only weakly
fragmented parts of the erupting flux tube is faster than the ongoing 
fragmentation processes. 

After flux emergence has come to an end, fragmentation
proceeds until all magnetic flux at the surface is in the form of small 
network elements with a size of about 100 km\fussn{*}
{This size is much larger than the resistive scale because surface
fields can temporarily achieve stable configurations and resist further
fragmentation. This is further
discussed in \sec{6.3}.}.
In the deep convection zone, all flux tends to become fragmented 
down to the diffusion
limit. The formation of filament bundles by convective flows and 
flux tube coalescence may lead to somewhat larger structures but their actual size
distribution is  difficult to determine. In any case, these structures
are very fragile: They are always subject to the various instabilities and
fragmentation processes and filament bundles are closely coupled to the
changing pattern of convective flows. We do not believe
that large active regions and sunspots are formed in this way: It seems
impossible to store even moderately large flux tubes within the convection
zone for a sufficiently long time against their inherent buoyancy and
instability. Large active regions probably are direct evidence for the
genuine dynamo process operating on a more ordered field in a less
unstable environment than the convection zone proper.

\b\b

\def\etopline{ 6.2 The relation of the basic forces}

\n {\bbf 6.2 The relation of the basic forces}

\b

\n We have argued in Ch.\ts{2} and Sec.\ts6.1 that the magnetic field in
a stellar convection zone consists of an ensemble of small
filaments whose dynamics can be described with aid of the approximation
of slender flux tubes discussed in Ch.\ts{3}. We found that the evolution
of a flux tube is determined by three basic forces, the {\it buoyancy force 
$F_B$,} the {\it curvature force $F_C$} and the {\it drag force $F_D$}, 
all of which are directed perpendicular to the flux tube axis while 
gravity and gas pressure determine the momentum balance in 
the direction along the field lines. 

Let us discuss the relative importance of these forces under the physical
conditions in a convection zone. If we presume temperature 
equilibrium between the flux tube and its environment (a reasonable
assumption for small filaments as we have
seen in the preceding section) we may use \es{3.26} and (3.28) to obtain
the order of magnitude of the various forces per unit length of the flux
tube:

$$ \eqalignno{
     F_B  &\sgs {B^2 \, a^2 \over 8\, \hpe} &(6.6) \cr && \cr
     F_C  &\sgs {B^2 \, a^2 \over 4\, R}    &(6.7) \cr && \cr
     F_D  &\sgs C_D\, \roe\, a\, v_\perp^2      &(6.8) \cr
             } $$

\n ($a$: flux tube radius, $\hpe$: external pressure scale height,
    $\roe$: external density, $v_\perp$: velocity component perpendicular
    to the flux tube due to large scale convection, differential rotation
    or the motion of the flux tube itself, $R$: radius of curvature).
Ignoring factors of order unity we may write for the ratio of buoyancy 
and curvature force to drag force, respectively:

$$ \eqalignno{
    {F_B\over F_D} \; &= \; \left( {a\over\hpe} \right)
                            \left( {B\over B_e} \right)^2 
                            \left( {v_c\over v_\perp} \right)^2 &(6.9) 
    \cr && \cr
    {F_C\over F_D} \; &= \; \left( {a\over R} \right) 
                            \left( {B\over B_e} \right)^2 
                            \left( {v_c\over v_\perp} \right)^2. &(6.10) 
    \cr      } $$

\n Here $B_e = v_c \sqrt{4\pi\roe}$ is the equipartition field strength with
respect to convective flows of typical velocity $v_c$. We see from these
ratios that for equipartition fields and $v_\perp \approx v_c$ {\it the
drag force dominates the dynamics of a thin flux tube:} $a/\hpe\ll 1$ and
$a/R\ll 1$ in fact are conditions for the applicability of the approximation
of slender flux tubes. For strongly fragmented fields in the deep convection
zone they are well met: $a<1$ km, $\hpe\approx 10^4$ km; the curvature force 
comes into play only for extreme distortions ($R\approx a$) of the filament.
Consequently, a magnetic field structured in this way {\it follows passively
any large-scale flow\/} with relative velocity of the order of the convective 
velocity, be it convection itself, differential rotation or, possibly, 
meridional circulation. Only structures with $a\approx\hpe$ or larger 
(flux tubes of sunspot size) or strong fields with $B>B_e$ (partially evacuated
flux tubes at the surface) can avoid being severely distorted by
large scale flows. 

On the other hand, the growing distortion of the tube leads to an increase
of its length and, if the mass content stays constant, to an increase of
the field strength. In the case $\beta\gg 1$ we find that the 
change of the magnetic field strength is proportional to the length
increment, \ie

$$ {\Delta B\over B} \sgs {\Delta l\over l}\,. \eqno (6.11) $$

\n Doubling the flux tube length doubles the field strength and the 
curvature force increases by a factor 4. Consequently, a flux tube with
constant mass content quickly becomes dynamically active and resists to further
distortion and stretching.  The curvature force
increases rapidly and a balance between curvature and drag force establishes
itself. However, the assumption of a constant mass content does not hold for two
important cases:

\medskip
\item {-} The distortion of flux tubes is strongest in the upper layers 
          of the convection zone where the flow velocities are large. Here
          the density is small and if parts of the tube are still located in the 
          deep layers they represent a large mass reservoir: Matter can flow along
          the tube to fill the volume created by stretching the tube, and
\smallskip
\item{-}  for small filaments formed by the various fragmentation processes 
          and instabilities we have seen in the preceding section that
          the diffusive time scale is smaller than the dynamical time scale which
          governs the distortion process. Consequently, matter can diffuse
          into the flux tube rapidly enough to fill the volume created
          by its distortion. This is particularly relevant for the 
          deep layers where no further mass reservoir
          exists for flux tubes contained within the convection zone. 
\medskip

\n Of course, both mechanisms can be relevant for the same flux tube. Their effect
is to keep the flux tube in a state of passiveness with respect to further 
stretching. We conclude that due to instability and fragmentation the magnetic 
field structures that reside in the convection zone for an extended period of 
time (\ie longer than about a month, the dynamical time scale in the deep layers)
are passive with respect to external flows (convection, differential rotation, 
meridional circulation), \ie they are strongly coupled to and distorted/fragmented
by these flows and probably never reach an equilibrium.

An exception from this rule are large flux tubes whose radius is of the order of 
the local scale height or larger. Here all three basic forces may become
comparable and the results concerning equilibrium structure and stability obtained
in Chs.\ts{4} and 5 can be applied. We have found that in most cases such a flux
tube is unstable and parts of it rapidly erupt towards the surface layers
while other parts sink down below superadiabatic layers of the convection zone.
Although the applicability of the approximation of slender flux tubes is 
somewhat doubtful for such large tubes we believe that the results
are qualitatively correct also in this case. 
Another exception from the rule of passive magnetic fields are the flux
tubes in the surface layers. They will be discussed in the following section.
          
\b\b

\def\etopline{ 6.3 The peculiar state of the surface fields}

\n {\bbf 6.3 The peculiar state of the surface fields}

\b

\n The observable surface layers of a star like the Sun 
are quite different from the state of the deep convection zone. They represent 
the transition region between the strongly superadiabatic top of 
the convection zone and the stably stratified photosphere where
most of the energy carried by the overshooting convective motions escapes 
into space by means of radiation. The gas is strongly stratified since the 
relatively low temperature leads to a small scale height while the temperature 
fluctuations become a significant fraction of the temperature itself. 

Under these conditions magnetic fields can be concentrated to field strengths
far beyond equipartition with convective motions.
Magnetic fields are swept to the granular downflow regions
and concentrated to about equipartition by the horizontal flows 
of granular convection. These flows at the same time are responsible for 
carrying heat to the downflow regions which compensates for the radiative losses.
Since the growing magnetic field retards the motions it throttles the
energy supply and the magnetic region cools down. This leads to an increase
of the magnetic field since the gas pressure in the magnetic region becomes 
smaller. Furthermore, hydrostatic equilibrium on the basis of a reduced 
temperature tends to reestablish itself via a downflow 
which gives rise to the {\it superadiabatic effect\/}
(Parker, 1978): An adiabatic downflow in a magnetic flux tube which is
thermally isolated from its surroundings leads to a cooling of the interior
with respect to the superadiabatically stratified surroundings and a partial
evacuation of the the upper layers ensues. Pressure equilibrium with the
surrounding gas is maintained by a contraction of the flux tube which increases
the magnetic pressure. In this way, the magnetic field can be locally
intensified to values which are only limited by the confining pressure of the
external gas.

It has been shown by a number of authors (Webb and Roberts, 1978; Spruit and
Zweibel, 1979; Unno and Ando, 1979) that the superadiabatic effect in the case of 
a vertical flux tube which is in magnetostatic and temperature equilibrium with 
its environment drives a {\it convective instability\/} in the form of a
monotonically increasing up- or downflow.  Consequently, the initial downflow
due to the radiative cooling is enhanced by this effect leading to an even
stronger amplification of the magnetic field, a process which is often referred
to as {\it convective collapse.} A more detailed discussion of this process
has been given elsewhere (Sch\"ussler, 1990).

Cooling due to suppression of convective motions and the superadiabatic effect
have the consequence that most of the observed magnetic flux in the solar 
photosphere has a field strength in excess of the equipartition value of a few
hundred Gauss. It is organized in a hierarchy of structures which have a 
magnetic pressure comparable to the gas pressure in their apparently non-magnetic
environment. This hierarchy extends from large sunspots 
(diameter $>$ \dex{5}{4}~km,
field strength up to 4000~Gauss) to small magnetic elements (diameter
$<$ 200~km, field strength about 2000~Gauss at optical depth unity). In contrast
to the circumstances in the deep convection zone which have been discussed in
the preceding section, under these conditions {\it buoyancy\/} becomes the
dominating force. Taking 500 Gauss for the equipartition field strength and
using \e{6.9} we find that $F_B/F_D$, the ratio of the buoyancy
to the drag force due to convective motions, is of the order of 10 for
magnetic elements ($a=100$ km, $B=2000$ Gauss) and amounts to about 3000 for 
large sunspots ($a=10^4$~km, $B=4000$ Gauss). By the dominance of buoyancy
the magnetic structures are forced to become straight and vertical in the surface 
layers: A very small inclination suffices to compensate any drag force exerted by
the external velocity fields (Sch\"ussler, 1986, 1987). 

Besides the ability to resist from being distorted and deformed by convective
motions another peculiarity of the surface fields is the existence
of configurations which are (at least temporarily) stable with respect to the 
interchange instability and other fragmentation processes. Buoyancy and the 
large field strength prevent the flux concentrations from being disrupted by
local convection. Large structures like sunspots (magnetic flux larger than
about $10^{19}$ mx) are stabilized
against the interchange instability by buoyancy (Meyer et al., 1977) while 
small magnetic elements (magnetic flux less than a few times $10^{17}$ mx) 
may be stabilized by whirl flows which surround them
(Sch\"ussler, 1984b). Such whirls form naturally in the narrow downflow
regions of convection in a strongly stratified medium (\eg Nordlund, 1985a).
These downflows are enhanced around small magnetic flux concentrations by 
thermal effects (Deinzer et al., 1984). 

The existence of two regimes of stable 
magnetic flux tubes connected by a range of magnetic configurations which are
subject to the interchange instability certainly should have an influence on the
observable size distribution of magnetic structures. On the one hand large
sunspots may live for an extended period of time during which they only show
a slow decay while, on the other hand, small features exist with sizes up to the 
limit where the stabilization due to a surrounding whirl flow becomes inefficient.
The verticality of the magnetic structures due to the dominating buoyancy
force facilitates their organization in a network pattern
defined by the downflow regions which are the loci of convergence
of the horizontal convective flows. The probability of encounters and
coalescence is much larger within such a network than for a more random
spatial distribution. For all these reasons we think that observed size
distributions of surface fields (\eg Spruit and Zwaan, 1981) are by no means 
representative for the conditions in the deep layers of the convection zone where 
most of the effects which determine the particular properties of surface fields
become irrelevant.

The special properties of the surface fields are restricted to
the photosphere and the rather narrow layer of strong superadiabaticity in the 
upper convection zone of about $1000$ km depth. As discussed in more detail
elsewhere (Sch\"ussler, 1987) the non-observation of a systematic dependence of
the dynamics of observed flux concentrations on their size together with
a consideration of the dominating forces strongly support the cluster model
of sunspots (Parker, 1979c; Spruit, 1981b). This model assumes that a spot
is a conglomerate of a large number of small magnetic filaments with a
diameter of 100 to 1000 km which are closely packed due to buoyancy in the 
uppermost layers of the convection zone to form the visible spot umbra.
Below some merging level (which is situated not much deeper than 1000~km)
the spot splits into its fibril components which
are passive with respect to the convective motions which continually stretch
and distort them; this causes the slow decay of the spot and the fragmentation 
of its magnetic flux into the network fields. 

We have seen that the observable surface fields probably are in many respects 
different from the conjectured properties of the magnetic fields in the
deep layers of the convection zone. The strong superadiabaticity of the
uppermost convective layers and the thermal effects associated with the
radiative release of the transported energy into free space entail a number
of peculiar effects which cannot be expected to operate in the main body of
the convection zone. Only at the surface can we expect the field strength to
significantly exceed the equipartition limit and only there do we find
mechanisms which can stabilize magnetic flux tubes from being quickly disrupted 
by instabilities. Consequently, it is improbable that a flux concentration 
observed at the surface maintains its identity as a single flux tube
throughout the whole or even a significant part of the convection zone. 
Visible sunspots and all other observed structures probably fragment into small 
filaments below about 1000~km depth. These filaments with sizes down to
the diffusion limit are passive with respect to
the convective motions, they are stretched and deformed, accumulated and
dispersed by the action of the dominating drag forces. Due to their inherent
stability, the surface fields can tolerate the distortion of their `roots'
for a certain amount of time until they are forced to react
accordingly, be it by slowly dispersing and eventually breaking apart as in the 
case of sunspots, be it by a continuous rearrangement of the small-scale fields in
the supergranular network.

\vfill\eject

\def\etopline{ 6.4 Consequences for the dynamo problem}

\n {\bbf 6.4 Consequences for the dynamo problem}

\b

\n The presently favored tool to describe the origin of solar activity is the
theory of dynamo action in a turbulent medium which started with the pioneering
work of Parker (1955). Beginning in the 1960s the next landmark was placed by
the Potsdam group (Krause, R\"adler, R\"udiger, Steenbeck and others; see
Krause and R\"adler, 1980, for an overview) who used a statistical approach
(mean field theory) in the kinematical limit (passive magnetic field, no
back-reaction on the turbulent flows). They found the so-called 
$\alpha${\it-effect\/} 
which can lead to dynamo action in turbulent fluids which lack mirror symmetry,
\eg due to the influence of rotation. Many large-scale properties of the solar
cycle could be successfully reproduced by
`$\alpha\omega$-dynamos' which are based on the combined induction effects of
turbulence ($\alpha$) and differential rotation ($\omega$).  Reviews of these
models have been given, among others, by Stix (1976, 1981a, 1982), Parker (1979a) 
and Yoshimura (1981).

It was again Parker (1975a) who raised doubt concerning the theory of turbulent
dynamo action {\it within\/} the convection zone. He argued that, if most of the
toroidal magnetic flux is in the form of large flux tubes as indicated by the
existence of sunspots and active regions, it cannot be stored in the
convection zone for times comparable with the solar cycle period:  Buoyancy
removes magnetic flux from the convection zone much too fast for the induction
mechanisms to operate efficiently, in particular for the differential rotation
to generate a sufficient amount of toroidal magnetic flux within a cycle period. 
The instabilities of
flux tubes due to superadiabaticity, flows along the field lines, rotation and
external flows investigated by Spruit and van Ballegooijen (1982), van
Ballegooijen (1983), Moreno-Insertis (1984, 1986), van Ballegooijen and
Choudhuri (1988), and in Ch.\ts{5} of this work aggravate the problem even
more.

The assumption of very small flux tubes or of a large turbulent viscosity
may reduce the velocity of buoyant rise drastically (Unno and Ribes, 1976;
Sch\"ussler, 1977; 1979; Kuznetsov and Syrovatskii, 1979) but, on the other
hand, leads to a strong coupling between the flux tubes and the convective
flows which would destroy their azimuthal orientation and quickly
raise them to the solar surface within the convective
time scale of about one month. Again, the flux cannot be contained within the
convection zone for a sufficiently long time. 

Recently it has been argued by Parker (1987a,d) that 
the magnetic flux observed to emerge in large complexes of
activity of relatively small latitude extension would fill as an equipartion field
a considerable part of the underlying convection zone and interfer significantly
with the convective energy transport. He proposed that the suppression of 
vertical convective heat transport by a band of horizontal (azimuthal) field
in a convection zone leads to an overlying  `thermal shadow', \ie a cool region 
of enhanced density, whose weight could balance the magnetic buoyancy and thus
keep the field down within the convection zone. He studied in some detail the
convective flows set up by a thermal shadow (Parker, 1987b) and the 
Rayleigh-Taylor instability caused by the pile-up of heat beneath the magnetic
layer (Parker, 1987c). He envisages that a thermal relaxation oscillation
evolves which leads to the intermittent eruption of magnetic flux in the
form of rising, hot plumes (Parker, 1988a). The dynamical instability of
a flux sheet of finite lateral extent with respect to sideways displacements
leads to a lateral velocity of the order of a few m/s which he connects
with the observed latitudinal motions of the solar activity belts (Parker, 1988b).

Parker's conjecture is based on a number of illustrative calculations of 
idealized problems in order to permit an analytical treatment. While the
thermal shadows may well have important dynamical effects it is not shown
that they can keep large amounts of magnetic flux in the superadiabatic
parts of the convection zone for time intervals comparable to the period
of the solar cycle. Actually, in view of the instabilities discussed by
Parker himself (Rayleigh-Taylor instability of top and bottom, lateral
dynamical instability) and the quick disruption of a magnetic layer even
in a stably stratified region (Cattaneo and Hughes, 1988), this seems to
be rather improbable. 

However we turn the problem, we run into difficulties if we assume the dynamo
to operate within the convection zone proper. If the magnetic field is
organized in large, dynamically active structures, these are unstable, buoyant and
rapidly lost from the deep layers of the convection zone.  If the field is
diffuse or consists of very small structures (the latter view is favored in
this work), it is passive with respect to the motion of the fluid (convection,
rotation, ...) and will be carried to the surface within the convective
time scale. Moreover, such a passive field is not in accordance with basic
features of solar activity:  In a large active region, magnetic flux erupts
coherently in large quantities and within a few days; large sunspots form which
comprise a significant fraction of the total flux erupting during the cycle;
Hale's polarity rules are obeyed nearly strictly, not in a statistical sense. On
the other hand, the relative velocities due to differential rotation and the
convective velocities are of the same order of magnitude as is well known from
surface observations (\eg Schr\"oter, 1985) and also shown by the rotational 
splitting of $p$-modes for the deeper layers of the convection zone (\eg 
Duvall et al., 1987; Brown et al., 1989; Dziembowski et al., 1989). 
How can passive fields be predominantly toroidal and obey the polarity rules 
under such conditions ?

We cannot avoid the conclusion that the original source region at least of the 
large, sunspot-forming active regions cannot be the convection zone proper.
It cannot be the radiative core of the Sun either since the time scale of
22 years for the magnetic cycle does not allow a penetration into the radiative
interior because of its large electrical conductivity which leads to a 
skin effect. Consequently, a number of authors (Spiegel and Weiss, 1980;
Galloway and Weiss, 1981; 
van Ballegooijen, 1982a,b; Schmitt and Rosner, 1983; Sch\"ussler, 1983; DeLuca
and Gilman, 1986; Durney, 1989; Durney et al., 1990) 
have proposed a boundary layer of overshooting convection
below the convection zone proper as a favorable site of the solar dynamo. There,
in a region of `mild' convection and turbulence, we may suppose that differential
rotation dominates all other velocity fields and generates predominantly 
toroidal magnetic fields. The failure of large-scale simulations to reproduce
the observed characteristics of solar activity (in particular, the direction of 
the latitude drift of the activity belts) by dynamically consistent 3D-models
of the convection zone (Gilman and Miller, 1981; Gilman, 1983; Glatzmaier,
1985a) led their proponents to the same conclusion (Glatzmaier, 1985b).

The subadiabatic stratification of an overshoot region alleviates the stability 
problems and a number of
mechanisms is available which may hold a growing toroidal  magnetic flux there 
for a time comparable with the cycle period (\cf Sch\"ussler, 1983). We may note
in particular the `turbulent diamagnetism' briefly discussed in Sec.\ts{2.2}, 
which is akin to the flux expulsion process and transports magnetic flux 
antiparallel to a gradient of turbulent intensity. Krivodubskii (1984, 1987)
has given quantitative estimates of the diamagnetic effect using models of the
solar convection zone. Since, at least in the kinematical limit, 
this mechanism operates equally well for vorticity (Sch\"ussler, 1984a) it 
leads to the formation and maintenance of a magnetic shear layer at the bottom 
of the convection zone (including the overshoot region), just the situation
we envisage as a favorable setup for the operation of the solar dynamo.

Besides differential rotation we need a regeneration mechanism for the poloidal
field component in order to close the dynamo cycle. Such a mechanism might be
provided by the usual
$\alpha$-effect due to cyclonic convection or by an analogous electromotive
force generated by waves propagating along the toroidal field in a rotating
system. Examples of such waves are slow magnetostrophic waves driven by the
magnetic Rayleigh-Taylor instability (Schmitt, 1984; 1985) or transversal
flux tube waves excited by differential rotation (van Ballegooijen, 1983) or
overstability in an external flow as described in Ch.\ts{5}. A boundary layer 
dynamo model may thus show many of the properties of the `classical'
$\alpha\omega$-dynamo models which have been so successful in describing basic 
features of the solar cycle. This
supports the conjecture that the mathematical description of the field
regeneration mechanism (\ie a mean current parallel to the mean field)
in the mean induction equation is basically correct. Note, however, that a 
kinematical approach is not justified for a boundary layer dynamo as envisaged 
above since the magnetic and kinetic energy densities are comparable for 
equipartition fields. A consistent theory
for an $\alpha$-effect in an overshoot region with a strong magnetic field,
shear flow and weak turbulence remains to be developed.
 
Eventually,
magnetic flux is released into the convection zone proper by instability or
buoyancy. The length and time scales of the formation of large active 
regions are well in agreement with the characteristics of the eruption of
a kink-unstable large flux tube originally situated near the bottom of the
convection zone (Moreno-Insertis, 1984; 1986). 
Moreover, we could think about a modification of Parker's thermal shadow scenario:
If the amount of magnetic flux is small enough that it `fits' into the 
subadiabatic overshoot layer there is almost no interference (except from
opacity effects, \cf Parker, 1984b) with the energy
flux which is mainly carried by radiation -- neither a shadow nor a significant
pile-up of heat evolve. If the magnetic flux layer intrudes significantly
into the superadiabatic convection zone proper a thermal shadow and a
pile-up of heat in the magnetic overshoot region are the consequence. Both
effects provoke Rayleigh-Taylor instability and the whole magnetic layer
is disrupted and erupts towards the surface: a large active region is born.
In this picture the thermal shadow does not primarily keep the flux submerged
but, on the contrary, is the agent of the eruption of magnetic flux. It
is tempting to speculate whether the larger efficiency of differential rotation
in generating azimuthal magnetic field near the equator might provide the excess
azimuthal flux that drives the magnetic layer unstable and produces large
active regions while the azimuthal flux in the polar regions always fits
into the subadiabatic region.

As shown by Choudhuri and Gilman (1987) and Choudhuri (1989) the influence
of rotation on the dynamics of rising loops might be quite significant.
In order to avoid the flux erupting in the polar regions either the flux
density in the overshoot region must be quite high, \ie of the order of
$10^5$ Gauss -- which alleviates the storage space problems pointed out
by Parker (1987a) but increases the buoyancy problems -- or the transport
is dominated by the drag of predominantly radial convective flows whose
motion the magnetic fields passively follow. Obviously, the r\^ole of
rotation in the transfer of magnetic fields in a convection zone deserves
further consideration.

Altogether, the results obtained here support the picture of a boundary
layer dynamo sketched above. We have not been able to find a plausible way
by which magnetic structures within the convection zone proper could escape
from being extremely distorted, dismembered down to the diffusion limit, and
eventually becoming completely passive with respect to the convective flows. 
Large active regions with their specific properties cannot arise from such
a kind of field. On the other hand, if a less fragmented magnetic structure
at some instant exists in the convection zone (\eg having been injected
from below) in most cases it will immediately become unstable by one of
the mechanisms discussed in Ch.\ts{5}, part of it will erupt at the surface
while another part sinks down into the subadiabatic boundary layer. After
flux emergence has come to an end, the magnetic structures are more and more
fragmented and shredded until they are merge into the extremely
filamented and distorted genuine convection zone fields.

It has been already remarked by Parker (1982b, see also Weiss, 1981c) that an 
ensemble of passive flux tubes can be described by a kinematic mean field theory 
in much the same way as in the theory developed by the Potsdam group 
(see Krause and R\"adler, 1980). We may speculate (\eg Durney, 1989)
that part of the field within the convection zone is maintained by a turbulent 
dynamo in the classical sense which contributes to the background fields
at the surface  while a boundary layer dynamo provides the source of the big 
active regions and the large-scale features of solar activity. 
Anyway, a major revision of the conventional picture of a convection zone dynamo 
seems to be in order. How the dynamic boundary layer dynamo 
operates is not understood yet and the detailed study of the dynamics of
concentrated fields in a convection zone and its adjacent lower overshoot 
layer has just started.

\vfill\eject

\def\otopline{ 7. Outlook }

\vglue 2cm

\def\etopline{ 7. Outlook }

\n {\bbbf 7. Outlook}

\b\m

\n As it turns out so often, most problems remain to be solved.
In view of the unsatisfactory state of our understanding of stellar convection
and turbulence, a closed and complete theory of magnetic fields in convection
zones is not in sight. The theory of photospheric magnetic fields is much more
advanced thanks to the availability of detailed measurements and the close 
connection between theoretical and observational work. Some information
about the internal magnetic field and large-scale velocity structure will be 
obtained in the future by helioseismology but, perhaps with the exception of 
differential rotation, we are not too optimistic about the prospects 
of obtaining much more stringent observational boundary conditions. Consequently,
besides the ongoing efforts of numerical simulation and the detailed analysis
of model problems this field of research will remain open for conjecture,
speculation and the presentation of more or less ingenious scenarios.

Numerical simulations will partially substitute unavailable observations.
As the development of ever faster computers and sophisticated numerical methods to
adequately use them proceeds, simulations will grow  more realistic as
three-dimensional, compressible MHD calculations with high spatial resolution 
become available and will provide an immensely useful tool for understanding the
magnetic field dynamics in the convection zone.
However, in contrast to some fashionable folklore
existing and forthcoming numerical simulations do not make other approaches
obsolete.  The dynamics of motions and magnetic fields in the solar convection
zone extends over huge ranges of temporal and spatial scales
which in both cases comprise more than ten decades. Since only a small part of
these can be covered by any simulation in the foreseeable future, 
artificial boundaries have to be
introduced, certain scales are ignored and others are included only in a
parametrized form.  Such parametrizations can only be made in a sensible way if
they are based on a sound understanding of processes which determine the
properties of flows and fields on the scales which they attempt to describe.

We have given some arguments in favor of the conjecture that magnetic
fields in stellar convection zones are strongly fragmented and can be
treated on the basis of the approximation of slender flux tubes.
In the deep parts of a convection zone the scale height is very large 
such that the approximation is justified even for structures containing the 
magnetic flux of a whole active region.
The investigation of flux tube dynamics in a realistic convection zone 
is a promising path for future research. Equilibrium structures, stability
and dynamical evolution of flux tubes in prescribed velocity fields can be
determined on the basis of the methods described in this work, 
guided by forthcoming 3D simulations of the large-scale convective flows
and observational results on differential rotation. As far as the linear stability
analysis is concerned, this requires a transformation of the formalism presented
in Ch.\ts5 to spherical geometry and the inclusion of (differential) rotation.
For realistic models of  convection zones, the equations describing flux tube 
equilibria and the perturbation equations will have to be treated numerically.
Work in this direction is in progress, being done in cooperation with
Antonio Ferriz-Mas.

A possibility to gain information about the size distribution of magnetic
structures in a convection zone is the application of methods taken from
statistical mechanics. The evolution of the properties of an ensemble of
flux tubes may be derived from a collisional Boltzmann equation which includes
the effect of large-scale flows, diffusion, fragmentation, coagulation and other
processes. Moreover, this approach opens a possibility to put the 
vague notion of a `flux tube dynamo' on a firm theoretical basis. A cooperation
with Tom Bogdan (Boulder), Antonio Ferriz-Mas and Michael Kn\"olker on such
a project is arranged.

Finally, the boundary layer or overshoot layer dynamo will remain a challenge.
Kinematical theory probably cannot be applied since kinetic and magnetic energy
density are of the same order of magnitude. Future work will focus on two
approaches, \ie the quantitative determination of a field regeneration process
and the angular momentum distribution in an overshoot layer, and the
investigation of nonlinear dynamo models with a given regeneration process
which take into account the particular geometry and thermodynamics of the
region as well as expulsion processes and magnetic instabilities. As for
most of the discussed problem areas here again 
comprehensive 3D simulations and idealized/simplified approaches will play
complimentary parts: The simulations help to identify the relevant processes and
allow us a glimpse at phenomena which are observationally unreachable. They
can guide us in picking the relevant pieces of physics to study in detail
without falling into the trap of oversimplified or prejudiced concepts. An
understanding of the physics governing these processes, of their general
properties and the validity of their description in the simulation can only
come from a detailed study in the spirit of theoretical physics.

\vfill\eject

\def\etopline{ References }
\def\otopline{ References }
\parindent = 20pt

\n {\bbf References}

\b

\ref {\bbrm A}bramovitz, M., Stegun, I.A.: 1965, {\it Handbook of Mathematical Functions}, Dover, New York
\ref Acheson, D.J.: 1978, {\it Phil. Trans. Roy. Soc. London}, {\bf A289}, 459
\ref Acheson, D.J.: 1979\sp 62, 23
\ref Acheson, D.J., Gibbons, M.P.: 1978\jfm 85, 743
\ref Achterberg, A.: 1982\aap 114, 233
\ref Achterberg, A.: 1988\aap 191, 167
\ref Anton, V.: 1984, Diplomarbeit, Universit\"at G\"ottingen

\m

\ref {\bbrm B}atchelor, G.K.: 1967, {\it An Introduction to Fluid Dynamics}, Cambridge Univ. Press
\ref Bernstein, I.B., Frieman, E.A., Kruskal, M.D., Kulsrud, R.M.: 1958, {\it Proc. Roy. Soc.} {\bf A244}, 17
\ref Bogdan, T.J.: 1984, {\it Phys. Fluids} {\bf 27}, 994
\ref Bogdan, T.J.: 1985\apj 299, 510
\ref Bogdan, T.J., Lerche, I.: 1985\apj 296, 719
\ref Bogdan, T.J., Gilman, P.A., Lerche, I., Howard, R.: 1988\apj 327, 451
\ref Brandenburg, A., Tuominen, I., Moss, D., R\"udiger, G.: 1990\sp 128, 243
\ref Brown, T.M., Christensen-Dalsgaard, J., Dziembowski, W.A., Goode, P., Gough, D.O., Morrow, C.A.: 1989\apj 343, 526

\m

\ref {\bbrm C}adez, V.M.: 1974, thesis, Institute of Physics, Beograd
\ref Cap, F.F.: 1976, {\it Handbook on Plasma Instabilities}, Vol. {\bf 1}, Academic Press, New York
\ref Cattaneo, F., Hughes, D.W.: 1988\jfm 196, 323
\ref Cattaneo, F., Hughes, D.W., Proctor, M.R.E.: 1988, {\it Geophys. Astrophys. Fluid Dyn.} {\bf 41}, 335 
\ref Chou, D.-Y., Fisher, G.H.: 1989\apj 341, 533
\ref Choudhuri, A.R.: 1988, {\it Geophys. Astrophys. Fluid Dyn.} {\bf 40}, 261  
\ref Choudhuri, A.R.: 1989\sp 123, 217
\ref Choudhuri, A.R., Gilman, P.A.: 1987\apj 316, 788

\m

\ref {\bbrm D}efouw, R.J.: 1976\apj 209, 266
\ref Deinzer, W., Hensler, G., Sch\"ussler, M., Weisshaar, E.: 1984\aap 139, 435
\ref DeLuca, E.E., Gilman, P.A.: 1986, {\it Geophys. Astrophys. Fluid Dyn.} {\bf 37}, 85 
\ref Durney, B.R.: 1989\sp 123, 197 
\ref Durney, B.R., De Young, D.S., Passot, T.P.: 1990\apj 362, 709
\ref Duvall, T.L., Dziembowski, W.A., Goode, P.R., Gough, D.O., Harvey, J.W., Leibacher, J.W.: 1984\nat 310, 22
\ref Duvall, T.L., Harvey, J.W., Pomerantz, M.A.: : 1987, in B.R. Durney and S. Sofia (eds.): {\it The Internal Solar Angular Velocity}, Reidel, Dordrecht, p. 19
\ref Dziembowski, W.A., Goode, P.R., Libbrecht, K.G.: 1989\apj 337, L53
\m

\ref {\bbrm F}erriz-Mas, A., Sch\"ussler, M.: 1989, {\it Geophys. Astrophys. Fluid Dyn.}, {\bf 48}, 217
\ref Ferriz-Mas, A., Sch\"ussler, M, Anton, V.: 1989\aap 210, 425

\m

\ref {\bbrm G}alloway, D.J.: 1978, in G. Belvedere, L. Patern\'o (eds.): {\it Workshop on Solar Rotation}, Catania, p. 352
\ref Galloway, D.J., Proctor, M.R.E., Weiss, N.O.: 1978, {\it J. Fluid Mech.} {\bf 87}, 243
\ref Galloway, D.J., Weiss, N.O.: 1981\apj 243, 945
\ref Garcia de la Rosa, J.I.: 1987\sp 112, 49
\ref Gilman, P.A.: 1970\apj 162, 1019
\ref Gilman, P.A.: 1983\apjs 53, 243
\ref Gilman, P.A., Miller, J.: 1981\apjs 46, 211
\ref Glatzmaier, G.A., Gilman, P.A.: 1982\apj 256, 316
\ref Glatzmaier, G.A.: 1985a\apj 291, 300
\ref Glatzmaier, G.A.: 1985b, {\it Geophys. Astrophys. Fluid Dyn.} {\bf 31}, 137
\ref Grappin, R., Frisch, U., L\'eorat, J., Pouquet, A.: 1982\aap 105, 6

\m

\ref {\bbrm H}opfinger, E.J., Browand, F.K., Gagne, Y.: 1982\jfm 125, 505
\ref Howard, R.: 1974\sp 38, 59
\ref Howard, R.: 1984, {\it Ann. Rev. Astron. Astrophys.}, {\bf 22}, 131
\ref Howard, R.: 1989\sp 123, 285
\ref Hughes, D.W.: 1985a, {\it Geophys. Astrophys. Fluid Dyn.} {\bf 32}, 273
\ref Hughes, D.W.: 1985b, {\it Geophys. Astrophys. Fluid Dyn.} {\bf 34}, 99
\ref Hughes, D.W.: 1987, {\it Geophys. Astrophys. Fluid Dyn.} {\bf 37}, 297
\ref Hughes, D.W., Cattaneo, F.: 1987, {\it Geophys. Astrophys. Fluid Dyn.} {\bf 39}, 297
\ref Hurlburt, N.E., Toomre, J.: 1988\apj 327, 920
\ref Hurlburt, N.E., Toomre, J., Massaguer, J.M.: 1984\apj 282, 557

\m

\ref {\bbrm K}nobloch, E.: 1981\apj 247, L93
\ref Knobloch, E., Rosner, R.: 1981\apj 247, 300
\ref Kraichnan, R.H.: 1976, {\it J. Fluid Mech.} {\bf 77}, 753
\ref Krall, N.A., Trivelpiece, A.W.: 1973, {\it Principles of Plasma Physics}, McGraw-Hill, New York
\ref Krause, F., R\"adler, K.-H.: 1980, {\it Mean-Field Magnetohydrodynamics and Dynamo Theory}, Pergamon, Oxford
\ref Krivodubskii, V.N.: 1984, {\it Sov. Astron.} {\bf 28}, 205
\ref Krivodubskii, V.N.: 1987, {\it Sov. Astron. Lett.} {\bf 13}, 338
\ref Kuznetsov, V.D., Syrovatskii, S.I.: 1979, {\it Sov. Astron.} {\bf 23}, 715

\m

\ref {\bbrm L}aBonte, B.J., Howard, R.: 1982\sp 75, 161
\ref LaBonte, B.J., Howard, R., Gilman, P.A.:1981\apj 150, 796
\ref Livingston, W., Holweger, H.: 1982\apj 252, 375

\m

\ref {\bbrm M}cIntosh, P.S.: 1981, in {\it The Physics of Sunspots}, L.E. Cram, J.H. Thomas (eds.), Sacramento Peak Observatory, p. 7
\ref McEwan, A.D.: 1973, {\it Geophys. Fluid Dyn.} {\bf 5}, 283
\ref McEwan, A.D.: 1976, {\it Nature} {\bf 260}, 126
\ref McWilliams, J.C.: 1984\jfm 146, 21
\ref Meneguzzi, M., Frisch, U., Pouquet, A.: 1981, {\it Phys. Rev. Lett.} {\bf 47}, 1060
\ref Meneguzzi, M., Pouquet, A.: 1989\jfm 205, 297
\ref Meyer, F., Schmidt, H.U., Weiss, N.O.: 1977, {\it Mon. Not. Roy. Astron. Soc.} {\bf 179}, 741
\ref Moffatt, H.K.: 1983, {\it Rep. Prog. Phys.} {\bf 46}, 621 
\ref Moreno-Insertis, F.: 1983\aap 122, 241
\ref Moreno-Insertis, F.: 1984, Dissertation, Universit\"at M\"unchen
\ref Moreno-Insertis, F.: 1986\aap 166, 291
\ref M\"uller, R., Roudier, Th.: 1984\sp 94, 33

\m

\ref {\bbrm N}ordlund, \AA.: 1983, in J.O. Stenflo (ed.):  {\it Solar and Stellar Magnetic Fields: Origins and Coronal Effects}, IAU-Symp. No. 102, Reidel, Dordrecht , p. 79
\ref Nordlund, \AA.: 1984, in S.L. Keil (ed.): {\it Small-scale Processes in Quiet Stellar Atmospheres}, Sacramento Peak Observatory, Sunspot, p. 174
\ref Nordlund, \AA.: 1985, in H.U. Schmidt (ed.): {\it Theoretical Problems in High Resolution Solar Physics}, MPA 212, Max-Planck Institut f\"ur Astrophysik, M\"unchen, p. 1
\ref Nordlund, \AA.: 1986, in W. Deinzer et al. (eds.): {\it Small Scale Magnetic Flux Concentrations in the Solar Photophere}, Abh. Akad. d. Wiss. G\"ottingen No. 38, Vandenhoeck \& Ruprecht, G\"ottingen, p. 83

\m

\ref {\bbrm O}rszag, S., Tang, C.-H.: 1979, {\it J. Fluid Mech.} {\bf 90}, 129

\m

\ref {\bbrm P}iddington, J.H.: 1975, {\it Astrophys. Space Sci.} {\bf 34}, 347
\ref Parker, E.N.: 1955\apj 122, 293
\ref Parker, E.N.: 1963\apj 138, 552
\ref Parker, E.N.: 1966\apj 145, 811
\ref Parker, E.N.: 1975a\apj 198, 205
\ref Parker, E.N.: 1975b\sp 40, 291
\ref Parker, E.N.: 1975c\apj 201, 494
\ref Parker, E.N.: 1978\apj 221, 368
\ref Parker, E.N.: 1979a, {\it Cosmical Magnetic Fields}, Clarendon, Oxford
\ref Parker, E.N.: 1979b, {\it Astrophys. Sp. Sci.} {\bf 62}, 135
\ref Parker, E.N.: 1979c\apj 230, 905
\ref Parker, E.N.: 1979d\apj 230, 914
\ref Parker, E.N.: 1979e\apj 232, 282
\ref Parker, E.N.: 1982a\apj 256, 292
\ref Parker, E.N.: 1982b\apj 256, 302
\ref Parker, E.N.: 1982c\apj 256, 736
\ref Parker, E.N.: 1982d\apj 256, 746
\ref Parker, E.N.: 1982e, {\it Geophys. Astrophys. Fluid Dyn.} {\bf 22}, 195
\ref Parker, E.N.: 1983a\apj 264, 635
\ref Parker, E.N.: 1983b\apj 264, 642
\ref Parker, E.N.: 1984a\apj 283, 343
\ref Parker, E.N.: 1984b\apj 286, 666
\ref Parker, E.N.: 1985a\apj 294, 47
\ref Parker, E.N.: 1985b\apj 294, 57
\ref Parker, E.N.: 1986, in W. Deinzer et al. (eds.): {\it Small Scale Magnetic Flux Concentrations in the Solar Photophere}, Abh. Akad. d. Wiss. G\"ottingen No. 38, Vandenhoeck \& Ruprecht, G\"ottingen, p. 13
\ref Parker, E.N.: 1987a\apj 312, 868
\ref Parker, E.N.: 1987b\apj 321, 984
\ref Parker, E.N.: 1987c\apj 321, 1009
\ref Parker, E.N.: 1987d, in B.R. Durney and S. Sofia (eds.): {\it The Internal Solar Angular Velocity}, Reidel, Dordrecht, p. 289
\ref Parker, E.N.: 1988a\apj 325, 880
\ref Parker, E.N.: 1988b\apj 326, 395
\ref Parker, E.N.: 1988c\apj 326, 407
\ref Parkinson, J.H.: 1983\nat 304, 518
\ref Pidatella, R.M., Stix, M.: 1986\aap 157, 338
\ref Pneuman, G.W., Raadu, M.A.: 1972\apj 172, 739
\ref Pouquet, A.: 1985, in H.U. Schmidt (ed.): {\it Theoretical Problems in High Resolution Solar Physics}, MPA 212, Max-Planck Institut f\"ur Astrophysik, M\"unchen, p. 34
\ref Proctor, M.R.E., Weiss, N.O.: 1982, {\it Rep. Progr. Phys.}, {\bf 45}, 1317

\m

\ref {\bbrm R}\"adler, K.H.: 1968, {\it Z. Naturforsch.} {\bf 23a}, 1851 (engl. transl. in Roberts, P.H., Stix, M.: {\it The Turbulent Dynamo}, Tech. Note 60, NCAR, Boulder, Colorado, 1971)
\ref Roberts, B., Webb, A.R.: 1978\sp 56, 5
\ref Roberts, P.H., Stewartson, K.: 1977, {\it Astron. Nachr.}, {\bf 298}, 311

\m

\ref {\bbrm S}chmitt, D.: 1985, Dissertation, Universit\"at G\"ottingen
\ref Schmitt, J.H.M.M., Rosner, R.: 1983\apj 265, 901
\ref Schmitt, J.H.M.M., Rosner, R., Bohn, H.U.: 1984\apj 282, 316
\ref Schr\"oter, E.H.: 1985\sp 100, ???
\ref Sch\"ussler, M.: 1977\aap 56, 439
\ref Sch\"ussler, M.: 1979\aap 71, 79
\ref Sch\"ussler, M.: 1980a\aap 89, 26
\ref Sch\"ussler, M.: 1980b\nat 288, 150
\ref Sch\"ussler, M.: 1983, in J.O. Stenflo (ed.):  {\it Solar and Stellar Magnetic Fields: Origins and Coronal Effects}, IAU-Symp. No. 102, Reidel, Dordrecht , p. 213
\ref Sch\"ussler, M.: 1984a, in T.D. Guyenne and J.J. Hunt (eds.): {\it The Hydromagnetics of the Sun}, ESA SP-220, p. 67
\ref Sch\"ussler, M.: 1984b\aap 140, 453
\ref Sch\"ussler, M.: 1986, in W. Deinzer et al. (eds.): {\it Small Scale Magnetic Flux Concentrations in the Solar Photophere}, Abh. Akad. d. Wiss. G\"ottingen No. 38, Vandenhoeck \& Ruprecht, G\"ottingen, p. 103
\ref Sch\"ussler, M.: 1987, in B.R. Durney and S. Sofia (eds.): {\it The Internal Solar Angular Velocity}, Reidel, Dordrecht, p. 303
\ref Sch\"ussler, M.: 1990, in J.O. Stenflo (ed.): {\it Solar Photosphere: Structure, Convection and Magnetic Fields}, IAU-Symp. No. 138, Kluwer, Dordrecht, p. 161
\ref Shaviv, G., Salpeter, E.E.: 1983\apj 184, 191
\ref Smirnow, W.I.: 1968, {\it Lehrgang der H\"oheren Mathematik}, Bd. II, Akademie-Verlag, Berlin, DDR
\ref Solanki, S.K.: 1990, in J.O. Stenflo (ed.): {\it Solar Photosphere: Structure, Convection and Magnetic Fields}, IAU-Symp. No. 138, Kluwer, Dordrecht
\ref Spiegel, E.A., Weiss, N.O.: 1980, {\it Nature} {\bf 287}, 616
\ref Spitzer, L.: 1957\apj 125, 525
\ref Spruit, H.C.: 1977a\sp 55, 3
\ref Spruit, H.C.: 1977b, thesis, University of Utrecht
\ref Spruit, H.C.: 1981a\aap 98, 155
\ref Spruit, H.C.: 1981b\aap 102, 129
\ref Spruit, H.C.: 1981c, in {\it The Physics of Sunspots}, L.E. Cram, J.H. Thomas (eds.), Sacramento Peak Observatory, p. 98
\ref Spruit, H.C.: 1982\aap 108, 348
\ref Spruit, H.C., Roberts, B.: 1983, {\it Nature} {\bf 304}, 401
\ref Spruit, H.C., van Ballegooijen, A.A.: 1982\aap 106, 58
\ref Spruit, H.C., Zweibel, E.G.: 1979\sp 62, 15
\ref Spruit, H.C., Zwaan, C.: 1981\sp 70, 207
\ref Stenflo, J.O.: 1989, {\it Astron. Astrophys. Rev.}, {\bf 1}, 3
\ref Stix, M.: 1981a\sp 74, 79
\ref Stix, M.: 1981b\aap 93, 339
\ref Stix, M.: 1976, in V. Bumba, J. Kleczek (eds.): {\it Basic Mechanisms of Solar Activity}, IAU-Symp. No. 71, Reidel, Dordrecht, p. 367
\ref Stix, M.: 1982, in W. Fricke, G. Teleki (eds.): {\it Sun and Planetary System}, 6th European Regional Meeting, Reidel, Dordrecht, p. 63

\m

\ref {\bbrm T}itle, A.M., Tarbell, T.D., Topka, K.P.: 1987\apj 317, 892
\ref Tsinganos, K.C.: 1980\apj 239, 746

\m

\ref {\bbrm U}nno, W., Ando, H.: 1979, {\it Geophys. Astrophys. Fluid Dyn.} {\bf 12}, 107
\ref Unno, W., Ribes, E.: 1976\apj 208, 222

\m

\ref {\bbrm V}an Ballegooijen, A.A.: 1982a\aap 106, 43
\ref Van Ballegooijen, A.A.: 1982b\aap 113, 99
\ref Van Ballegooijen, A.A.: 1983\aap 118, 275
\ref Van Ballegooijen, A.A., Choudhuri, A.R.: 1988\apj 333, 965

\m

\ref {\bbrm W}ebb, A.R., Roberts, B.: 1978\sp 59, 249
\ref Weiss, N.O.: 1966, {\it Proc. Roy. Soc.} {\bf A293}, 310 
\ref Weiss, N.O.: 1981a, {\it J. Fluid Mech.} {\bf 108}, 247
\ref Weiss, N.O.: 1981b, {\it J. Fluid Mech.} {\bf 108}, 273
\ref Weiss, N.O.: 1981c, in C. Jordan (ed.): {\it Solar Activity}, Proc. 3rd European Solar Meeting, Oxford, p. 35
\ref Willson, R.C.: 1984, {\it Sp. Sci. Rev.} {\bf 38}, 203
\ref Wittmann, A., Bonet Navarro, J.A., W\"ohl, H.: 1981, in L.E. Cram, J.H. Thomas (eds.): {\it The Physics of Sunspots}, Sacramento Peak Observatory, p. 424

\m

\ref {\bbrm Y}oshimura, H.: 1981, in F. Moriyama, J.C. H\'enoux, Proc. of the Japan-France Seminar on Solar Physics, p. 19

\m

\ref {\bbrm Z}el'dovich, Ya.B.: 1957, {\it Sov. Phys. JETP} {\bf 4}, 460
\ref Zwaan, C.: 1978\sp 60, 213

\bye